\documentclass[twocolumn]{aastex631}
\usepackage{amsmath}
\usepackage{rotating}
\usepackage{soul} 
\usepackage[hang,flushmargin]{footmisc}
\received{July 9, 2022}
\revised{April 26, 2023}
\accepted{May 30, 2023}
\submitjournal{ApJ}

\shorttitle{V2493 Cygni}
\graphicspath{{./}{figures/}}
\DeclareOldFontCommand{\bf}{\normalfont\bfseries}{\mathbf} 
\providecommand{\DIFadd}[1]{{\bf #1}} 
\providecommand{\DIFdel}[1]{} 
\providecommand{\DIFaddbegin}{} 
\providecommand{\DIFaddend}{} 
\providecommand{\DIFdelbegin}{} 
\providecommand{\DIFdelend}{} 
\providecommand{\DIFaddFL}[1]{\DIFadd{#1}} 
\providecommand{\DIFdelFL}[1]{\DIFdel{#1}} 
\providecommand{\DIFaddbeginFL}{} 
\providecommand{\DIFaddendFL}{} 
\providecommand{\DIFdelbeginFL}{} 
\providecommand{\DIFdelendFL}{} 
\RequirePackage{listings} 
\RequirePackage{color} 
\lstdefinelanguage{DIFcode}{ 
  moredelim=[il][\color{white}\tiny]{\%DIF\ <\ }, 
  moredelim=[il][\sffamily\bfseries]{\%DIF\ >\ } 
} 
\lstdefinestyle{DIFverbatimstyle}{ 
	language=DIFcode, 
	basicstyle=\ttfamily, 
	columns=fullflexible, 
	keepspaces=true 
} 
\lstnewenvironment{DIFverbatim}{\lstset{style=DIFverbatimstyle}}{} 
\lstnewenvironment{DIFverbatim*}{\lstset{style=DIFverbatimstyle,showspaces=true}}{} 

\begin{document}

\title{Post-outburst evolution of bonafide FUor V2493 Cyg: A Spectro-photometric monitoring. }

\correspondingauthor{Arpan Ghosh}
\email{arpan@aries.res.in}

\author{Arpan Ghosh}
\affiliation{Aryabhatta Research Institute of Observational Sciences (ARIES)  Manora Peak, Nainital 263 001, India}
\affil{School of Studies in Physics and Astrophysics, Pandit Ravishankar Shukla University, Raipur 492010, Chhattisgarh, India},
\author{Saurabh Sharma}
\affil{Aryabhatta Research Institute of Observational Sciences (ARIES) Manora Peak, Nainital 263 001, India},
\author{Joe P. Ninan}
\affil{ Department of Astronomy and Astrophysics, Tata Institute of Fundamental Research (TIFR), Mumbai 400005, Maharashtra, India}
\author{Devendra K. Ojha}
\affil{ Department of Astronomy and Astrophysics, Tata Institute of Fundamental Research (TIFR), Mumbai 400005, Maharashtra, India}
\author{Bhuwan C. Bhatt}
\affil{Indian Institute of Astrophysics, II Block, Koramangala, Bangalore 560 034, India}
\author{D. K. Sahu}
\affil{Indian Institute of Astrophysics, II Block, Koramangala, Bangalore 560 034, India}
\author{Tapas Baug}
\affil{Satyendra Nath Bose National Centre for Basic Sciences (SNBNCBS), Salt Lake, Kolkata-700 106, India}
\author{R. K. Yadav}
\affil{National Astronomical Research Institute of Thailand, Chiang Mai, 50200, Thailand}
\author{Puji Irawati}
\affil{National Astronomical Research Institute of Thailand, Chiang Mai, 50200, Thailand}
\author{A. S. Gour}
\affil{School of Studies in Physics and Astrophysics, Pandit Ravishankar Shukla University, Raipur 492010, Chhattisgarh, India}
\author{Neelam Panwar}
\affil{Aryabhatta Research Institute of Observational Sciences (ARIES) Manora Peak, Nainital 263 001, India}
\author{Rakesh Pandey}
\affil{Aryabhatta Research Institute of Observational Sciences (ARIES) 
Manora Peak, Nainital 263 001, India}
\affil{School of Studies in Physics and Astrophysics, Pandit Ravishankar Shukla University, Raipur 492010, Chhattisgarh, India}
\author{Tirthendu Sinha}
\affil{Aryabhatta Research Institute of Observational Sciences (ARIES) 
Manora Peak, Nainital 263 001, India}
\affil{Kumaun University, Nainital 263001, India}
\author{Aayushi Verma}
\affil{Aryabhatta Research Institute of Observational Sciences (ARIES) 
Manora Peak, Nainital 263 001, India}




\begin{abstract}
We present here the results of eight  years of our near-simultaneous optical/near-infrared spectro-photometric monitoring of bonafide FUor candidate `V2493 Cyg' starting from 2013 September to 2021 June. 
During our optical monitoring period (between October 16, 2015 and December 30, 2019), the V2493 Cyg is slowly dimming with an average dimming rate of $\sim$26.6 $\pm$ 5.6  mmag/yr in V band. 
Our optical photometric colors show a significant reddening of the source post the second outburst pointing towards a gradual expansion of the emitting region post the second outburst. The mid infra-red colors, on the contrary, exhibits a blueing trend which can be attributed to the brightening of the disc due to the outburst. Our spectroscopic monitoring shows a dramatic variation of the H$\alpha$ line as it transitioned from
absorption feature to the emission feature and back. Such transition can possibly be explained by the variation
in the 
wind structure in combination with  
accretion. Combining our time evolution
spectra of the Ca II infra-red triplet lines with the previously published spectra of V2493 Cyg, we find that the accretion region has stabilised compared to the early days of the outburst. 
The evolution of the O I $\lambda$7773 \AA~ line also points towards the stabilization of the circumstellar disc post the second outburst.

\end{abstract}


\section{Introduction} \label{sec:intro}

Enhanced episodic accretion (or outburst) is an important phenomenon in the evolution of a low-mass star (mostly $\leq$ 1 M$_{\odot}$, and in some systems 2-3 M$_{\odot}$,
\citealt{2016ARA&A..54..135H})spanning from the embedded protostars to the main-sequence (MS) stages \citep{2014prpl.conf..387A}. The timescale of these outbursts ($\sim$months to
decades) is very short if we compare it to the millions of years spent on the formation stages or billions of years
spent on the MS stages of the low-mass stars. This makes these events extremely rare. However,
according to \citet{2013MNRAS.430.2910S}, each low-mass star is expected to 
experience statistically $\sim$50 short duration outbursts anytime 
during their early pre-main-sequence (PMS) stages, i.e., 
Class\,{\sc 0} and Class\,{\sc ii} stages  \citep[see also,][]{2015ApJ...800L...5S}.
Although, these outbursts are small in temporal scales, but are 
capable of delivering significant fraction of mass from the circumstellar disc
onto the central PMS star \citep{2006ApJ...650..956V}. 
The episodically accreting young stars are broadly 
classified into two groups, i.e,  FUors and EXors, which usually show a  
brightening of 4-5 mags (lasting for several decades) and 2-3 mags (spanning for few months to few years), respectively, in optical wavelengths. The FUors show only absorption lines in their spectra, whereas the EXors show mostly emission lines \citep{1996ARA&A..34..207H,1998apsf.book.....H}. 
Because of the rarity of the outburst events, the physical origin behind the enhanced accretion rates in the low-mass PMS stars is not fully understood. 
In recent times, various models have been proposed to explain this, e.g., thermal instability, magneto-rotational instability, combination of magneto-rotational instability and gravitational instability, disc fragmentation,  external perturbations, etc. \citep[see for details,][]{2014prpl.conf..387A}. Thus,
a continuous spectroscopic and photometric monitoring of the evolution of the outburst sources 
can provide important insights on these models on enhanced accretion in the life of low-mass PMS stars.

\begin{figure*}
\centering
\includegraphics[width=0.75\textwidth]{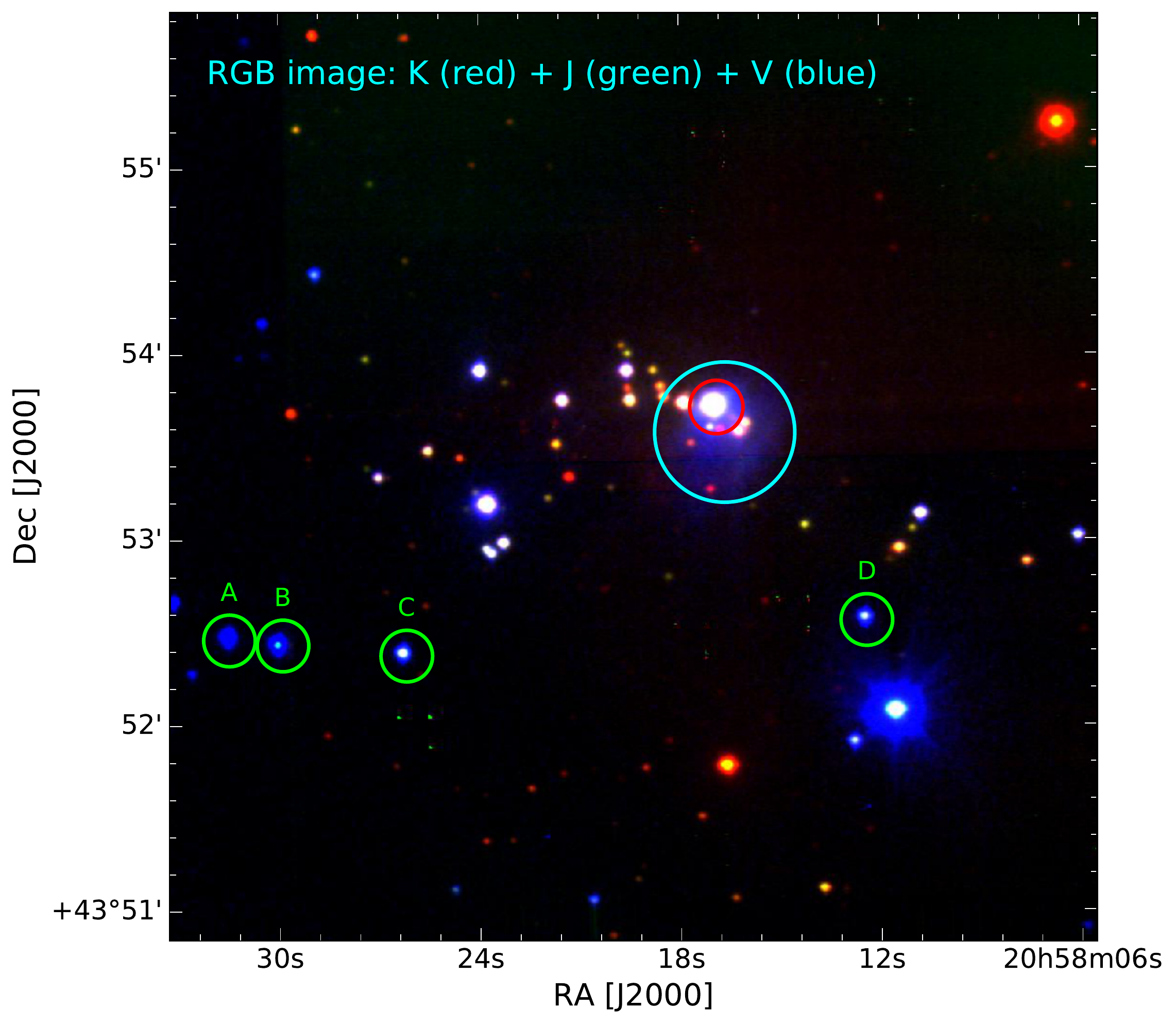}
\caption{\label{img} RGB color composite image of V2493 Cyg obtained by using the TIRSPEC $K_s$ (red), $J$ (green), and 1.3m DFOT $V$ (blue) band images.
The source V2493 Cyg is marked by a red circle. The cyan circle marks the extent of the reflection nebula that first appeared after the first outburst and has remained till date. The local standard stars that have been used for calibration of V2493 Cyg are marked by green circles.}
\end{figure*}

V2493 Cyg (V2493 Cygnus; $\alpha_{J2000}$=20$^{h}$58$^{m}$16$^{s}$.90, $\delta_{J2000}$=+43$^\circ$53$^\prime$42$^{\prime\prime}$.8; $l$=85.113248$^\circ$, $b$=-1.200676$^\circ$) is one such source, categorized as a bonafide FUor \citet{2011ApJ...730...80M}, located in the dark cloud known as
   the ``Gulf of Mexico" in the Cygnus star-forming arm of our Galaxy. The name V2493 Cyg was designated by General catalog of variable stars \citep{2011PZ.....31....2K}. It is also referred to as HBC 722 \citep{1988cels.book.....H}, LkHA 188-G4 \citep{1979ApJS...41..743C} and PTF10qpf based on the identification of the source by Palomar Transient Factory \citep{2011ApJ...730...80M}. Henceforth, we will use the name V2493 Cyg throughout this paper to designate this source.  
  The distance of this source was previously estimated as 520 $\pm$50 pc   \citep{2006BaltA..15..483L,2013ApJ...764...22G}.  
   However, from the precise parallax measurements of Gaia data release (DR) 3 \citep{2021AJ....161..147B}, this source is found to be located at a distance of 760 $\pm$ 8 pc. Henceforth,
   we will adopt the distance obtained using Gaia DR3. 
   From the shape of the spectral energy distribution (SED), \citet{2011A&A...527A.133K}  and \citet{2011ApJ...730...80M} have shown
   that the V2493 Cyg was a Class\,{\sc ii} young stellar object (YSO)
   before it transitioned to its first outburst stage. 
   V2493 Cyg experienced its  first outburst during the summer of 2010. It reached its maximum
   brightness in September 2010 
   with $\Delta$V$\sim$4.7 mag \citep{2010A&A...523L...3S}. It then
   started to fade, reaching an intermediate quiescent stage in October 
   2011 ($\Delta$V$\sim$1.45 mag from the initial outburst peak).
   Subsequently, it again started to brighten, reaching a maximum brightness
   in 2012 April with $\Delta$V$\sim$3.3 mag \citep{2012A&A...542A..43S}. 
   Post the second outburst, \citet{2013ApJ...764...22G} and
   \citet{2015AJ....149...73B} investigated for periodicity in V2493 Cyg.
   They have found periods of $\sim$ 6 and $\sim$10 days. They
   attributed these periodicities to the rotating clumps in the disc. 
   \citet{2015ApJ...807...84L} studied the evolution of V2493 Cyg using its high resolution spectra taken between 2010 December and 2014 November. They have found evidence of the wind driven by 
   the accretion in V2493 Cyg. Their monitoring also revealed an anti-correlation between the spectroscopic features originating from the disc and from the wind. They have 
   attributed this to the possible re-building of the 
   inner disc. 
  The extent of the disc at optical and near-infrared (NIR) wavelengths was found to be 39 R$_{\odot}$ and 76 R$_{\odot}$ respectively using the optical and  NIR spectra.  \citet{2016A&A...596A..52K} performed a detailed modeling of the 2010 and 2011 outburst of V2493 Cyg based on their detailed optical, NIR and millimeter monitoring and spectroscopic observations. 
  Their investigation revealed that the initial outburst of 2010 was powered by the rapid infall of piled-up material from 
  the innermost part of the disc onto the star. The 2011 outburst was believed to be due to a slower ionization front that started to expand outward. The study also demonstrated that episodic 
  accretion can be observed in young stellar objects with very low mass discs.

  Since the year 2013, no detailed investigation on V2493 Cyg has been carried out, which is still in the second outburst stage.
  Thus, a long-term monitoring of V2493 Cyg can help us characterize the complex physical processes going on in this current evolutionary stage. 
  We have monitored V2493 Cyg from 2013 September to 2021 June, both
  photometrically and spectroscopically in optical to NIR wavelengths, using different telescopes. Figure \ref{img} shows the color composite image of V2493 Cyg generated from our observations during our monitoring period. A reflection nebula appeared after the first outburst which obscured V2493 Cyg at optical wavelengths. The source is visible at NIR K$_s$ band whereas the extent of the reflection nebula is traced out by the optical V band as shown in Figure
  \ref{img}. We have combined our results with
  the previously published results. This helps us to provide further insights on the nature of these types of sources. 
   In Section \ref{pt2}, we describe the details of the observations and the techniques employed in the data
   reduction. In Section \ref{pt3}, we describe the photometric and spectrometric evolution of V2493 Cyg during our monitoring period.
   In Section \ref{pt4}, we explain the evolution of V2493 Cyg based on various physical models and conclude our study at the end.

\section{Observation and Data reduction} \label{pt2} 

\begin{table*}
\centering
\tiny
\caption{Log of Photometric and Spectroscopic Observations.}
\label{tab:obs_log}
\begin{tabular}{p{1.25in}p{.55in}p{.30in}p{.90in}p{2.70in}}
\hline 
Telescope/Instrument & Date  & JD & Filters/Grisms & Exposure(s) =  int. time (s)  $\times$ no. of frames (f) / \\
                     &       &    &                & int. time (s)  $\times$ no. of frames (f) $\times$ no. of dither positions (d)\textsuperscript{\textdagger}  \\
\hline 
2.0m HCT TIRSPEC & 2013 Sep 27 & 2,456,563 & $J,H,K_S$             &  20 s $\times$ 3 f  $\times$ 5 d,  15 s $\times$ 3 f $\times$ 5 d,  20 s $\times$ 3 f $\times$ 5 d  \\  
$"$                & 2013 Nov 14 & 2,456,611 & $H,K_S$               &  15 s  $\times$ 3 f $\times$ 5 d,  20 s $\times$ 3 f $\times$ 5 d  \\
"                & 2013 Nov 30 & 2,456,627 & $J,H,K_S$             &  20 s  $\times$ 3 f $\times$ 5 d,  15 s $\times$ 3 f $\times$ 5 d,  20 s $\times$ 3 f $\times$ 5 d   \\
"                & 2013 Dec 20 & 2,456,647 & $ K_S$                &  20 s  $\times$ 3 f $\times$ 5 d\\
"                & 2014 Mar 07 & 2,456,725 & H,K                   & 100 s  $\times$ 6 f, 100 s $\times$ 6 f \\
"                & 2014 Mar 25 & 2,456,742 & $J,H,K_S$             &  20 s  $\times$ 3 f $\times$ 5 d,  15 s $\times$ 3 f $\times$ 5 d, 20 s $\times$ 3 f $\times$ 5 d  \\
"                & 2014 May 04 & 2,456,782 & $J,H,K_S$             &  20 s  $\times$ 3 f $\times$ 5 d,  15 s $\times$ 3 f $\times$ 5 d, 20 s $\times$ 3 f $\times$ 5 d  \\
"                & 2014 May 29 & 2,456,807 & $J,H$                 &  15 s  $\times$ 3 f $\times$ 5 d,  20 s $\times$ 3 f $\times$ 5 d   \\
"                & 2014 Jun 04 & 2,456,813 & K                     & 100 s  $\times$ 6 f \\
"                & 2014 Jun 05 & 2,456,814 & $J,H,K_S$             &  20 s  $\times$ 3 f $\times$ 5 d,  15 s $\times$ 3 f $\times$ 5 d, 20 s $\times$ 3 f $\times$ 5 d  \\
"                & 2014 Jul 02 & 2,456,841 & J,H,K,$J,H,K_S$       & 100 s  $\times$ 6 f, 100 s $\times$ 6 f, 100 s $\times$ 6 f, 20 s $\times$ 3 f $\times$ 5 d, 15 s $\times$ 3 f $\times$ 5 d, 20 s $\times$ 3 f $\times$ 5 d  \\
"                & 2014 Jul 03 & 2,456,842 & $J,H,K_S$             &  20 s  $\times$ 3 f $\times$ 5 d,  15 s $\times$ 3 f $\times$ 5 d,  20 s $\times$ 3 f $\times$ 5 d  \\
"                & 2014 Aug 24 & 2,456,894 & H,K,$J$               & 100 s  $\times$ 6 f, 100 s $\times$ 6 f,  20 s $\times$ 3 f $\times$ 5 d \\
"                & 2014 Aug 25 & 2,456,895 & Y,J,$J,H,K_S$         & 100 s  $\times$ 6 f, 100 s $\times$ 6 f,  20 s $\times$ 3 f $\times$ 5 d,  15 s $\times$ 3 f $\times$ 5 d, 20 s $\times$ 3 f $\times$ 5 d  \\
"                & 2014 Sep 25 & 2,456,926 & K,$J,H,K_S$           & 100 s  $\times$ 6 f,  20 s $\times$ 3 f $\times$ 5 d,  15 s $\times$ 3 f $\times$ 5 d,  20 s $\times$ 3 f $\times$ 5 d  \\
"                & 2014 Sep 26 & 2,456,927 & Y,J,H,K               & 100 s  $\times$ 6 f, 100 s $\times$ 6 f, 100 s $\times$ 6 f, 100 s $\times$ 3 f $\times$ 2 d  \\
"                & 2014 Oct 26 & 2,456,957 & $J,H,K_S$             &  20 s  $\times$ 3 f $\times$ 5 d,  15 s $\times$ 3 f $\times$ 5 d, 20 s $\times$ 3 f $\times$ 5 d  \\
"                & 2014 Nov 21 & 2,456,983 & Y,J, $J,H,K_S$        & 100 s  $\times$ 6 f, 100 s $\times$ 6 f,  20 s $\times$ 3 f $\times$ 5 d, 15 s $\times$ 3 f $\times$ 5 d, 20 s $\times$ 3 f $\times$ 5 d \\
"                & 2014 Nov 22 & 2,456,984 & Y,J,H,K               & 100 s  $\times$ 6 f, 100 s $\times$ 6 f, 100 s $\times$ 6 f, 100 s $\times$ 6 f  \\
"                & 2014 Nov 28 & 2,456,990 & $J,H,K_S$             &  20 s  $\times$ 3 f$\times$ 5 d,  15 s $\times$ 3 f $\times$ 5 d,  20 s $\times$ 3 f $\times$ 5 d  \\
"                & 2014 Dec 25 & 2,457,017 & $J$                   &  20 s  $\times$ 3 f$\times$ 5 d\\ 
"                & 2015 Feb 12 & 2,457,066 & $J,K_S$               &  20 s  $\times$ 3 f$\times$ 5 d, 15 s $\times$ 3 f $\times$ 5 d, 20 s $\times$ 3 f $\times$ 5 d  \\
"                & 2015 Apr 14 & 2,457,127 & $K_S$                 &  20 s  $\times$ 3 f$\times$ 5 d\\  
"                & 2015 Apr 27 & 2,457,141 & Y,J,H,K               & 100 s  $\times$ 6 f, 100 s $\times$ 6 f, 100  s $\times$ 6 f, 100 $\times$  6 f \\
TIRSPEC+HFOSC    & 2015 May 16 & 2,457,159 & $V,R,J,H,K_S$         &  60 s  $\times$ 3 f,  30 s $\times$ 3 f,   20 s $\times$ 3 f $\times$ 5 d, 15 s $\times$ 3 f $\times$ 5 d, 20 s $\times$ 3 f $\times$ 5 d  \\
"                & 2015 May 26 & 2,457,170 & Y,J,H,K               & 100 s  $\times$ 6 f, 100 s $\times$ 6 f, 100  s $\times$ 6 f, 100 s $\times$ 6 f \\         
HFOSC            & 2015 Jun 10 & 2,457,184 & $V,R,I_c$             &  60 s  $\times$ 3 f, 30 s  $\times$ 3 f, 30 s  $\times$ 3 f      \\
TIRSPEC          & 2015 Jun 23 & 2,457,197 & J,H,K                 & 100 s  $\times$ 6 f, 100 s $\times$ 6 f, 100 s $\times$ 6 f \\ 
"                & 2015 Jul 27 & 2,457,231 & J,H,K $J,H,K_S$       & 100 s  $\times$ 6 f, 100 s $\times$ 6 f, 100 s $\times$ 6 f,  20 s  $\times$ 3 f $\times$ 5 d, 15 s $\times$ 3 f $\times$ 5 d, 20 s $\times$ 3 f $\times$ 5 d \\
"                & 2015 Aug 17 & 2,457,252 & Y,J,H,K,$J,H,K_S$     & 100  s $\times$ 6 f , 100 s $\times$ 6 f, 100 s $\times$ 6 f, 100 s $\times$ 6 f, 20 s $\times$ 3 f $\times$ 5 d, 15 s $\times$ 3 f $\times$ 5 d, 20 s $\times$ 3 f $\times$ 5 d   \\
"                & 2015 Oct 13 & 2,457,309 & Y,J,$J,H,K_S$         & 100  s $\times$ 6 f , 100 s $\times$ 6 f,  20 s $\times$ 3 f $\times$ 5 d, 15 s $\times$ 3 f $\times$ 5 d, 20 s $\times$ 3 f $\times$ 5 d   \\
HFOSC            & 2015 May 17 & 2,457,160 & $Gr7,Gr8$             & 1800 s , 1800 s     \\
"                & 2015 May 26 & 2,457,169 & $Gr8$                 & 1200 s \\
"                & 2015 Jun 10 & 2,457,184 &$Gr7,Gr8$              & 1200 s , 1200 s \\
"                & 2015 Aug 18 & 2,457,253 &$V,R,Gr7,Gr8$          & 60 s $\times$ 3 f, 30 s $\times$ 3 f, 1200 s, 1200 s \\
"                & 2015 Sep 27 & 2,457,293 &$V,R,Gr7,Gr8$          & 60 s $\times$ 3 f, 30 s $\times$ 3 f, 1200 s, 1200 s \\
"                & 2015 Oct 12 & 2,457,308 &$V,R,Gr7,Gr8$          & 60 s $\times$ 3 f, 30 s $\times$ 3 f, 1200 s, 1200 s \\
1.3m DFOT ANDOR $512\times512$ & 2015 Oct 16 & 2,457,312 & $V,R_c,I_c$   & 60 s $\times$ 3 f, 30 s $\times$ 3 f, 30 s $\times$ 3 f \\
"                & 2015 Nov 02 & 2,457,329 & $R_c,I_c$             &       30 s $\times$ 3 f, 30 s $\times$ 3 f  \\
HFOSC                 & 2015 Nov 10 & 2,457,337 & $V,R,Gr7$        & 60 s $\times$ 3 f, 30 s $\times$ 3 f, 1200 s \\
ANDOR $512\times512$  & 2015 Nov 18 & 2,457,345 & $V,R_c,I_c$      & 60 s $\times$ 3f, 30 s $\times$ 3f, 30 s $\times$ 3 f  \\
"                & 2016 Jan 01 & 2,457,390 & $V,R_c,I_c$           & 60 s $\times$ 3f,   30 s $\times$ 3f, 30 s $\times$ 3 f  \\
HFOSC            & 2016 Jan 20 & 2,457,409 & $Gr7,Gr8$             & 1200 s,1200 s  \\
"                & 2016 Mar 03 & 2,457,451 & $V,R$                 & 60 s $\times$ 3 f, 30 s $\times$ 3 f   \\
"                & 2016 Mar 15 & 2,457,463 & $V,R,Gr7,Gr8$         & 60 s $\times$ 3 f, 30 s $\times$ 3 f,  900 s, 900 s \\ 
"                & 2016 Jun 17 & 2,457,557 & $V,R,,Gr7,Gr8$        & 60 s $\times$ 3 f, 30 s $\times$ 3 f, 1200 s,1200 s \\
"                & 2016 Jul 24 & 2,457,594 & $V,R,Gr7,Gr8$         & 60 s $\times$ 3 f, 30 s $\times$ 3 f, 1800 s,1200 s \\
"                & 2016 Sep 26 & 2,457,658 & $V,R,Gr7,Gr8$         & 60 s $\times$ 3 f, 30 s $\times$ 3 f, 1800 s,1800 s \\
"                & 2016 Oct 05 & 2,457,667 & $V,R,Gr8$             & 60 s $\times$ 3 f, 30 s $\times$ 3 f,  700 s\\
ANDOR $512\times512$  & 2016 Oct 08 & 2,457,670 & $V,R_c,I_c$      & 60 s $\times$ 3 f, 30 s $\times$ 3 f, 30 s $\times$ 3 f  \\
HFOSC                 & 2016 Oct 13 & 2,457,675 & $V,R,I_c,Gr7,Gr8$  & 60 s $\times$ 3 f, 30 s $\times$ 3 f, 30 s $\times$ 3 f, 1800 s, 1200 s  \\
ANDOR $512\times512$  & 2016 Nov 15 & 2,457,708 & $V,R_c,I_c$        &   60 s $\times$ 3 f, 30 s $\times$ 3 f, 30 s $\times$ 3 f \\
HFOSC                 & 2016 Nov 17 & 2,457,710 & $V,R,I_c,Gr7,Gr8 $ &  60 s $\times$ 3 f, 30 s $\times$ 3 f, 30 s $\times$ 3 f, 1200 s, 1200 s \\
"                 & 2016 Dec 04 & 2,457,727 &     $V,R,I_c,Gr7,Gr8$   & 60 s $\times$ 3 f, 30 s $\times$ 3 f, 30 s $\times$ 3 f, 1200 s, 1200 s \\
"                 & 2016 Dec 08 & 2,457,731 & H,K,$Gr7,Gr8$ &  100 s $\times$ 6 f, 100 s $\times$ 6 f, 1800 s, 1800 s  \\
"                 & 2017 Apr 25 & 2,457,869 & $Gr7,Gr8 $    & 1800 s , 1800     s \\
"                 & 2017 Apr 28 & 2,457,872 & $Gr7$         & 1800 s \\
TIRSPEC               & 2017 Jul 02 & 2,457,937 & Y,J,H,K   &  100 s $\times$ 6 f , 100 s $\times$ 6 f, 100 s $\times$ 6 f, 100 $\times$ 3 f  \\ 
HFOSC                 & 2017 Jul 03 & 2,457,938 & $Gr8$         & 1800 s  \\
TIRSPEC               & 2017 Aug 20 & 2,457,983 & $J,H,K_S$     &   20 s  $\times$ 3 f $\times$ 5 d,  15 s $\times$ 3 f $\times$ 5 d,  20 s $\times$ 3 f $\times$ 5 d \\
HFOSC                 & 2017 Aug 22 & 2,457,985 & $Gr8$         & 1800 s  \\
TIRSPEC               & 2017 Sep 14 & 2,458,011 & Y,J,H,K,$J,H,K_S$ & 100 s $\times$ 6 f, 100 s $\times$ 6 f, 100 s $\times$ 6 f, 100 s $\times$ 6 f, 20 s $\times$ 3 f $\times$ 5 d, 15 s $\times$ 3 f $\times$ 5 d, 20 s $\times$ 3 f $\times$ 5 d  \\
HFOSC                 & 2017 Sep 17 & 2,458,014 &  $Gr7,Gr8$   & 1800 s, 1800 s  \\  
ANDOR $512\times512$ & 2017 Oct 08 & 2,458,035 & $V,R_c,I_c$ &  60 s $\times$ 3 f, 30 s $\times$ 3 f, 30 s $\times$ 3 f  \\
1.3m DFOT ANDOR $2k\times2k$   & 2017 Oct 13 & 2,458,040 & $V,R_c,I_c$ &  60 s $\times$ 3 f, 30 s $\times$ 3 f, 30 s $\times$ 3 f \\
"   & 2017 Nov 11 & 2,458,069 & $V,R_c,I_c$ &   60 s $\times$ 3 f, 30 s $\times$ 3 f, 30 s $\times$ 3 f \\
"   & 2017 Dec 10 & 2,458,098 & $V,R_c,I_c$ &   60 s $\times$ 3 f, 30 s $\times$ 3 f, 30 s $\times$ 3 f \\
HFOSC                 & 2018 Aug 13 & 2,458,344 & $V,R,I_c,Gr7,Gr8$    & 60 s $\times$ 3 f, 30 s $\times$ 3 f, 30 s $\times$ 3 f, 1800 s, 2700 s \\
ANDOR $2k\times2k$    & 2018 Oct 06 & 2,458,398 & $V,R_c,I_c$ & 60 s $\times$ 3 f, 30 s $\times$ 3 f, 30 s $\times$ 3 f  \\
"                     & 2018 Oct 18 & 2,458,410 & $V,R_c,I_c$ &   60 s $\times$ 3 f, 30 s $\times$ 3 f, 30 s $\times$ 3 f  \\
HFOSC                 & 2018 Nov 06 & 2,458,429 & $Gr7,Gr8$   & 1200 s, 1200 s \\
TIRSPEC               & 2018 Nov 07 & 2,458,429 & Y,J,H,K     &  100 s $\times$ 6 f, 100 s $\times$ 6 f, 100 s $\times$ 6 f, 100 s $\times$ 6 f  \\
HFOSC                 & 2018 Nov 21 & 2,458,444 & $Gr7,Gr8$   & 1800 s, 1200 s \\
ANDOR $512\times512$  & 2018 Nov 25 & 2,458,448 & $V,R_c,I_c$ &   60 s $\times$ 3 f, 30 $\times$ 3 f, 30 $\times$ 3f  \\
ANDOR $2k\times2k$    & 2018 Nov 26 & 2,458,449 & $V,R_c,I_c$ &   60 s $\times$ 3 f, 30 $\times$ 3 f, 30 $\times$ 3f  \\ 
\hline
\textsuperscript{\textdagger} For the NIR photometric observations
\end{tabular}
\end{table*}

\setcounter{table}{0}
\begin{table*}
\centering
\tiny
\caption{Contd.}
\begin{tabular}{@{}rrrrr@{}}
\hline
Telescope/Instrument & Date  & JD & Filters/Grisms & Exposure(s) =  int. time (s)  $\times$ no. of frames (f) / \\
                     &       &    &                & int. time (s)  $\times$ no. of frames (f) $\times$ no. of dither positions (d)\textsuperscript{\textdagger} \\
\hline

ANDOR $2k\times2k$           & 2018 Dec 03 & 2,458,456 & $V,R_c,I_c$ &  $60\times3,30\times3,30\times3$ 60 s $\times$ 3 f, 30 $\times$ 3 f, 30 $\times$ 3f  \\
HFOSC                        & 2018 Dec 12 & 2,458,465 & $Gr7,Gr8$   & 1500 s, 900 s \\ 
ANDOR $512\times512$         & 2018 Dec 20 & 2,458,473 & $V,R_c,I_c$ &   60 s $\times$ 3 f, 30 s $\times$ 3 f, 30 $\times$ f  \\
TIRSPEC                      & 2018 Dec 24 & 2,458,477 & $J,H,K_S$   &   20 s $\times$ 3 f $\times$ 5 d, 15 s $\times$ 3 f $\times$ 5 d, 20 s $\times$ 3 f $\times$ 5 d  \\
HFOSC                        & 2018 Dec 25 & 2,458,478 & $Gr7,Gr8$   & 1500 s, 900 s \\
2.4m TNT MRES                & 2019 Apr 28 & 2,458,602 & Echelle Spectra & 3600 s \\
ANDOR $512\times512$         & 2019 May 20 & 2,458,624 &$V,R_c,I_c$ &$60\times3,30\times3,30\times3$ \\
TIRSPEC                      & 2019 Jun 14 & 2,458,649 & $J,H,K_S$ & 20 s $\times$3 f $\times$ 5 d, 15 s $\times$3 f $\times$ 5 d, 20 s $\times$ 3 f $\times$ 5 d  \\
"                            & 2019 Oct 12 & 2,458,769 & $J,H,K_S$ & 20 s $\times$3 f $\times$ 5 d, 15 s $\times$3 f $\times$ 5 d, 20 s $\times$ 3 f $\times$ 5 d  \\
HFOSC                        & 2019 Oct 28 & 2,458,785 &$Gr8 $ & 1800 s \\
ANDOR $2k\times2k$           & 2019 Nov 04 & 2,458,792 &$V,R_c,I_c$ & 60 s$\times$3 f, 30 s $\times$ 3 f, 30 s$\times$3 f \\
"                            & 2019 Nov 05 & 2,458,793 &$R_c,I_c$   & 30 s $\times$ 3 f, 30 s $\times$3 f \\
"                            & 2019 Nov 06 & 2,458,794 &$V,R_c,I_c$ & 60 s $\times$ 3 f, 30 s $\times$3 f, 30 s $\times$ 3 f \\
"                            & 2019 Nov 25 & 2,458,813 &$V,R_c,I_c$ & 60 s $\times$ 3 f, 30 s $\times$3 f, 30 s $\times$ 3 f \\
"                            & 2019 Nov 29 & 2,458,817 &$V,R_c,I_c$ & 60 s $\times$ 3 f, 30 s $\times$3 f, 30 s $\times$ 3 f \\
"                            & 2019 Dec 08 & 2,458,826 &$V,r_c,I_c$ & 60 s $\times$ 3 f, 30 s $\times$3 f, 30 s $\times$ 3 f \\
"                            & 2019 Dec 16 & 2,458,834 &$V,R_c,I_c$ & 60 s $\times$ 3 f, 30 s $\times$3 f, 30 s $\times$ 3 f \\
"                            & 2019 Dec 30 & 2,458,848 &$V,R_c,I_c$ & 60 s $\times$ 3 f, 30 s $\times$3 f, 30 s $\times$ 3 f \\
3.6m DOT TANSPEC             & 2020 Oct 25 & 2,459,148 & Cross-dispersed spectra & 120 s $\times$5 f \\
"                            & 2020 Nov 10 & 2,459,164 & Cross-dispersed spectra & 120 s $\times$5 f \\ 
HFOSC                        & 2021 Jun 14 & 2,459,380 &$Gr7$,$Gr8$ & 1800 s, 1800 s \\
\hline
\textsuperscript{\textdagger} For the NIR photometric observations
\end{tabular}
\end{table*}

\subsection{Photometric data}
\subsubsection{Present data}\label{Phot_data}
We have observed V2493 Cyg in optical broadband $V$, {\bf $R_C$,} and $I_C$ Johnson-Cousins filters
using the ANDOR $512\times512$ and ANDOR $2K\times2K$ CCD cameras on the 1.3m Devasthal Fast Optical Telescope (DFOT) of Aryabhatta Research Institute of Observational Sciences (ARIES){\footnote{\url{https://www.aries.res.in/~1.3m/imager.html}}} and the Hanle Faint Object Spectrograph Camera (HFOSC) of 2m Himalayan
{\it Chandra} Telescope
(HCT){\footnote{\url{https://www.iiap.res.in/iao/hfosc.html}}} at Hanle, India. In the NIR wavelengths, we have observed V2493 Cyg in $J,H,$ and $K_S$ bands using the TIFR Near Infrared Spectrometer and Imager (TIRSPEC) mounted on 
HCT \citep{2014JAI.....350006N}. Table \ref{tab:obs_log} provides the complete log of observations. We have photometrically observed V2493 Cyg at 70 different epochs between 2013 and 2019 with 26 epochs using the 1.3m DFOT (optical bands), 16 epochs using HFOSC (optical bands) and 28 epochs using the TIRSPEC on 2m HCT (NIR bands).

For the image processing, we have used the standard procedures as outlined
in \citet[][]{2020MNRAS.498.2309S}.   
As V2493 Cyg is surrounded by a reflection nebula, Point Spread Function (PSF) photometry does not fully remove the inter-night nebular contamination. 
Therefore, we have followed the approach of \citet{2013ApJ...778..116N}  by which they did photometry of another similar FUor/EXor source `V1647 Ori' which is also surrounded by a nebula. In 
the case of V1647 Ori, it was found out that there is a strong correlation between fluctuation in magnitude with that of atmospheric seeing. This resulted from the contamination of the nebular 
flux into the aperture of V1647 Ori which is a  function of atmospheric seeing.  
Briefly, we have generated a set of images for each night by convolving each frame with a two-dimensional Gaussian kernel of different standard deviations to simulate different atmospheric seeing conditions. Magnitudes of V2493 Cyg were then calculated by performing differential photometry on the set of images for each night interpolated to a seeing of 1.5 arcsec  (median seeing during our observations).

Previously, \citet{2010A&A...523L...3S} have listed 7 stars as local standards
near V2493 Cyg. Out of them, we have selected 4 stars (2 stars in the NIR wavelengths) for our analysis, that are
located in the common field of view (FOV) of the detectors used. We have also observed the Landolt standard field \citep{1992AJ....104..340L} SA 98 ($\alpha_{J2000}$: 06$^{h}$52$^{m}$14$^{s}$, $\delta_{J2000}$: -00$\degr$18$\arcmin$59$\arcsec$) on the night of  2019 November 4. This is done to calculate the present $V$, $R_C$, and $I_C$ magnitudes of the selected local standards. We have also checked the long term behaviour of the selected local standards in 
ASAS-SN\footnote{\url{https://www.astronomy.ohio-state.edu/asassn/index.shtml}} survey to verify the stability of magnitudes of our selected local standards. The calibration equations derived
from the Landolt standard stars by the least-squares linear regression are as follows:

\begin{equation}
\begin{split}
    V &= v + (2.46\pm0.01) + (0.04\pm0.01)\times(v - i_c)\\ &+ 0.19\pm0.01\times X_V
\end{split}
\end{equation}

\begin{equation}
\begin{split}
    R_C &= r_c + (2.06\pm0.01) + (0.05\pm0.01)\times(v - r_c)\\ &+ 0.17\pm0.01\times X_R
\end{split}
\end{equation}
    .

\begin{equation}
\begin{split}
    I_C &= i_c + (2.59\pm0.01) + (-0.03\pm0.01)\times(v - i_c)\\ &+ 0.11\pm0.01\times X_I
\end{split}
\end{equation}

where, $v$, $r_c$ and $i_c$ are the instrumental magnitudes and $V$, $R_C$ and $I_C$ are the
standard magnitudes. X is the airmass term in the corresponding filters. In the NIR wavelengths, we have calibrated the local standard stars using the color transformation equations provided by \citet{2014JAI.....350006N}. The rest of the procedures to calibrate V2493 Cyg is similar to that has been described in \citet{2022ApJ...926...68G}. Table \ref{tab:std_log} lists the coordinates and V, R$_C$, I$_C$,
J, H and K$_S$ magnitudes of the local standard stars used in the present study. Table \ref{tab:phot_tab} provides the calibrated $V$, $R_C$, $I_C$, J, H and K$_S$ magnitudes of V2493 Cyg at 54 epochs starting from 2013 September upto 2019 December. 

\begin{table*}
\centering
\tiny
\caption{Coordinates and magnitudes of the local standard stars.}
\label{tab:std_log}
\begin{tabular}{@{}rrrrrrrrr@{}}
\hline
Names  & $\alpha$$_{2000}$  & $\delta$$_{2000}$ & V$\pm$ $\sigma$ & R$_C \pm$ $\sigma$ & I$_C$ $\pm$ $\sigma$  & J$\pm$ $\sigma$ & H$\pm$ $\sigma$  & K$_S$ $\pm$ $\sigma$\\
   & (degrees)          &  (degrees)        & (mag)           & (mag)           & (mag)                 &  (mag)          & (mag)            & (mag)\\
\hline
2MASS J20583004+4352257   &  314.625108        & +43.874258        & 15.02$\pm$0.01  & 14.10$\pm$0.01  & 13.20$\pm$0.01        &  $-$            &  $-$             &  $-$\\
2MASS J20583004+4352257   &  314.631458        & +43.874539        & 15.45$\pm$0.01  & 14.79$\pm$0.01  & 14.15$\pm$0.01        &  13.47$\pm$0.01 & 12.97$\pm$0.01   &  12.87$\pm$0.01\\
USNO-B1.0 1338-00391522   &  314.610204        & +43.873175        & 15.87$\pm$0.01  & 14.90$\pm$0.01  & 13.92$\pm$0.01        &  12.99$\pm$0.01 & 12.40$\pm$0.01   &  12.13$\pm$0.01\\
2MASS J20581249+4352347   &  314.552492        & +43.876633        & 15.92$\pm$0.01  & 15.12$\pm$0.01  & 14.48$\pm$0.01        &  13.70$\pm$0.01 & 13.10$\pm$0.01   &  12.96$\pm$0.01\\
\hline
\end{tabular}
\end{table*}

\subsection{Archival data}

We have obtained $B$, $V$, $R$ and $I$ photometric data provided by \citet{2010A&A...523L...3S} in their study.
Photometric data in the above bands are also downloaded from the archive of American Association
for the Variable Star Observers (AAVSO)\footnote{\url{https://www.aavso.org/}}. 

    We have also acquired the photometric data provided by the Palomar Transient Factory (PTF) in their $g$ and $R$ bands\footnote{\url{https://www.ptf.caltech.edu/}}. The details of PTF survey are provided in
    \citet{2009PASP..121.1395L}. 

    Multi-epoch photometric data of V2493 Cyg is also available in the Zwicky Transient Factory (ZTF) archive. We have downloaded the ZTF DR6  $zg$ and $zr$ band data from
    the NASA/IPAC Infrared Science Archive\footnote{\url{https://irsa.ipac.caltech.edu/Missions/ztf.html}}.
    The details of the ZTF is available in \citet{Bellm_2018}.

    Archival mid-infrared (MIR) data in W1 (3.4 $\mu$m) and W2 (4.6 $\mu$m) channels from the Near-Earth Object Wide-field Infrared Survey Explorer (NEOWISE) survey is obtained from the NASA/IPAC Infrared
    Science archive\footnote{\url{https://irsa.ipac.caltech.edu/Missions/wise.html}}. The details of the NEOWISE survey are available at \citet{2014ApJ...792...30M}

\begin{table*}
\centering
\DIFaddbeginFL \tiny
\DIFaddendFL \scriptsize
\caption{Photometric magnitudes of V2493 Cyg in different filters using the present observations.}
\label{tab:phot_tab}
\begin{tabular}{@{}rrrrrrr@{}}
\hline
Julian Day      & V                 & R                & I$_C$               & J               & H               & K$_S$     \\
                & (mag)             & (mag)            & (mag)               & (mag)           & (mag)           & (mag)     \\
\hline  
2456563         &  $-$              & $-$              & $-$                 & 9.35$\pm$0.01  & 8.37$\pm$0.01  & 8.04$\pm$0.01 \\ 
2456611         &  $-$              & $-$              & $-$                 & $-$            & 8.52$\pm$0.01  & 7.97$\pm$0.01  \\       
2456627         &  $-$              & $-$              & $-$                 & 9.26$\pm$0.01  & 8.45$\pm$0.01  & 7.94$\pm$0.01  \\
2456647         &  $-$              & $-$              & $-$                 & $-$             & $-$           & 7.94$\pm$0.01  \\   
2456742         &  $-$              & $-$              & $-$                 & 9.34$\pm$0.01  & 8.50$\pm$0.01  & 8.03$\pm$0.01  \\ 
2456782         &  $-$              & $-$              & $-$                 & 9.36$\pm$0.01  & 8.57$\pm$0.01  & 7.96$\pm$0.01  \\
2456807         &  $-$              & $-$              & $-$                 & 9.33$\pm$0.01  & 8.49$\pm$0.01  & $-$             \\      
2456814         &  $-$              & $-$              & $-$                 & 9.36$\pm$0.01  & 8.52$\pm$0.01  & 8.13$\pm$0.02  \\
2456841         &  $-$              & $-$              & $-$                 & 9.36$\pm$0.01  & 8.49$\pm$0.01  & 7.96$\pm$0.01  \\  
2456842         &  $-$              & $-$              & $-$                 & 9.32$\pm$0.01  & 8.45$\pm$0.01  & 7.96$\pm$0.01  \\          
2456894         &  $-$              & $-$              & $-$                 & 9.30$\pm$0.01  & $-$             & $-$              \\     
2456895         &  $-$              & $-$              & $-$                 & 9.31$\pm$0.01  & 8.48$\pm$0.01  & 7.94$\pm$0.01   \\              
2456926         &  $-$              & $-$              & $-$                 & 9.24$\pm$0.01  & 8.43$\pm$0.01  & 7.86$\pm$0.01    \\                 
2456957         &  $-$              & $-$              & $-$                 & 9.24$\pm$0.01  & 8.30$\pm$0.01  & 7.80$\pm$0.01  \\                  
2456983         &  $-$              & $-$              & $-$                 & 9.22$\pm$0.01  & 8.45$\pm$0.01  & 7.89$\pm$0.01  \\                    
2456990         &  $-$              & $-$              & $-$                 & 9.31$\pm$0.01  & 8.48$\pm$0.01  & 7.92$\pm$0.01  \\                
2457017         &  $-$              & $-$              & $-$                 & 9.34$\pm$0.01  & $-$            & $-$             \\                    
2457066         &  $-$              & $-$              & $-$                 & 9.23$\pm$0.01  & $-$            & 7.89$\pm$0.02  \\                 
2457127         &  $-$              & $-$              & $-$                 & $-$            & $-$            & 7.89$\pm$0.02  \\                 
2457159         &  13.43$\pm$0.01   & 12.25$\pm$0.01   & $-$                 & 9.29$\pm$0.01  & 8.42$\pm$0.01  & 7.75$\pm$0.01  \\             
2457184         &  13.41$\pm$0.01   & 12.29$\pm$0.01   & 11.10$\pm$0.01      & $-$            & $-$            & $-$         \\    
2457231         &  $-$              & $-$              & $-$                 & 9.39$\pm$0.01  & 8.35$\pm$0.01  & 7.81$\pm$0.01  \\            
2457252         &  $-$              & $-$              & $-$                 & 9.33$\pm$0.01  & 8.40$\pm$0.01  & 7.87$\pm$0.01  \\             
2457253         &  13.29$\pm$0.03   & 12.16$\pm$0.01   & $-$                 & $-$            & $-$            & $-$           \\ 
2457293         &  13.39$\pm$0.01   & 12.34$\pm$0.01   & $-$                 & $-$            & $-$            & $-$           \\ 
2457308         & 13.42$\pm$0.02    & $-$              & $-$                 & $-$            & $-$            & $-$  \\ 
2457309         &  $-$              & $-$              & $-$                 & 9.25$\pm$0.01  & 8.48$\pm$0.01  & 7.86$\pm$0.01  \\     
2457312         &  13.50$\pm$0.01   & 12.45$\pm$0.01   & 11.44$\pm$0.02      & $-$             & $-$             &  $-$             \\
2457329         &  $-$              & 12.29$\pm$0.01   & 11.17$\pm$0.01      & $-$             & $-$             &  $-$             \\
2457337         &  13.36$\pm$0.02   & 12.22$\pm$0.02   & $-$                 & $-$             & $-$             &  $-$             \\  
2457345         &  13.53$\pm$0.01   & 12.34$\pm$0.01   &   11.13$\pm$0.01    & $-$             & $-$             &  $-$             \\
2457390         &  13.57$\pm$0.01   & 12.39$\pm$0.01   &   11.12$\pm$0.01    & $-$             & $-$             &  $-$             \\
2457451         &  13.34$\pm$0.01   & 12.22$\pm$0.03   &  $-$                & $-$             & $-$             &  $-$             \\ 
2457463         &  13.40$\pm$0.01   & 12.34$\pm$0.01   &  11.15$\pm$0.01     & $-$             & $-$             &  $-$             \\  
2457557         &  13.25$\pm$0.01   & 12.24$\pm$0.01   &  $-$                & $-$             & $-$             &  $-$             \\ 
2457594         &  13.42$\pm$0.01   & 12.27$\pm$0.01   &  $-$                & $-$             & $-$             &  $-$             \\ 
2457658         &  13.31$\pm$0.01   & 12.25$\pm$0.01   &  $-$                & $-$             & $-$             &  $-$             \\   
2457667         &  13.43$\pm$0.01   & 12.23$\pm$0.01   &  $-$                & $-$             & $-$             &  $-$             \\   
2457660         &  13.66$\pm$0.01   & 12.51$\pm$0.01   &   11.27$\pm$0.01    & $-$             & $-$             &  $-$             \\
2457670         &  13.61$\pm$0.01   & 12.45$\pm$0.01   &   11.24$\pm$0.01    & $-$             & $-$             &  $-$             \\
2457675         &  13.55$\pm$0.01   & 12.37$\pm$0.01   &   11.15$\pm$0.01    & $-$             & $-$             &  $-$             \\   
2457708         &  13.62$\pm$0.01   & 12.49$\pm$0.01   &   11.25$\pm$0.01    & $-$             & $-$             &  $-$             \\
2457710         &  13.63$\pm$0.01   & 12.37$\pm$0.01   &   11.18$\pm$0.01    & $-$             & $-$             &  $-$              \\
2457727         &  13.41$\pm$0.01  & 12.27$\pm$0.01  &   11.12$\pm$0.01  & $-$                 & $-$             &  $-$             \\
2457976         &  13.65$\pm$0.02  & 12.40$\pm$0.01  &   11.16$\pm$0.01   & $-$             & $-$             &  $-$             \\
2457986         &  $-$             & $-$              &   $-$             & 9.24$\pm$0.01   & 8.42$\pm$0.01   &  7.74$\pm$0.01  \\ 
2458011         &  $-$             & $-$              &   $-$             & 9.22$\pm$0.01   & 8.46$\pm$0.01   &  7.80$\pm$0.01  \\
2458344         &  13.73$\pm$0.01  & 12.77$\pm$0.01   &  11.14$\pm$0.01   & $-$             & $-$             &  $-$             \\      
2458448         &  13.57$\pm$0.01  & 12.26$\pm$0.01  &   11.10$\pm$0.01   & $-$             & $-$             &  $-$             \\
2458473         &  13.38$\pm$0.02  & 12.28$\pm$0.01  &   11.15$\pm$0.01   & $-$             & $-$             &  $-$             \\
2458477         &  $-$             & $-$              &   $-$             & 9.16$\pm$0.01   & 8.42$\pm$0.004  &  7.76$\pm$0.01  \\
2458624         &  13.73$\pm$0.01  & 12.49$\pm$0.01  &   11.23$\pm$0.01   & $-$             & $-$             &  $-$             \\
2458649         &  $-$             & $-$              &   $-$             & 9.26$\pm$0.01   & 8.37$\pm$0.01   & 7.63$\pm$0.02\\
2458040         &  13.62$\pm$0.01  & 12.46$\pm$0.01  &   11.19$\pm$0.01   & $-$             & $-$             &  $-$             \\
2458069         &  13.58$\pm$0.01  & 12.43$\pm$0.01  &   11.18$\pm$0.01   & $-$             & $-$             &  $-$             \\
2458098         &  13.54$\pm$0.01  & 12.36$\pm$0.01  &   11.12$\pm$0.01   & $-$             & $-$             &  $-$             \\
2458398         &  13.48$\pm$0.01  & 12.35$\pm$0.01  &   11.09$\pm$0.01   & $-$             & $-$             &  $-$             \\
2458410         &  13.49$\pm$0.01  & 12.38$\pm$0.01  &   11.09$\pm$0.01   & $-$             & $-$             &  $-$             \\
2458449         &  13.54$\pm$0.01  & 12.39$\pm$0.01  &   11.12$\pm$0.01   & $-$             & $-$             &  $-$             \\
2458456         &  13.48$\pm$0.01  & 12.34$\pm$0.01  &   11.08$\pm$0.01   & $-$             & $-$             &  $-$             \\
2458769         &  $-$             & $-$             &   $-$              & 9.37$\pm$0.01   & 8.58$\pm$0.01   &  7.95$\pm$0.02   \\
2458792         &  13.76$\pm$0.01  & 12.56$\pm$0.01  &   11.23$\pm$0.01   & $-$             & $-$             &  $-$             \\
2458793         &  $-$             & 12.56$\pm$0.01  &   11.22$\pm$0.01   & $-$             & $-$             &  $-$             \\
2458794         &  13.81$\pm$0.01  & 12.61$\pm$0.01  &   11.27$\pm$0.01   & $-$             & $-$             &  $-$             \\
2458813         &  13.81$\pm$0.01  & 12.59$\pm$0.01  &   11.27$\pm$0.01   & $-$             & $-$             &  $-$             \\
2458817         &  13.70$\pm$0.01  & 12.48$\pm$0.01  &   11.17$\pm$0.01   & $-$             & $-$             &  $-$             \\
2458826         &  13.71$\pm$0.01  & 12.50$\pm$0.01  &   11.18$\pm$0.01   & $-$             & $-$             &  $-$             \\
2458834         &  13.92$\pm$0.01  & 12.76$\pm$0.01  &   11.43$\pm$0.01   & $-$             & $-$             &  $-$             \\
2458848         &  13.74$\pm$0.01  & 12.53$\pm$0.01  &   11.21$\pm$0.01   & $-$             & $-$             &  $-$             \\

\end{tabular}
\end{table*}

\subsection{Spectroscopic data} 

\subsubsection{Medium Resolution single order Optical Spectroscopy}\label{pt6}

V2493 Cyg is monitored spectroscopically starting from 2015 May 17 to 2021 June 14 at 28 different epochs using the 
HFOSC. 
HFOSC provides a medium resolution spectra with R$\sim$2000 with 1 arcsec slit
from $\sim$4000\AA~ to 9000\AA~  using grisms Gr 7 and Gr 8. 

Standard IRAF\footnote{IRAF is distributed by National Optical
Astronomy Observatories, USA which is operated by the Association of
Universities for Research in Astronomy, Inc., under cooperative
agreement with National Science Foundation for performing image
processing.} tasks are used for the spectroscopic data reduction.
The task {\sc apall} is used to extract the
spectrum in one dimensional format. The extracted spectrum is then
wavelength calibrated using the task {\sc identify} with the help of
calibration lamps (FeAr and FeNe)
taken immediately after the source spectrum. We have followed this
with the task {\sc continuum} to
continuum normalize our spectra. We have used this normalized
spectra to measure the equivalent widths (W$_{\lambda}$) of important
spectral lines by using the {\sc splot} package of IRAF. Three independent measurements of
W$_{\lambda}$ per line is made and the median value of those
measurements is taken as the final W$_{\lambda}$ of that particular
line.

\subsubsection{Medium Resolution Echelle Spectroscopy}

We have obtained medium resolution echelle spectrum of V2493 Cyg on 2019 April 28 by using the Medium Resolution Echelle
Spectrograph (MRES), that is mounted on the 2.4m Thai National Telescope (TNT) located at Thai national observatory. MRES has wavelength coverage
 from 3900\AA~ to 8800\AA~ and it provides a spectral resolution of R$\sim$16,000-19,000. Details about the MRES instrument  can be found in \citet{2022ApJ...926...16Y}.  The lunar illumination
during our observation date was $\sim$33\% and the sky was hazy due to forest fires resulting in low signal
to noise (SNR). We can only use the spectrum covering the wavelength range of 3900\AA~  to 6600\AA~ for our present analysis. Wavelength beyond this range is dominated by multiple telluric lines and bands and hence is not used
for the analysis.

 We have used the standard calibration frames, e.g., bias, flat and Th-Ar lamp, for image cleaning and wavelength calibration. The spectrum is extracted using the 
 {\sc echelle} module of the IRAF package. The procedure to calibrate the wavelength and normalize the
 spectrum is similar to that has been described in Section \ref{pt6}.

\subsubsection{NIR Spectroscopy}
We have spectroscopically monitored V2493 Cyg in the NIR wavelengths starting from 2014 March 07 till 2020 November 11 using the TIRSPEC mounted on 2m HCT and TIFR-ARIES Near Infrared Spectrograph
(TANSPEC) \citep{2018BSRSL..87...58O,2022PASP..134h5002S} mounted on 3.6m Devasthal Optical Telescope (DOT). TIRSPEC provides a wavelength coverage from 1 $\mu$m
to 2.5 $\mu$m at a spectral resolution of $\sim$1200. The details about the instrument is provided in \citet{2014JAI.....350006N}. 
The TIRSPEC spectroscopic data is processed using the TIRSPEC pipeline\footnote{\url{http://indiajoe.github.io/TIRSPEC/Pipeline/}}. The sky conditions were poor on the nights of 2017 September 14
and 2018 November 7 leading to poor SNR in the extracted spectra. The output extracted spectrum obtained is continuum normalised using the {\sc continuum} task of IRAF. 

  We have also obtained the NIR spectra of V2493 Cyg using the TANSPEC with its 1$^{\prime\prime}.0$ slit providing a R$\sim$1500 on the nights 
of 2020 October 25 and 2020 November 11. The spectra are obtained following the technique that is described in detail in \citet{2022ApJ...926...68G}. The obtained spectra are then
wavelength calibrated by using the recently developed TANSPEC data reduction pipeline\footnote{\url{https://github.com/astrosupriyo/pyTANSPEC}} \citep{2022ascl.soft12014G}. The telluric correction to the spectra is done in a similar manner as described in \citet{2022ApJ...926...68G}. The resulting spectra are then continuum normalised using the Python {\sc specutils} module\footnote{\url{https://specutils.readthedocs.io/en/stable/}} \citep{2021zndo...4603801E}.

In summary, we have monitored V2493 Cyg spectroscopically at 50 epochs with HFOSC (28),  MRES (1), TIRSPEC (19) and TANSPEC (2) starting from 2014 July till 2021 June. Table 
\ref{tab:obs_log} contains the complete log of spectroscopic observations.

\begin{figure*}
\centering
\includegraphics[width=0.95\textwidth]{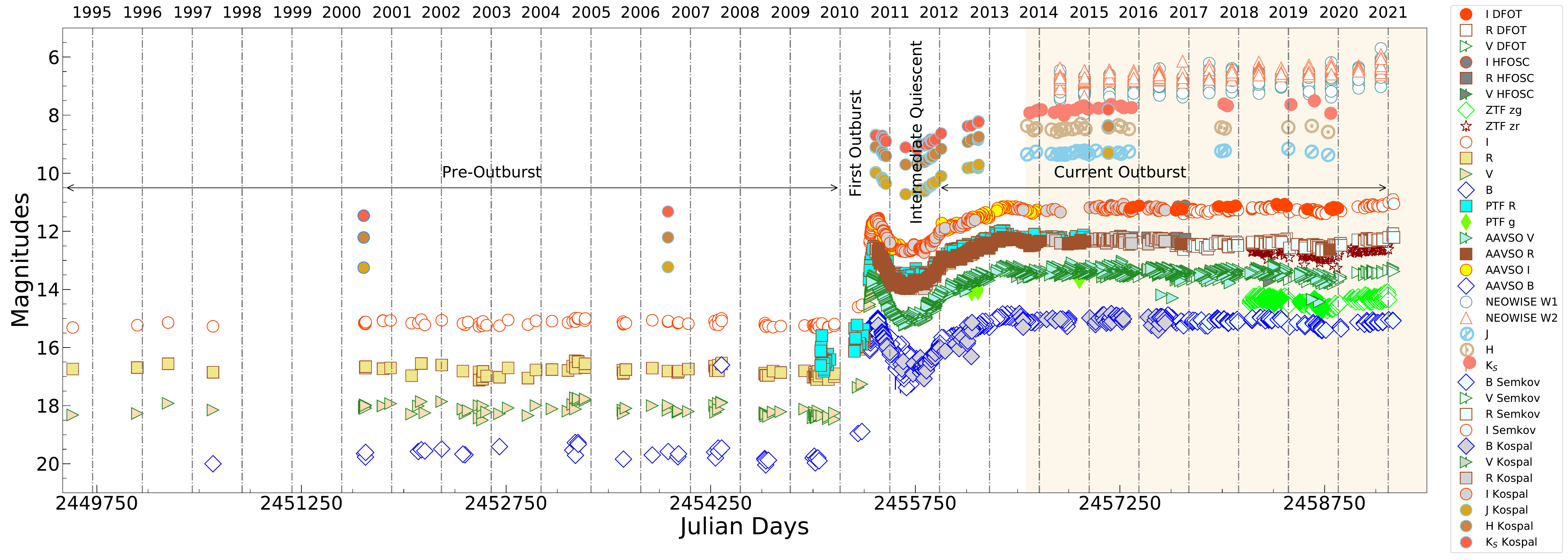}
\includegraphics[width=0.95\textwidth]{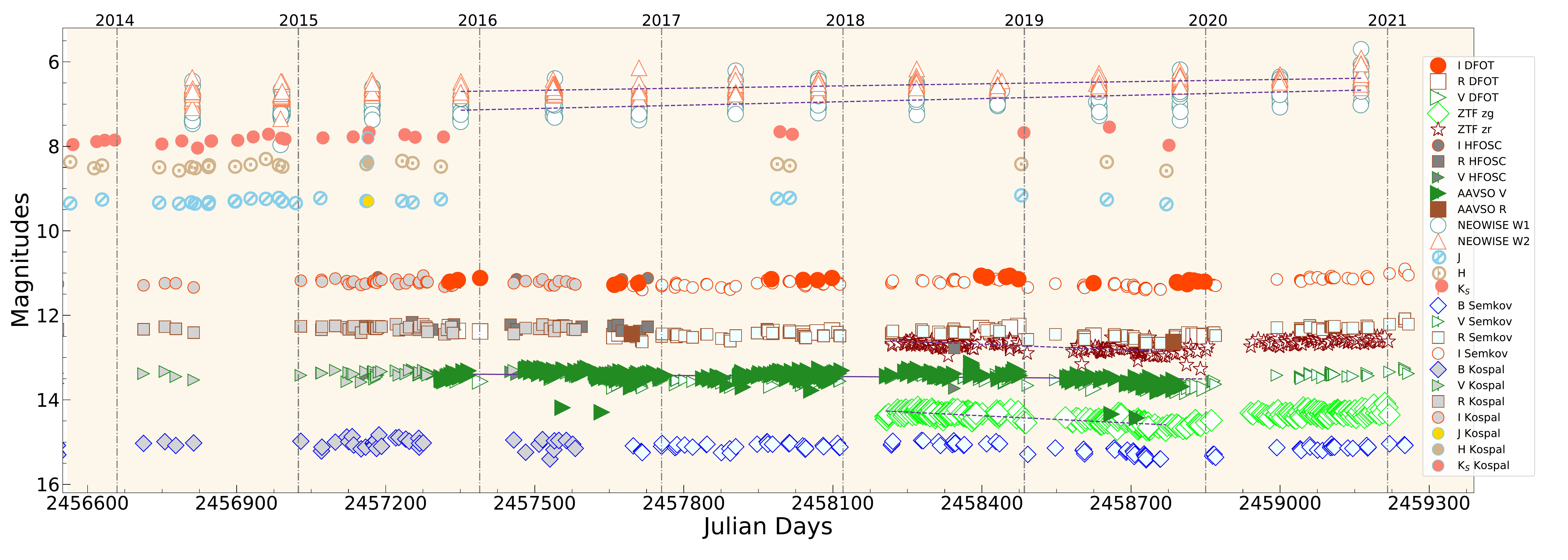}
\includegraphics[width=0.95\textwidth]{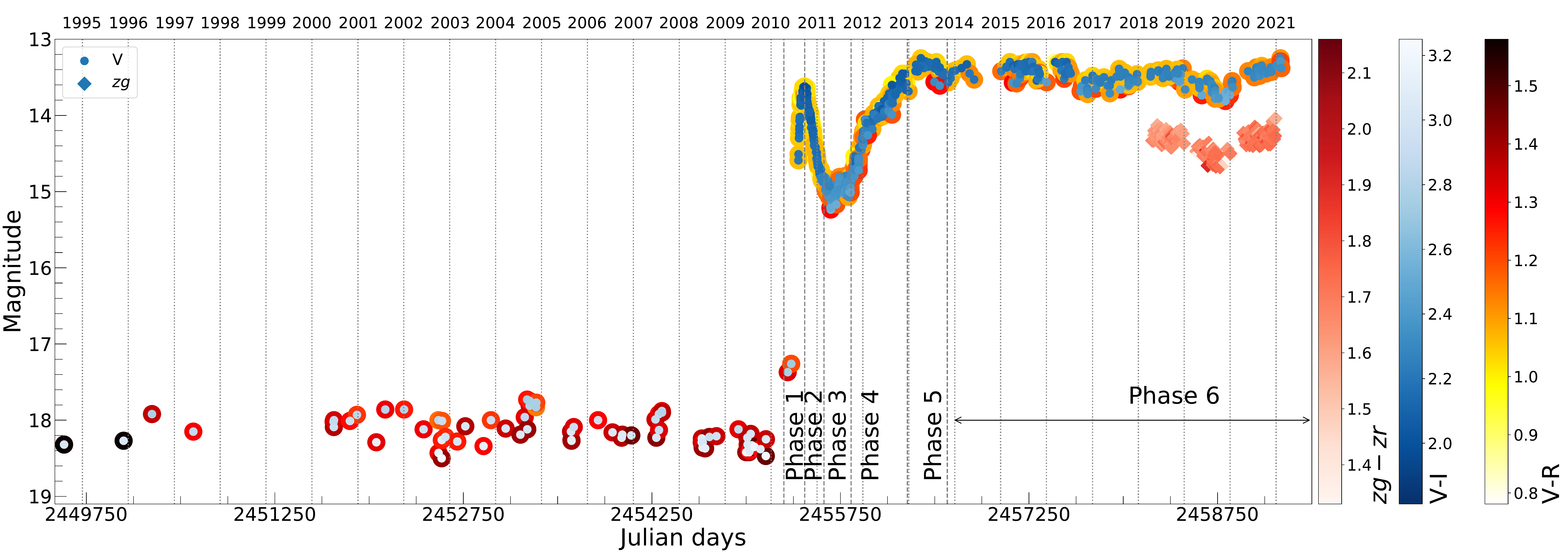}
\caption{\label{f1} Top panel: Light Curve of V2493 Cyg  in Johnson-Cousins $B$, $V$, $R_C$ and $I_C$, ZTF $zg$ and $zr$, PTF $g'$ and $R$, $J$, $H$ and $K_S$ and $NEOWISE$ $W1$ and $W2$ bands 
showing pre-outburst, first outburst, intermediate quiescent phase and the current outburst phase. The shaded region is our monitoring period. Middle panel shows the zoomed-in LC of V2493 Cyg
which covers our monitoring period corresponding to the time period between 2013 September 9 and 2021 January 4.  We call our monitoring period as phase 6. Bottom panel: V band LC of
V2493 Cyg coded with that of V$-$R and V$-$I$_C$ colors. The V$-$R color is depicted at the outer ring (annulus), while the color in the center depicts V$-$I$_C$ color. The diamonds represent the
$zg$ band LC coded with $zg-zr$ color. We find that all the colors are slowly becoming redder. The different phases, starting from the Phase 1 to Phase 5, are adopted from the study of
\citet{2015AJ....149...73B}.}
\end{figure*}

\section{Results and Analysis} \label{pt3}

\subsection{Light Curve} \label{lc}   

\begin{figure*}
\centering
\includegraphics[width=0.95\textwidth]{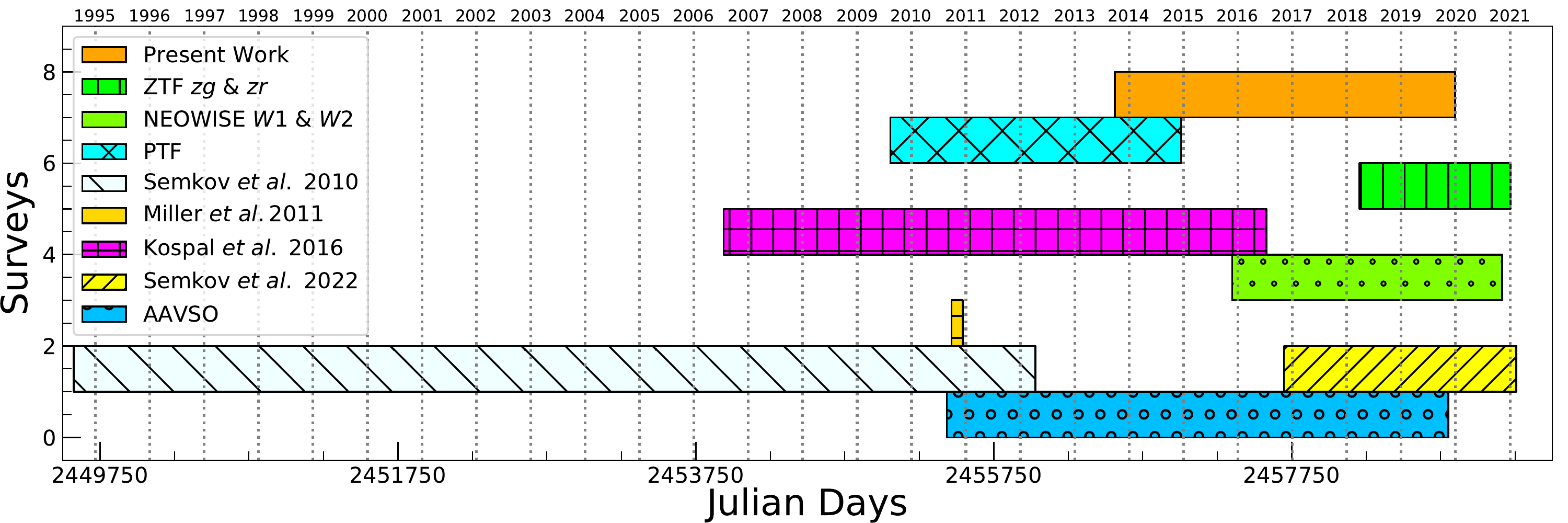}
\caption{\label{gantt} Timescales of the different studies and surveys at optical and NIR wavebands with which the evolution of V2493 Cyg was monitored. Our monitoring program 
is denoted by the orange horizontal bar and it includes monitoring in both optical and NIR wavebands. \citet{2010A&A...523L...3S,2021Symm...13.2433S} monitored V2493 Cyg extensively in optical
wavebands as shown. The NIR monitoring was primarily carried out by \citet{2011ApJ...730...80M,2016A&A...596A..52K} which covered both the 2010 and 2012 outburst. }
\end{figure*}

    Top panel of Figure \ref{f1} shows the light curve (LC) of V2493 Cyg in optical $B$, $V$, $R_C$ and $I_C$, PTF $g'$ and
    $R$, ZTF $zg$ and $zr$, NIR $J$, $H$ and $K_S$ and NEOWISE MIR W1 and W2 bands. Data in the $V$, $R$ and $I_C$  bands have the
    longest coverage. We have 
    combined our $V$, $R_C$, $I_C$, $J$, $H$ and $K_S$ band data\footnote{Optical photometric data obtained with the 1.3m DFOT    
    telescope and NIR photometric data from 2m HCT telescope.} 
    with that of the archival data from $AAVSO$ and the previously published data taken from \citet{2010A&A...523L...3S,2021Symm...13.2433S,2011ApJ...730...80M, 2016A&A...596A..52K}.
    The photometric monitoring timescales of the different surveys both in optical and NIR bands of V2493 Cyg begins with the 2MASS observations on 2000 June 10 (JD=2451706). V2493 Cyg was then observed in NIR bands during its 2010 outburst by \citet{2011ApJ...730...80M} and  \citet{2010ATel.2854....1L} are shown in Figure \ref{gantt}. 
    Our monitoring period is denoted by the orange horizontal bars. We have monitored V2493 Cyg photometrically both in optical and NIR wavebands.

    \citet{2015AJ....149...73B} have studied the photometric evolution of V2493 Cyg from 2010 September
   to 2013 May. During this period, V2493 Cyg  transitioned to its first outburst state to an
   intermediate quiescent stage and finally to its second outburst state. They have subsequently divided this period 
   into 5 phases based on the LC evolution. We call our optical photometric monitoring period which is carried out between 2015 October 16 and 2021 December 4 to be phase 6 which is actually a period after the peak of second outburst. Middle panel of Figure \ref{f1} shows the LC of V2493 Cyg during our monitoring period.  
   During phase 6, in the period between October 16, 2015 and December 30, 2019\footnote{This time period coincides with our monitoring of V2493 Cyg with 1.3m DFOT} the V-band brightness of V2493 Cyg dimmed at an average rate of $\sim$2.3 mmag/month with a drop by $\sim$0.6 mag from its peak outburst V-band magnitude in August 2010. 
  Since the dimming rate is too small, we can call phase 6 to be `Long Plateau' phase also. 
  Such a long plateau phase, which is also observed in other bonafide FUor sources like FU Ori, etc., is attributed to the time required to 
  deplete the inner disc completely after an \citep[cf.,][]{2014prpl.conf..387A}. 
We note that the individual WISE measurements for a given JD showed considerable scatter. We have therefore used the median
  WISE magnitudes for a given JD in our MIR analysis. We have taken the standard deviation of the magnitude variations within a day as a conservative limit of measurement error.
In the same duration, the source seems to brighten-up by $\sim$0.7 mag and $\sim$0.4 mag in MIR $W1$ and $W2$ bands, respectively with an average rise rate of 
$\sim$ 94$\pm$ 18 mmag/yr and 62$\pm$7 mmag/yr.

\begin{table}
\centering
\caption{Reddening invariant colors of V2493 Cyg during our monitoring period. N$\sigma$ is the ratio of color change to the quadrature
        sum of measurement errors.}
\label{tab:cc_table1}
\begin{tabular}{@{}cccccc@{}}
\hline 
Days    & $Q_{VRI}$  & N$\sigma$     & Days   & $Q_{VRI}$ & N$\sigma$\\
(JD)    & (mag)               &               & (JD)   & (mag)              & \\
\hline
2457312 & 0.15                & $-$           & 2458098  & 0.03    & 8.0 \\
2457345 & 0.06                & 6.1           & 2458398  & 0.03    & 8.6 \\
2457390 & 0.02                & 8.1           & 2458410  & $-0.01$ & 10.1 \\
2457660 & 0.07                & 5.9           & 2458449  & 0.05    & 7.3 \\
2457670 & 0.09                & 4.1           & 2458456  & 0.04    & 8.2 \\
2457708 & 0.04                & 8.1           & 2458792  & 0.01    & 10.3 \\
2457976 & 0.11                & 1.7           & 2458794  & 0.01    & 10.4 \\
2458448 & 0.07                & 5.8           & 2458813  & 0.05    & 7.2\\ 
2458473 & 0.11                & 1.6           & 2458817  & 0.05    & 6.9 \\  
2458624 & 0.11                & 2.8           & 2458826  & 0.01    & 10.1 \\    
2458040 & 0.01                & 9.7           & 2458834  & 0.01    & 8.7\\
2458069 & 0.02                & 8.6           & 2458848  & 0.04    & 8.1\\
\hline
\end{tabular}
\end{table}

\begin{figure}
\centering
\includegraphics[width=0.45\textwidth]{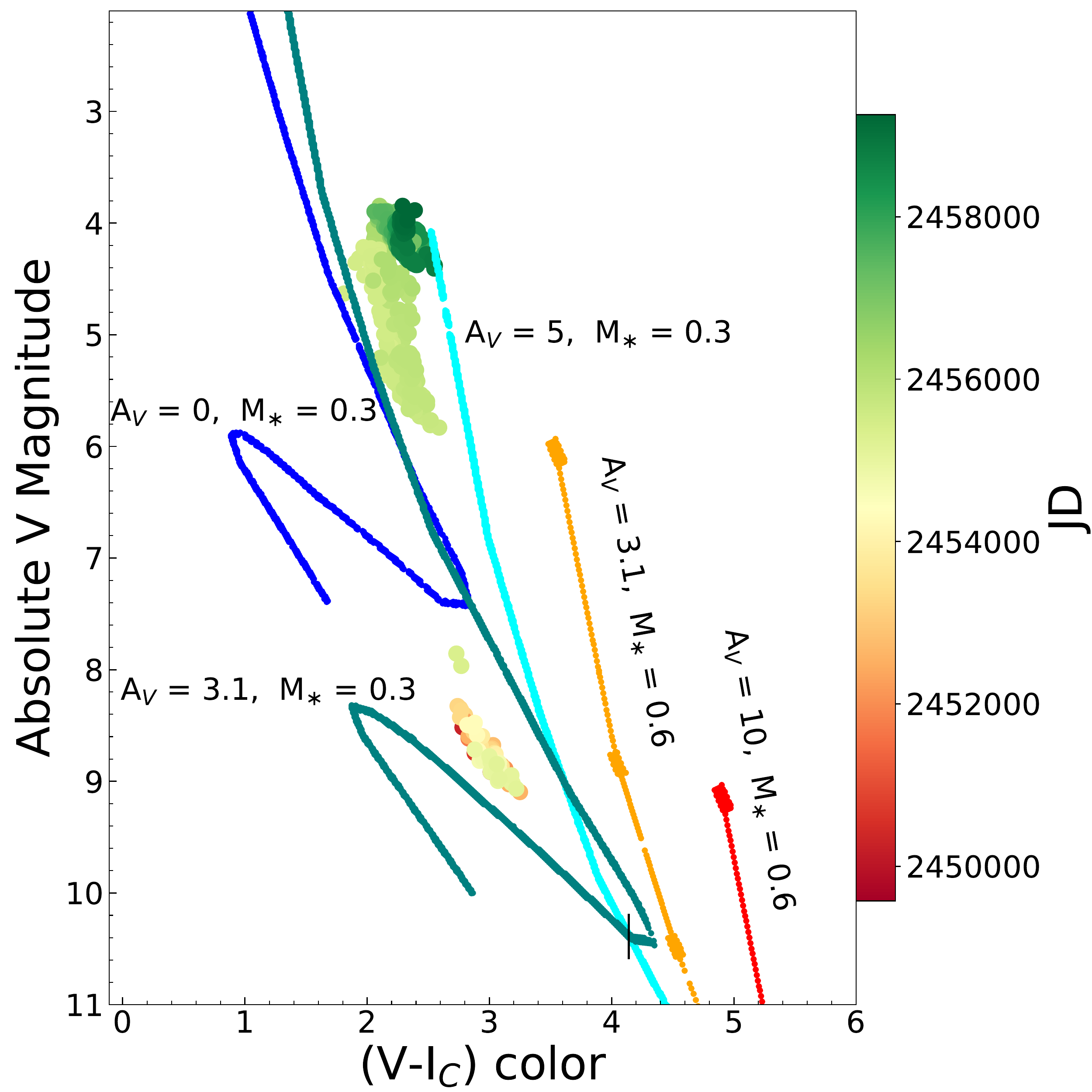}
\caption{\label{optical_cmd} The plot shows the evolution of V2493 Cyg in the absolute magnitude V vs (V-I$_C$) color magnitude diagram (CMD). The model iso-mass curves for mass 0.3 M$_{\odot}$, A$_V$ = 0 mag, A$_V$ = 3.1 mag and A$_V$ = 5 mag and for mass 0.6 M$_{\odot}$, A$_V$ = 3.1 mag and A$_V$ = 10 mag are also plotted \citep{2022ApJ...936..152L}. The phase 6 locus of V2493 Cyg in the CMD plane shows that it currently lies above the $\eta$ $>$ 5 regime (marked by the vertical black red line), the regime where the 
circumstellar disc outshines the central PMS star so that all the spectral features that we observe are typical of the viscous accretion disc. }
\end{figure}

\begin{figure*}
\centering
\includegraphics[width=0.45\textwidth]{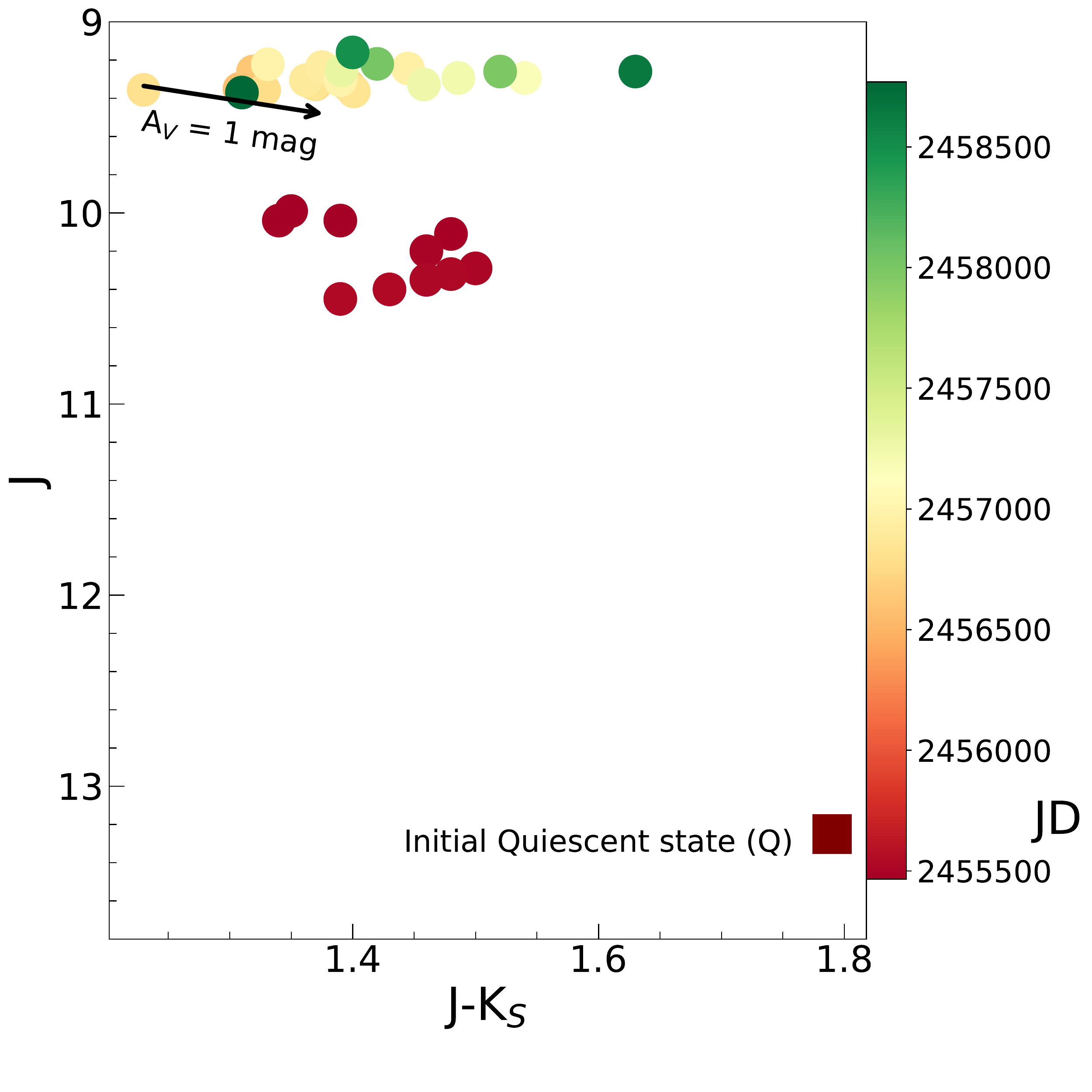}
\includegraphics[width=0.45\textwidth]{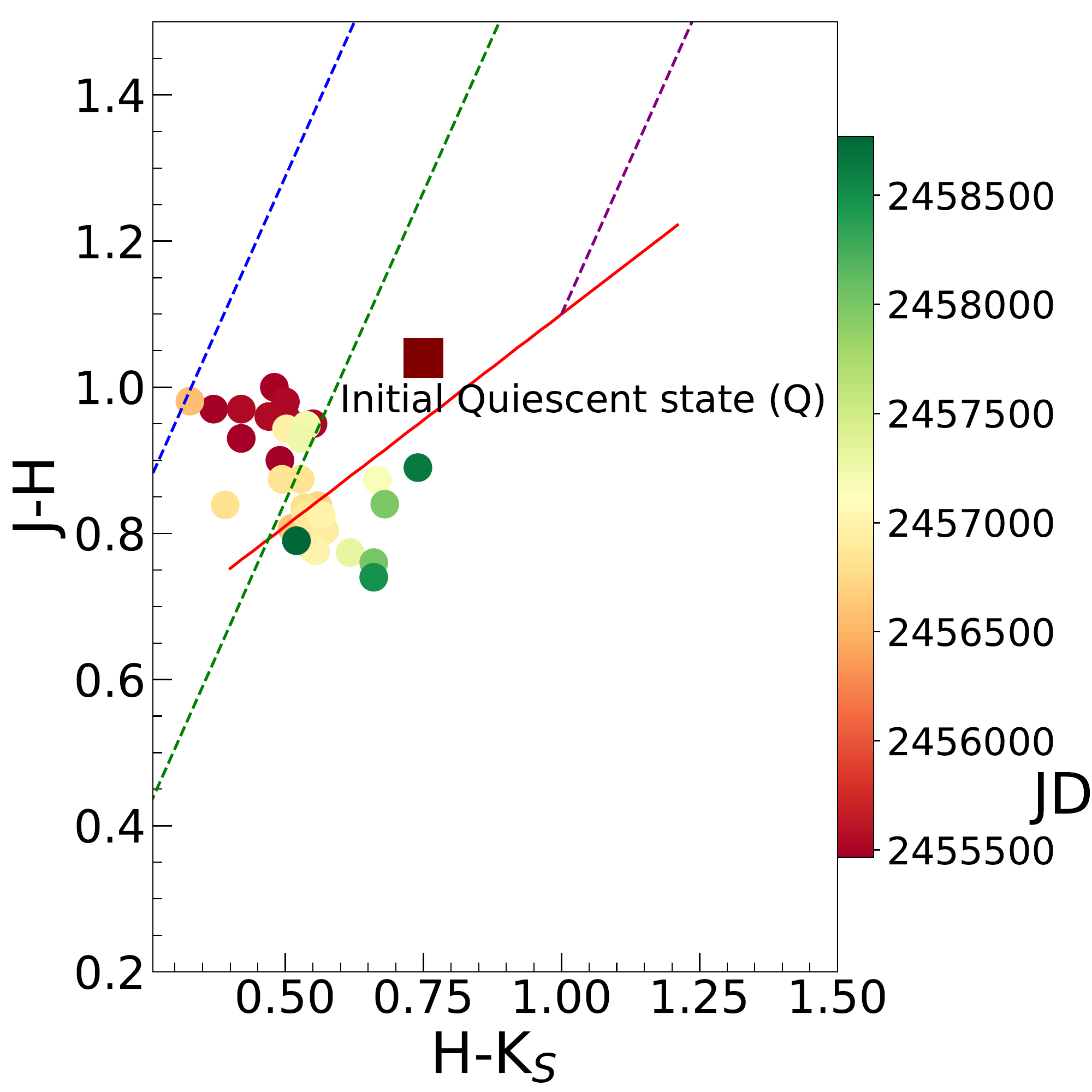}
\caption{\label{nir_color} Left panel shows the $J-K_S$ color evolution of V2493 Cyg with J magnitude and time. The square represents the quiescent phase location of V2493 Cyg as obtained from the 2MASS survey on 2000 June 10. The dots represent the evolution of V2493 Cyg from Phase 2 to Phase 6 part of the LC. The red dots represent the evolution of V2493 Cyg during its transition to intermediate quiescent phase after the first outburst. Right panel shows the position of V2493 Cyg in the J-H/H-K$_S$ CC diagram during our monitoring period corresponding to the post second outburst phase. The square represents the quiescent phase locus of V2493 Cyg in the CC plane. The solid red line represents the locus of the classical T-Tauri (CTT) stars \citep{1997AJ....114..288M}. The blue, green and maroon dashed lines represent the reddening vectors \citep{1985ApJ...288..618R}.    }
\end{figure*}

\begin{figure*}
\centering
\includegraphics[width=0.45\textwidth]{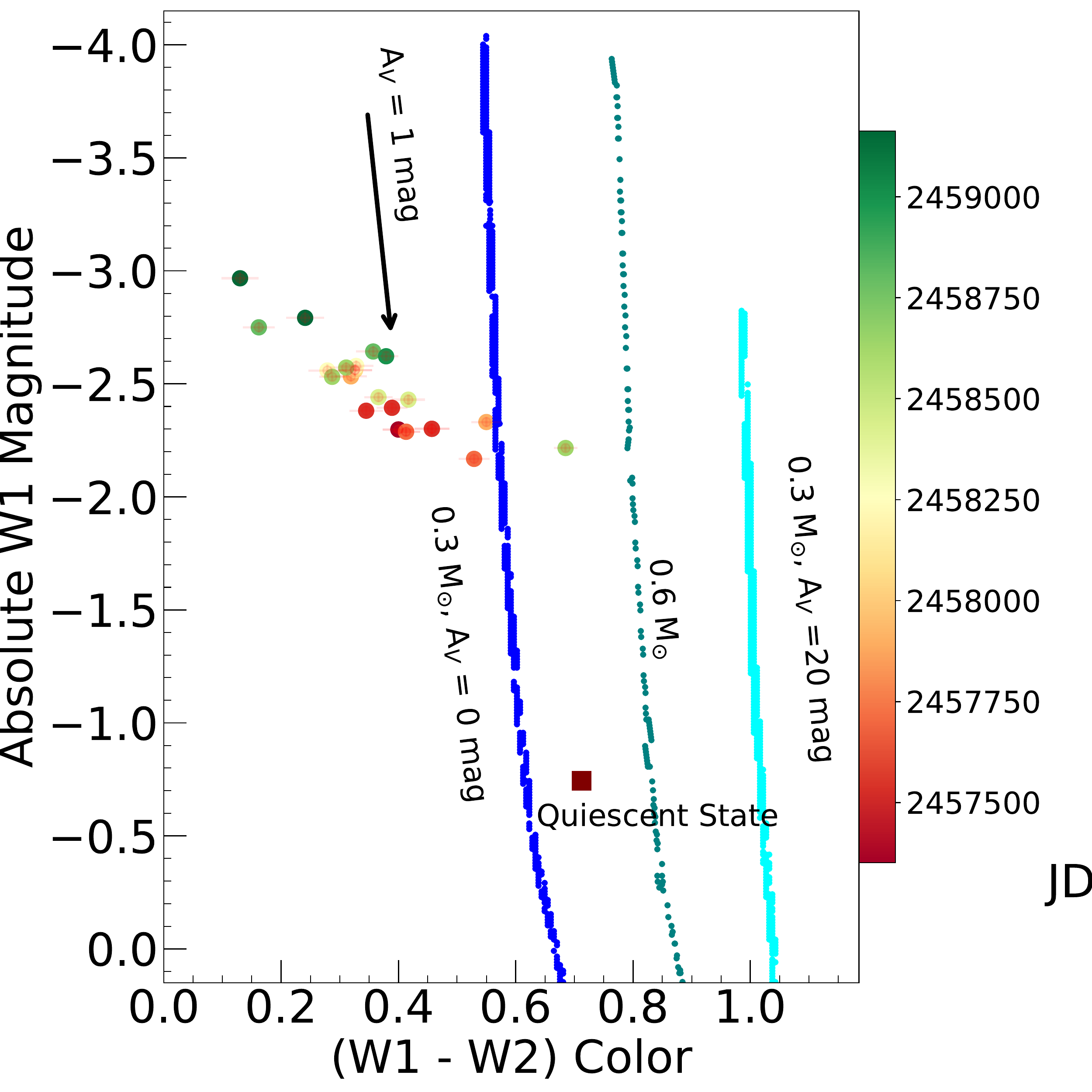}
\includegraphics[width=0.45\textwidth]{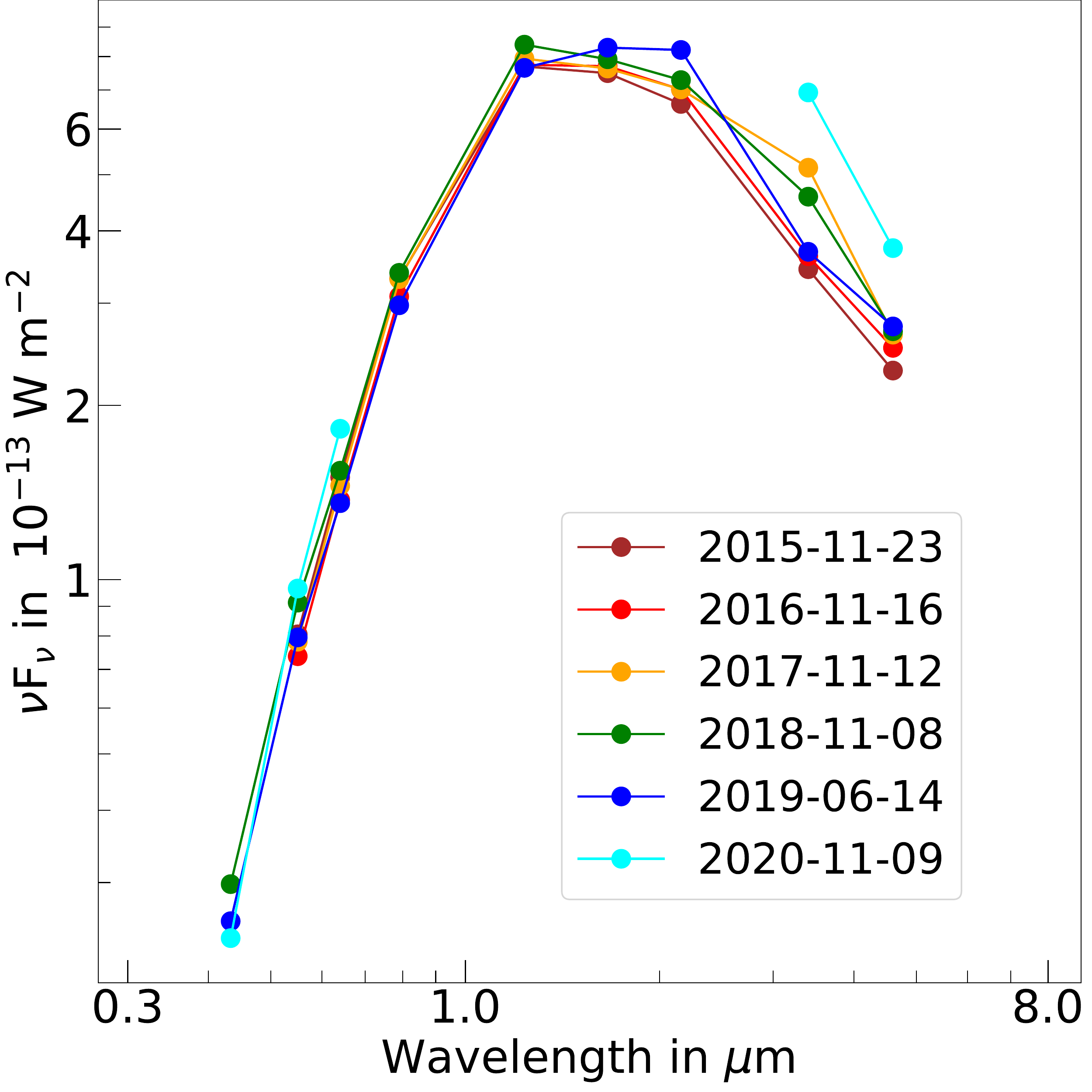}
\DIFaddendFL \caption{\label{sed} Left panel shows the W1 absolute magnitude vs $W1-W2$ color evolution of V2493 Cyg with magnitude and time. The quiescent phase locus of V2493 Cyg is depicted 
by the brown 
square. The model iso-mass curves for mass 0.3 M$_{\odot}$, A$_V$ = 0 mag and A$_V$ = 20 mag and for mass 0.6 M$_{\odot}$ are also plotted \citep{2022ApJ...936..152L}. The right panel shows the spectral energy distribution of V2493 Cyg during our monitoring period.   }
\end{figure*}

\subsection{Color Evolution } \label{color}

   The bottom panel of Figure \ref{f1} shows the complete evolution of ($V-I_C$) and ($V-R_C$) colors coded onto the $V$-band magnitude of V2493 Cyg from 2015 October 16 to 2019 December 30, which is within phase 6 defined earlier.
Both the ($V-R_C$) and ($V-I_C$) colors have consistently become redder in the phase 6, implying the presence of a cooler flux than that was present during the previous phases. 
   During phase 5, V2493 Cyg showed a weak bluer trend with small scale fluctuations in its color \citep{2015AJ....149...73B}. 
   Therefore, the present evolution possibly points towards a different physical process occurring in the V2493  Cyg as compared to the previous phases. The  V$-$R$_C$ and V$-$I$_C$ colors have 
   become redder by 0.14 and 0.32 mag, respectively during phase 6. The color evolution of V2493 Cyg is similar to that of V899 Mon, in a sense that its transition fluxes are redder compared to the outburst stages, therefore implying that the outburst flux possibly has emerged from the circumstellar disc \citep{2015ApJ...815....4N}.
   \citet{2021Symm...13.2433S}, have also noted this color evolution in their monitoring during the Phase 6 period. The reddening of the optical colors can be a result of 
   gradual decline in accretion rate in V2493 Cyg such that the central PMS star is exiting the maximum phase of the FUor outburst \citep{2021Symm...13.2433S}. Alternatively, the reddening of the 
   optical colors can also be 
   possibly explained due to the expansion of the emitting region around the star. This expansion of the emitting area is possibly a result of an ionization front expanding outwards 
   \citep{2016A&A...596A..52K}. 
   The color evolution of V2493 Cyg can be further analysed based on the theoretical model developed by \citet{2022ApJ...936..152L}. 
   According to their model, the PMS stars accreting in the ``FU Ori'' regime occupy a certain locus in the CMD plane as compared to the stars still accreting via magnetospheric regime. 
   They introduced a parameter $\eta$ which is the ratio of total H- and K- band viscous disc flux to that of the PMS photosphere. For a given stellar mass (M$_{\ast}$) the 
   accretion rate was varied and the values of parameter $\eta$ was computed for which the PMS photosphere dominates over the viscous accretion disc and vice versa. 
   Based on the values of parameter $\eta$ the CMD plane can be divided into regions of magnetospheric accretion regime and boundary layer accretion regime for a given M$_{\ast}$. 
   For $\eta$ = 0, the model developed by \citet{2022ApJ...936..152L}, degenerates into classical T Tauri star magnetospheric accretion regime.  
   Comparing our absolute V vs (V-I$_C$) CMD\footnote{We have obtained the values of isomass curves of \citet{2022ApJ...936..152L} using the online available plot digitizer tool ( \url{https://apps.automeris.io/wpd/}). The obtained values  are in the Gaia photometric system. The obtained photometric magnitudes and colors were then transformed to the Johnson Cousins photometric system using the transformation equations between the Gaia photometric system and the Johnson Cousins photometric system (\url{https://gea.esac.esa.int/archive/documentation/GDR2/Data_processing/chap_cu5pho/sec_cu5pho_calibr/ssec_cu5pho_PhotTransf.html} )}, (Figure \ref{optical_cmd}) with Figure 15 of \citet{2022ApJ...936..152L}, we find after its first outburst V2493 Cyg has mostly resided over $\eta$ $>$ 5 region in the CMD plane. The $\eta$ $>$ 5 region signifies the region in the optical CMD plane where the circumstellar disc outshines the central PMS star and all the observed spectral 
   features originate from the disc itself. It is to be noted that in $\eta$ $>$ 5 region, the magnetospheric accretion ceases and the central PMS star accretes via the boundary layer 
   accretion \citet{2022ApJ...936..152L}. Therefore, the reddening of the optical colors likely points towards the expansion of the emitting region around V2493 Cyg rather than the 
   PMS star exiting the ``FUor stage''.\\ 
    
    The color change observed in the present study can be 
    further investigated using 
    the reddening invariant colors following the recipe of \citet{2004ApJ...616.1058M}.
    The reddening invariant colors take the generic form $Q_{xyz}$ = (x$-$y) $-$ E(x$-$y)/E(y$-$z)
    $\times$ (y$-$z), where x, y and z are the observed magnitudes in respective filters. A color change `$\Delta$
    $Q_{xyz}$' that is statistically distinct from 0 indicates changes in the SED that cannot be explained by changes in the extinction alone, rather an intrinsic change 
    (changes in temperature of the system) might have occurred in the SED. We have tabulated the values of
    reddening invariant colors for R$_V$ = 3.1 in Table \ref{tab:cc_table1}. We see that there is a 
    $>$ 5$\sigma$ change in the $Q_{VRI}$ colors throughout the phase 6 period indicating an intrinsic change in the SED of V2493 Cyg. 

    The color change in the NIR wavelengths is less pronounced as compared to the optical bands. Left panel of Figure \ref{nir_color} shows the evolution of V2493 Cyg in the NIR color magnitude 
    diagram (CMD). The square box represents the pre-outburst quiescent phase locus of V2493 Cyg. The red dots represent the loci of V2493 Cyg during the 2010 outburst stage
    and its transition towards the intermediate quiescent phase as obtained from 
    \citet{2010ATel.2854....1L} and \citet{2011ApJ...730...80M}. The green dots represent the evolution of V2493 Cyg during our monitoring period. In the J/J$-$K$_S$ CMD, we see that there is a gradual 
    slow transition towards redder color with very little change in J magnitude. 
    Right panel of Figure \ref{nir_color} shows the evolution of 
    V2493 Cyg in the NIR color-color (CC) plane of J$-$H/H$-$K$_S$. Position of V2493 Cyg
    in the NIR CC plane shows no significant extinction towards the source. We note that V2493 Cyg 
    moved along the reddening vector in the Phase 1 and 2 periods. During our monitoring period, 
    which corresponds to Phase 5 and Phase 6, we note that the locus of V2493 Cyg is near to the 
    Classical T Tauri (CTT) locus. From the NIR CC plane, we also note that there is no significant variation in the extinction of V2493 Cyg during the Phase 5 and 
    Phase 6.
    Thus, the NIR color evolution also reinforces our assumptions based on 
    our reddening invariant color analysis in the optical regime.

    We have also investigated the color evolution in MIR $W1$ and $W2$ bands as obtained from
     the NEOWISE archive and is shown as a MIR CMD in the left panel of Figure \ref{sed}. 
    Interestingly, the MIR colors of V2493 Cyg has become bluer during our monitoring period. The blueing of the MIR colors can be possibly explained by the model developed by \citet{2022ApJ...936..152L}. According to their model, at the outburst stage, the viscous disc temperatures are higher, therefore heating up the disc which results in the blueing of the MIR
    colors. The gradual blueing might possibly hint at gradual increase of the warm continuum emission \citep{2022ApJ...936..152L}. 

\begin{figure*}
\centering
\includegraphics[width=0.95\textwidth]{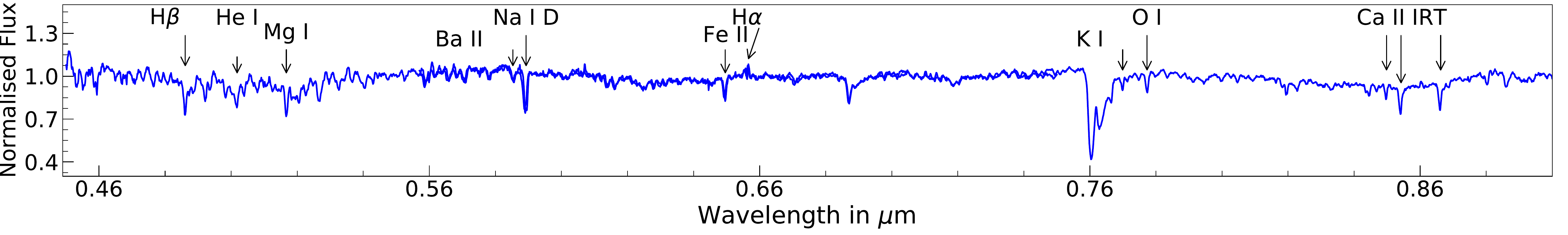}
\includegraphics[width=0.95\textwidth]{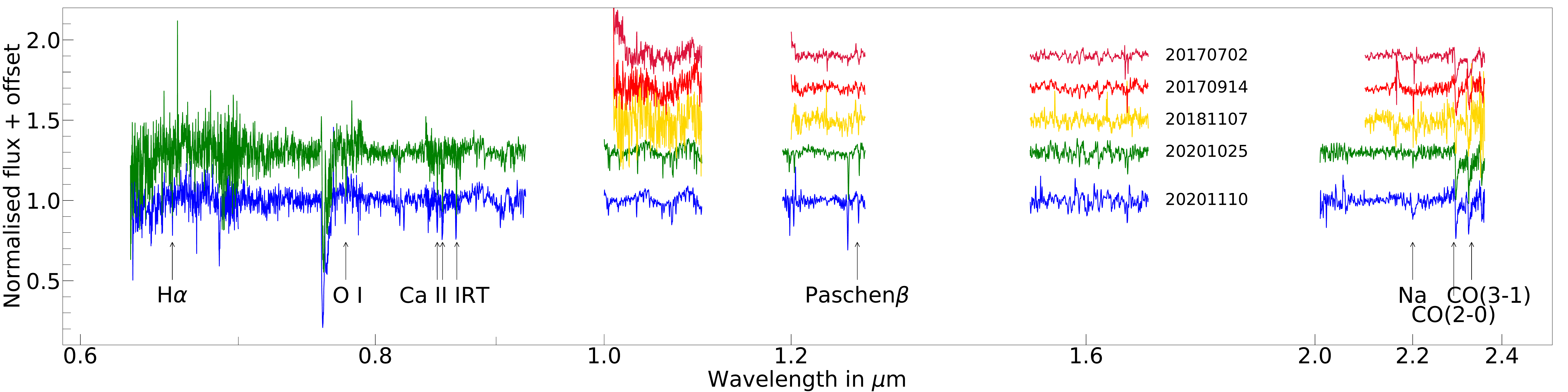}
\caption{\label{f2a}
Sample of the normalised spectra of V2493 Cyg obtained during our monitoring period using HFOSC on 2m HCT ($\sim$0.4-0.9 $\mu$m; top panel), TIRSPEC ($\sim$1.0-2.4 $\mu$m; bottom panel) on 2m HCT and TANSPEC ($\sim$0.65-2.4 $\mu$m; bottom panel) on 3.6m DOT. The lines that are used for the present study have been marked.
Complete sets of normalised spectra are provided as a figure set in the online journal.
}
\end{figure*}

\begin{figure*}
\centering
\includegraphics[width=0.95\textwidth]{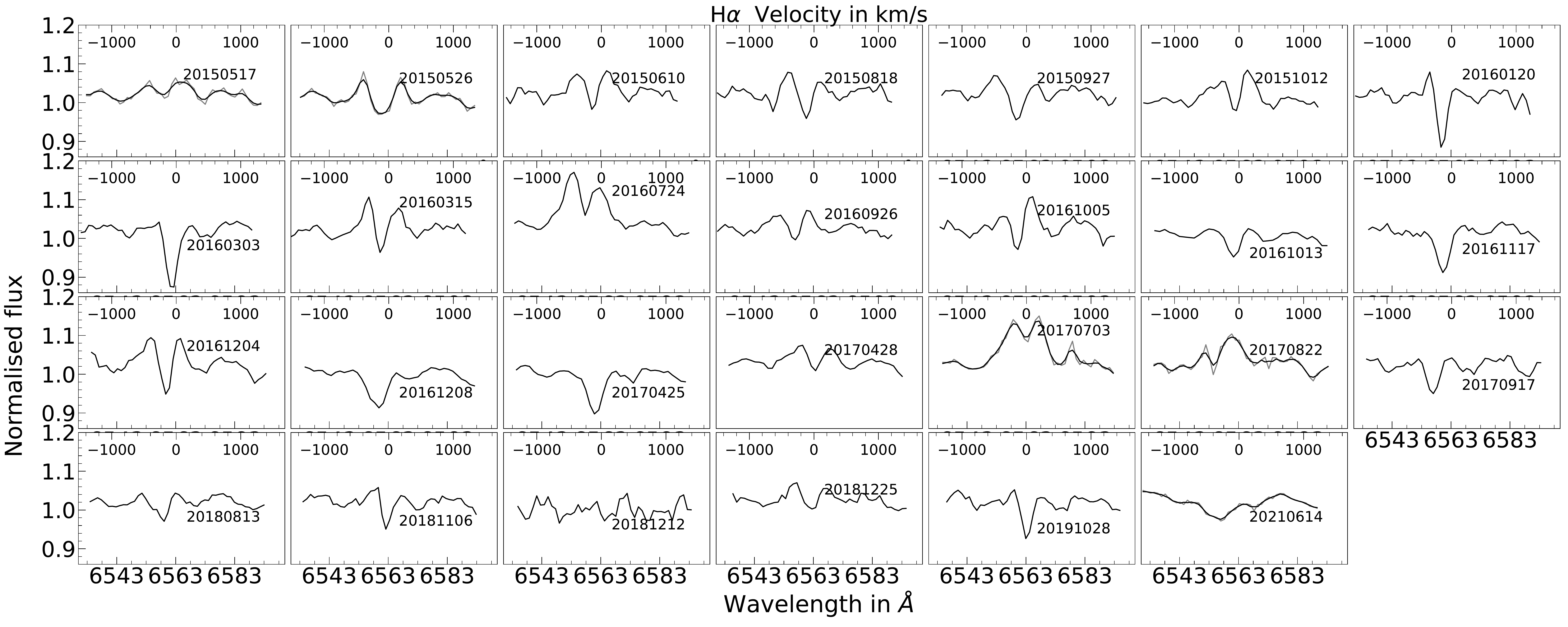}
\includegraphics[width=0.95\textwidth]{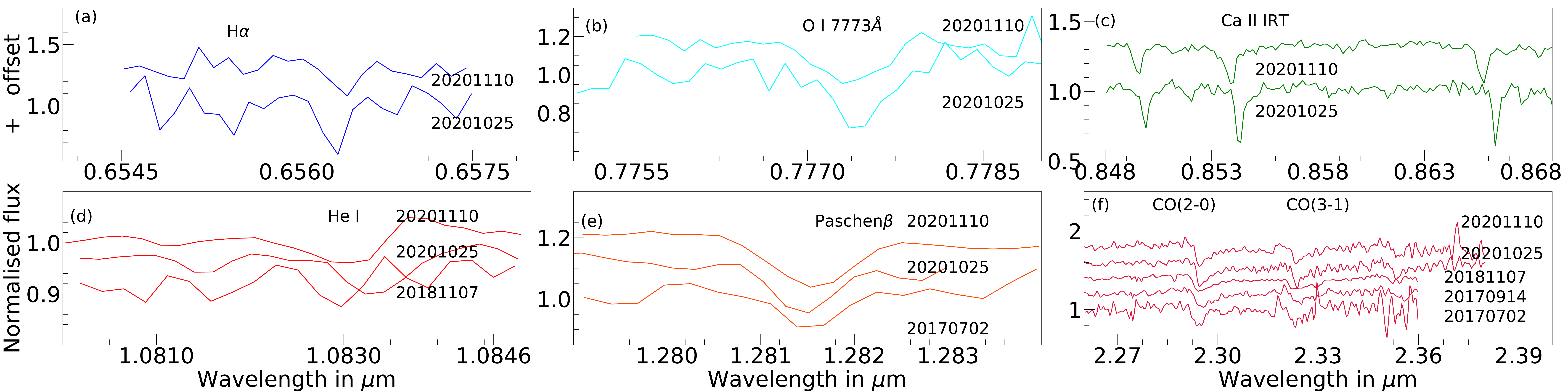}
\caption{\label{f6} The time evolution of the different line profiles in the spectra of V2493 Cyg during our monitoring period. Complete sets of plots are provided as a figure set in the online journal.}
\end{figure*}

\subsection{The Evolution  of Photometric Spectral Energy Distribution} \DIFaddbegin \label{SEDs}
\DIFaddend

 The right-panel of  Figure \ref{sed} shows the multi-epoch SEDs of V2493 Cyg during our monitoring period, i.e, Phase 6. We have constructed the SEDs using the multi-wavelength data (optical to  MIR
   wavelengths, i.e., 0.44 $\mu m$ ($B$), 0.48 $\mu m$ ($zg$), 0.55  $\mu m$ ($V$), 0.64 $\mu m$ ($zr$), 0.65 $\mu m$ ($R$), 0.80 $\mu m$ ($I$), 3.4  $\mu m$ ($W1$) and 
   4.6 $\mu m$ ($W2$), taken
   from our observations and from the data archives of ZTF and NEOWISE). The details of the ZTF
   filter system are available in 
   \citet{Bellm_2018}. Currently, we do not have any magnitude conversion system from the ZTF filter system
   to the Johnson-Cousins system. However, the ZTF system is calibrated using the Panoramic Survey
   Telescope and Rapid Response System (Pan-STARRS; PS1) data. We have therefore used the
   \citet{2012ApJ...750...99T} relations to transform ZTF photometric data to the Johnson-Cousins
   photometric system. The magnitudes are then converted to the corresponding flux values using the
   online tool provided by the Gemini
   Observatory\footnote{\url{https://www.gemini.edu/observing/resources/magnitudes-and-fluxes/conversions-between-magnitudes-and-flux}}. We hereby note that the data points used for the SED generation are 
   not simultaneous but from nearby epochs.    

      During Phase 6, the optical colors of the V2493 Cyg has become consistently redder (cf., Section \ref{color}).
     This implies a decrease in the strength of the physical process that triggered the 
     second outburst.  
     The multi-epoch SEDs exhibit a change in the shape particularly in the MIR regime which is in accordance with the blueing of the MIR colors due to the 
     brightening of the disc by viscous heating.

\subsection{Spectral Features}

  Figure \ref{f2a} shows our continuum normalised optical-NIR medium resolution (0.4-2.4 $\mu$m) spectra of V2493 Cyg  during our monitoring period. 
  We can easily identify the absorption features, i.e., H$\beta$ $\lambda$4861\AA~, HeI $\lambda$ 5015\AA~/ Fe II
  $\lambda$5018\AA~, Mg I at $\lambda$5167\AA~/ Fe II at $\lambda$5169\AA~, Ba II
  $\lambda$5853\AA~, Na I D resonance lines $\lambda$5890/6 \AA~, Fe II $\lambda$6496\AA~, H$\alpha$ $\lambda$ 6563\AA~, \DIFdelbegin \DIFdel{Li
  $\lambda$6707\AA~, }\DIFdelend \DIFaddbegin \DIFaddend K I $\lambda$7699\AA~, O I $\lambda$7773\AA~, \DIFdelbegin \DIFdel{O I
  $\lambda$8446\AA~ }\DIFdelend \DIFaddbegin \DIFaddend and Ca II infrared triplet (IRT) $\lambda$8498, $\lambda$8542 and $\lambda$8662 \AA~ in the optical part of the spectrum. 
  These absorption features are mostly blue-shifted which indicate that the powerful winds are 
  coming out from V2493 Cyg \DIFdelbegin \DIFdel{. 
  }\DIFdelend (The wind velocities are discussed in detail in Section \ref{outflow_velocity} ). One of the important spectral feature that is not detected in our HFOSC spectrum is the Li I 6707 \AA~ line. 
  In FUors, this feature is observed in absorption. The presence of Li I confirms the PMS stage of FUors as lithium gets destroyed in the main
  sequence stage \citep{1998apsf.book.....H}. The absorption feature in Li I is also indicative of the shell like features as observed in other bona-fide FUors \citep{2021ApJ...917...80S}. 
  The absence of Li I 6707 \AA~ is likely because the spectral feature is too weak to be detected in our HFOSC spectrum. In the NIR  wavelengths, we can identify several distinct absorption features, e.g., the CO (CO(2$-$0) and CO(3$-$1)) bandheads,  \DIFdelbegin \DIFdel{Na $\lambda$2.2 $\mu$m metallic line}\DIFdelend Paschen$\beta$ (5-3) $\lambda$1.28 $\mu$m  line \DIFdelbegin \DIFdel{, etc}\DIFdelend and He I 1.083 $\mu$m line. With the TANSPEC spectra, we can additionally identify few more lines in absorption, like the Br (11-4) $\lambda$1.68 $\mu$m line, Paschen$\gamma$ (6-3) $\lambda$1.09 $\mu$m, the Ca II IRT, O I $\lambda$7773\AA~ and H$\alpha$.  
  We have computed the equivalent widths `W$_{\lambda}$' of the important spectral lines and are listed in the Table \ref{tab:eqw_log}. 
  \DIFdelbegin \DIFdel{These W$_{\lambda}$ provide the information about the density of the emitting/absorbing regions, thus, giving us an idea about the physical processes such as outflow and accretion happening at the surface of the young star.
  }\DIFdelend 
  The evolutions of the profiles of the above lines during our spectroscopic monitoring period, i.e., 2014 March 7 to 2021 June 14, are shown in the Figure \ref{f6}. Few of the optical lines, i.e., H$\alpha$, Ca II IRT and O I lines, and \DIFdelbegin \DIFdel{few }\DIFdelend NIR lines, i.e., CO bandheads, \DIFdelbegin \DIFdel{Na $\lambda$2.208 $\mu$m}\DIFdelend Paschen lines, He I 1.083 $\mu$m which are important diagnostic tools for the  accretion and outflow mechanisms,  are discussed below.

  \subsubsection{Evolution of the H$\alpha$ line}\label{halpha}

   The H$\alpha$ line \DIFdelbegin \DIFdel{originates close to the regions of the magnetospheric accretion, therefore, }\DIFdelend observed in YSOs  originates from a variety of processes like accretion, outflowing winds, etc. 
   Therefore, the time series analyses of the H$\alpha$ line profile can help us to constrain the physical processes responsible for their  origin and variations \DIFdelbegin \DIFdel{\citep{2001ApJ...550..944M,2003ApJ...592..266M,2005ApJ...625..906M,2006MNRAS.370..580K}. }\DIFdelend \citep{1990ApJ...349..168H}.  
  During our monitoring period, the line that showed the most riveting evolution is the H$\alpha$ line. 
  Previously, \citet{2011ApJ...730...80M} have found the H$\alpha$ line to be in emission in their low resolution
  optical spectra taken on 2010 September 16 and 2010 November 2. However, during their monitoring period of
  40 days, the H$\alpha$ emission flux decreased by almost a factor of 0.5 from $\sim$2.8$\times$10$^{30}$ erg s$^{-1}$ to
  1.5$\times$10$^{30}$ erg s$^{-1}$. The high resolution spectra of V2493 Cyg taken between 2010 to 2014 
  revealed the P Cygni profile of H$\alpha$ line \citep{2011ApJ...730...80M,2015ApJ...807...84L}. The time
  evolution of H$\alpha$ line from these spectra also showed a decrease in the emission peak of the P Cygni
  profile, particularly from 2013 onwards (cf., Figure 1 of \citet{2015ApJ...807...84L} ). \DIFdelbegin \DIFdel{This points that the magnetospheric accretion with little or no wind feature was the physical 
  process that triggered the first outburst \citep{2006MNRAS.370..580K}.}\DIFdelend This evolution of the H$\alpha$ profile likely points towards the changing of the physical 
  conditions responsible for the outflowing winds from regions close to the accretion \citep{1990ApJ...349..168H,2006MNRAS.370..580K}.
  \DIFaddend We have started our spectroscopic monitoring of V2493 Cyg from 2015 May 17. The H$\alpha$ line was present on 2015 May 17 with a
  primary peak at $\sim$6563\AA~ along with a secondary blue-shifted peak which was about half of the
  primary peak's strength. This profile remained till 2015 June 10. The H$\alpha$ line profile changed in the spectrum obtained on 2015 August 18 where the primary peak became 
  half of the secondary blue-shifted emission peak. The spectrum obtained on 2015 September 27 displayed
  a blue-shifted absorption trough which then changed to a 
  self absorption profile as observed in the spectrum obtained 
  on 2015 October 12. Spectra obtained on following epochs on 2016 January 20 and 2016
  March 03 shows that the P Cygni profile changed to a absorption trough. Our spectra obtained on 2016 March 
  15 onwards till 2016 July 24 showed a H$\alpha$ profile that is similar to the profile observed on 2015
  August 18. The spectrum obtained on 2016 September 26 displayed a weak P Cygni profile which became prominent
  in the spectrum taken on 2016 October 5. Subsequently, P Cygni profile disappeared and
  an absorption trough appeared on 2016 October 13 and it remained till
  2017 April 25. The absorption trough changed to a line profile displaying emission feature in the 
  spectrum obtained on 2017 April \DIFdelbegin \DIFdel{28 which }\DIFdelend 28. The observed emission profile  is similar to the profiles previously observed on 2015 August 18 and
  2016 March 15. The line profile of H$\alpha$ as observed on the next two dates of observation, i.e, 2017 July 3
  and 2017 August 22, is similar to the profile as observed on 2015 May 17. We then note the reappearance of the
  absorption feature starting from 2017 September 17 and remained in absorption till the end of our spectroscopic monitoring period, i.e,
  upto 2021 June 14 with one exception on 2018 December 12 where the H$\alpha$ line was absent.

  The variation in the line profile of H$\alpha$ during our monitoring 
  period indicates towards changes in the structure and/or speed of the outflowing winds originating 
  from V2493 Cyg. Several of the H$\alpha$ line profiles that we have observed during our monitoring period
  match with the atlas of H$\alpha$ profiles as published by \citet{1996A&AS..120..229R}. The variation of
  the H$\alpha$ profile in T Tauri stars for a variety of scenarios has been
  previously studied in detail by \citet{1990ApJ...349..168H} and \citet{2006MNRAS.370..580K}. Emission profile in
  H$\alpha$ as seen in the spectrum of \citet{2011ApJ...730...80M} corresponds to the Type I line profiles
  of the \citet{1996A&AS..120..229R} atlas. This type of line profile is reproduced in a   
  accretion dominated regime with accretion rates in excess of 10$^{-8}$ M$_{\odot}$/yr
  with relatively weak disc winds and the system is viewed in a high inclination \citep{2006MNRAS.370..580K}. It is worthwhile to
  mention here that \citet{2011A&A...527A.133K} calculated the inclination of the V2493 Cyg to be
  73$_{-15}^{+6}$ deg which is regarded as a high inclination angle.
  The line profile observed on 2015 June 10 and similar profiles observed in 
  other dates during our monitoring period are categorized as Type II-B in the H$\alpha$ atlas. This type of
  profile is produced by the disc wind magnetosphere hybrid model of \citet{2006MNRAS.370..580K}. The main 
  feature of this line profile is high wind accelerations. The line profile observed on 2016 March 15 and 
  similar line profiles on the other dates of our monitoring period are classified as Type II-R (Type II-R line profile is defined as a line profile that consists of a secondary red peak whose strength is in
  excess of half the strength of the primary peak), according to
  the classification scheme of \citet{1996A&AS..120..229R}. This type of line profile is also reproduced by the
  hybrid model of \citet{2006MNRAS.370..580K}. According to this hybrid model, Type II-R is characterized by the decrease in wind acceleration. With the help of these hybrid models, we can also put a constrain on
  the temperature of the V2493 Cyg system to be 7500 K. We have also observed the P Cygni profile in H$\alpha$ which is attributed to outflowing winds from regions close to accretion. The 
  evolution of H$\alpha$ profile into a P Cygni feature from a self absorption feature can be possibly explained by the variations in the collimation of the outflowing winds as shown by
  the models of \citet{2006MNRAS.370..580K}.
  We have also observed H$\alpha$ to be in
  absorption on multiple dates during our monitoring period. The absorption features in H$\alpha$ also help us to put a constrain on the temperature of the system. 
  The temperature of the region producing absorption features in H$\alpha$ is below 5000 K as 
  evident from Model 4 in Table 1 of \citet{1990ApJ...349..168H}. The change in line profile from
  emission to absorption therefore indicates a change in temperature of the \DIFdelbegin \DIFdel{region }\DIFdelend \DIFaddbegin outflowing winds producing H$\alpha$ 
  \DIFdelbegin \DIFdel{. This could also be seen in the }\DIFdelend \citep{1990ApJ...349..168H,2006MNRAS.370..580K}. This change in the outflowing wind structure can be possibly attributed to the small scale photometric \DIFdelbegin \DIFdel{variation }\DIFdelend variations in the LC of V2493 Cyg \DIFdelbegin \DIFdel{.
  The evolution of H$\alpha$ line profile as seen by combining our spectra with that of \citet{2011ApJ...730...80M}, therefore possibly points towards the evolution of V2493 Cyg system from 
  magnetospheric accretion with weak disc wind regime towards a magnetospheric accretion regime with disc
  winds having variable acceleration. }\DIFdelend even 
  though there is no significant change in the accretion rate as evident from the SEDs (c.f. Section \ref{SEDs}). 

\subsubsection{Ca II IR Triplet lines}

 The Ca II IRT lines are found to be in absorption throughout our
 monitoring period. Previously, Ca II $\lambda$8498\AA~ line displayed a P Cygni profile
from 2010 to 2012 before transitioning to the absorption feature on 2013 April 21 (cf., bottom panel
of Figure 1 of \citet{2015ApJ...807...84L}). Combining our spectroscopic monitoring with that
of \citet{2015ApJ...807...84L}, we see that there is a
gradual transition of Ca II $\lambda$8498\AA~ line from the P Cygni profile to the absorption feature. 
This points towards the \DIFdelbegin \DIFdel{gradual transition of V2493 Cyg from the accretion dominating regime 
to a heavy outflow dominating regime.}\DIFdelend heavy outflow during our monitoring period. 

The profile of the Ca II IRT line $\lambda$8542\AA~ is very similar to the isothermal wind models (Model 2 to
4) developed by \citet{1990ApJ...349..168H}. The models take into account a variety of system temperatures ranging from 20000 K to 5000 K. The H$\alpha$ line profile variations as discussed in Section \ref{halpha}
are produced by systems having temperature in the range 5000 - 7500 K. Therefore, the models named as Model 3(I)
and Model 4(I) are the likely models that can explain the observed profiles. This also helps us to constrain
the density of the region to be around 10$^{12}$ cm$^{-3}$.

\begin{figure*}
\centering
\includegraphics[width=0.75\textwidth]{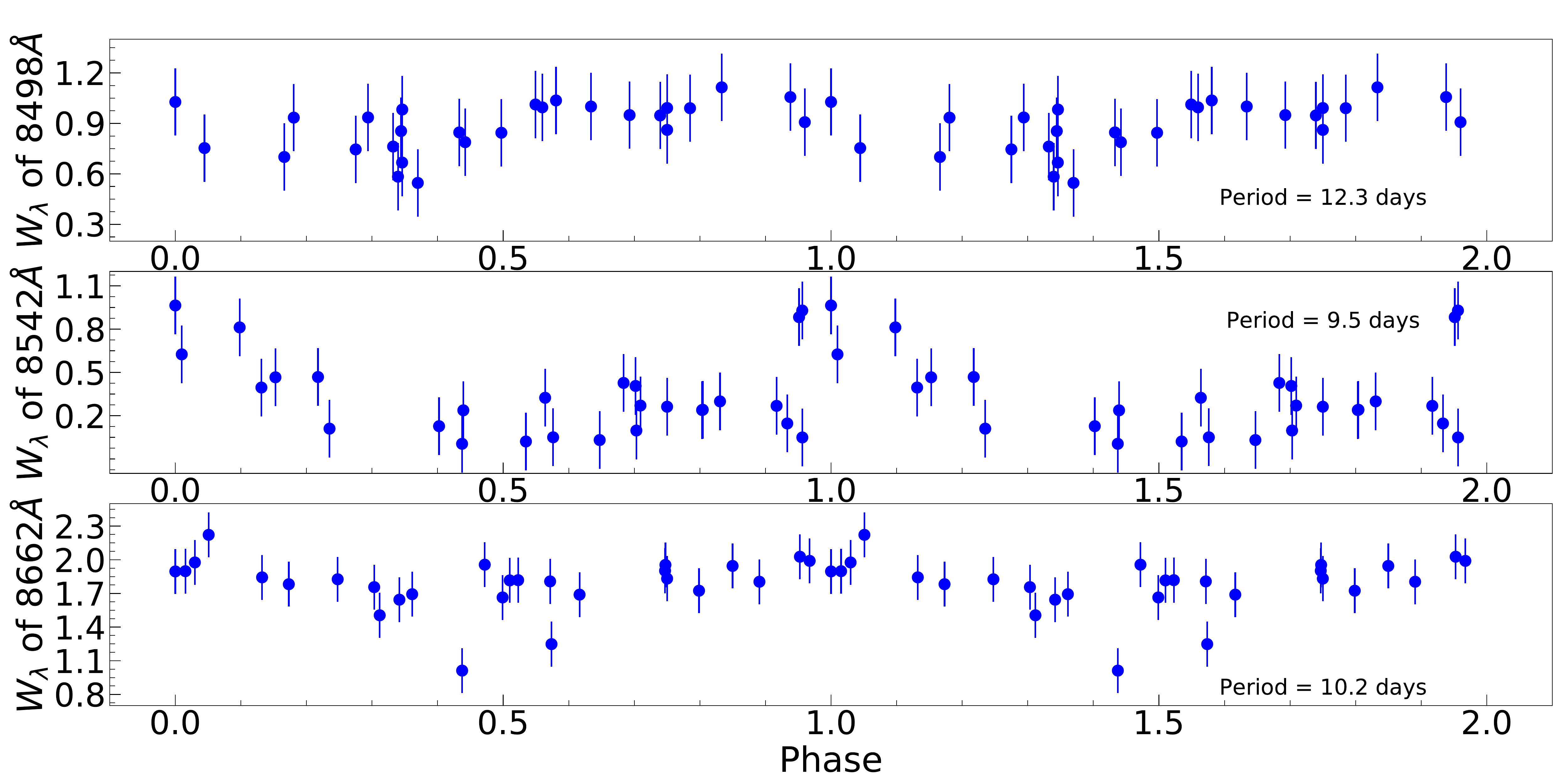}
\caption{\label{period_IRT}  Folded phase plot of the W$_{\lambda}$ of the Ca II IRT lines. 
The periods obtained are 12.3 days, 9.5 days and 10.2 days respectively for
$\lambda$8498\AA~, $\lambda$8542\AA~ and $\lambda$8662\AA~ lines. The folding is done using
the data obtained during our entire monitoring period. The obtained periodicities are 
calculated with a 2 $\sigma$ confidence label. }
\end{figure*}

 The mean values of  W$_{\lambda}$ for the Ca II IRT lines are estimated as 0.8$\pm$0.1, 2.2$\pm$0.3 and 1.8$\pm$0.2 \AA~. The low scatter values 
 indicate that the density of the absorbing regions remained almost constant during our monitoring period. 
 \DIFdelbegin \DIFdel{We note that  }\DIFdelend 
  The ratio of the
  W$_{\lambda}$ \DIFdelbegin \DIFdel{of the three Ca II IRT lines are in the ratio of 1.0:2.0:2.4. }\DIFdelend between 8542 \AA~ and 8662 \AA~ during our monitoring period is 2.2:1.8 = 1.2:1.  This value is \DIFdelbegin \DIFdel{inconsistent with the optically thin line formation scenario which requires 
    the ratio of the triplet to be 1:9}\DIFdelend lower than the intensity of the
  atomic transition strengths of these lines which is 1.8 :\DIFdelbegin \DIFdel{5 \citep{2022ApJ...926...68G}. }\DIFdelend 1, even after incorporating the error bars. Hence, the regions producing the 
  absorption features of Ca II are not optically thin. This observed feature of Ca II spectral lines can be likely due to the high mass-loss 
  winds from V2493 Cyg. The observed blue-shifted absorption features in Na I D lines in the spectrum lend further credence to this scenario.
  This is because the Na I atom is easily ionised and is observed only in densest or high mass-loss rate wind models \citep{1990ApJ...349..168H}.  
  We also searched for periodicities in the W$_{\lambda}$ of the Ca II IRT lines. The periodicity search was done using the Period\footnote{
\DIFdelend  \url{http://www.starlink.ac.uk/docs/sun167.htx/sun167.html } } software \citep{2014ASPC..485..391C}, 
    which works upon the principle of Lomb$-$Scargle (LS) periodogram \citep{1976Ap&SS..39..447L, 1982ApJ...263..835S}, to determine the period in the W$_{\lambda}$ of the triplet. The advantage of the LS
    method is that it is effective even in case of the data set being non$-$uniformly sampled. We have also
    used the NASA Exoplanet Archive Periodogram\footnote{\url{https://exoplanetarchive.ipac.caltech.edu/docs/tools.html}} service for cross verification. The periods determined using both the softwares matched well. The periods obtained for the
    IRT lines are 12.3 days, 9.5 days and 10.2 days, respectively. Figure \ref{period_IRT} shows the phase
    folded data of the IRT lines covering our entire monitoring period. However, we note that the periodicities
    are calculated with a 2$\sigma$ level of confidence, hence we \DIFdelbegin \DIFdel{need more data to lend more confidence to our
    calculated period. The possible implication of these periodicities in the W$_{\lambda}$ of the triplet lines
    points towards a change in  the continuum flux as pointed out by \citet{2013ApJ...778..116N} in the case 
    of V1647 Ori, another well known episodically accreting young star.}\DIFdelend refrain from interpreting about its possible implications. 
\DIFaddend 

\begin{table*}
\centering
\caption{Variation of the equivalent widths (in \AA~) of the various lines identified in our study. The typical error in estimation of equivalent width is $\sim$0.2\AA~ by using the relation provided by \citet{1988IAUS..132..345C}}
\label{tab:eqw_log}
\begin{tabular}{c@{ }c@{ }c r@{ }r@{ }r@{ }r@{ }r@{ }r@{ }r@{ }r@{ } r@{ }r@{ }}
\hline   
Date       &Julian date & H$\beta$ & Fe II    & Mg I/FeII         & Ba II     & Na I D & Fe II     & \DIFdelbeginFL \DIFdelFL{Li I }
\DIFdelendFL K I      & O I       &\DIFdelbeginFL \DIFdelFL{O I }
\DIFdelendFL \multicolumn{3}{c}{Ca II IRT} \\
           &            &          & 5018\AA~ & 5167\AA~/5169\AA~ & 5853\AA~  &        & 6496\AA~  & \DIFdelbeginFL 
\DIFdelendFL 7699\AA~ & 7773\AA~  &     \DIFdelbeginFL \DIFdelFL{8446\AA~ }
\DIFdelendFL &     &  \\ 
\hline
2015-05-17 & 2457160    & $-$      & $-$      & $-$               & 0.8       & 3.7    &  1.4      & \DIFdelbeginFL \DIFdelFL{0.9 }
\DIFdelendFL 0.6      & 1.3       & \DIFdelbeginFL \DIFdelFL{1.4 }
\DIFdelendFL 0.9 & 2.2 & 1.5 \\
2015-05-26 & 2457169    & $-$      & $-$      & $-$               & 1.4       & 3.1    &  1.4      & \DIFdelbeginFL \DIFdelFL{1.1 }
\DIFdelendFL 0.6      & 1.6       & \DIFdelbeginFL \DIFdelFL{1.6 }
\DIFdelendFL 1.1 & 2.5 & 1.8 \\
2015-06-10 & 2457184    & 3.3      & 2.7      & 2.5               & 1.3       & 3.2    &  1.6      & \DIFdelbeginFL \DIFdelFL{0.5 }
\DIFdelendFL 0.6      & 1.7       & \DIFdelbeginFL \DIFdelFL{1.7 }
\DIFdelendFL 0.9 & 2.2 & 1.8 \\
2015-06-10 & 2457184    & $-$      & $-$      & $-$               & $-$       & 3.3    &  $-$      & $-$      & $-$       & $-$ & $-$ & $-$ \DIFdelbeginFL 
\DIFdelFL{$-$ }
\DIFdelFL{$-$ }\DIFdelendFL \\
2015-08-18 & 2457253    & 2.9      & 2.5      & 2.6               & 1.0       & 3.5    &  1.5      & \DIFdelbeginFL \DIFdelFL{0.9 }
\DIFdelendFL 0.6      & 1.6       & \DIFdelbeginFL \DIFdelFL{1.7 }
\DIFdelendFL 1.4 & 3.4 & 1.9 \\
2015-09-27 & 2457293    & 4.7      & 2.1      & 2.6               & 1.3       & 3.2    &  2.1      & \DIFdelbeginFL \DIFdelFL{0.5 }
\DIFdelendFL 0.6      & 1.6       & \DIFdelbeginFL \DIFdelFL{1.6 }
\DIFdelendFL 1.5 & 2.7 & 2.4 \\
2015-10-12 & 2457308    & 3.1      & 2.3      & 2.2               & 1.2       & 3.1    &  1.4      & 0.6      & \DIFdelbeginFL \DIFdelFL{0.6 }
\DIFdelendFL 1.3       & \DIFdelbeginFL \DIFdelFL{1.2 }
\DIFdelendFL 0.9 & 2.0 & 1.9 \\
2015-11-10 & 2457337    & 3.3      & 3.2      & 2.4               & 0.6       & 3.2    &  1.4      & \DIFdelbeginFL \DIFdelFL{0.8 }
\DIFdelFL{$-$ }
\DIFdelendFL $-$      & $-$       & $-$ & $-$ & $-$ \\
2016-01-20 & 2457408    & 2.3      & 3.0      & 2.5               & 1.4       & 3.5    &  1.7      & \DIFdelbeginFL \DIFdelFL{0.6 }
\DIFdelendFL 0.5      & 1.3       & \DIFdelbeginFL \DIFdelFL{1.3 }
\DIFdelendFL 1.0 & 1.9 & 1.7 \\
2016-03-03 & 2457450    & $-$      & $-$      & $-$               & 0.8       & 3.1    &  1.5      & \DIFdelbeginFL \DIFdelFL{0.8 }
\DIFdelendFL 0.6      & 1.3       & \DIFdelbeginFL \DIFdelFL{1.1 }
\DIFdelendFL 0.9 & 2.6 & 2.3 \\
2016-03-15 & 2457463    & 3.4      & 2.8      & 2.4               & 0.9       & 3.2    &  1.4      & \DIFdelbeginFL \DIFdelFL{0.9 }
\DIFdelendFL 0.5      & 1.3       & \DIFdelbeginFL \DIFdelFL{2.6 }
\DIFdelendFL 1.0 & 2.0 & 2.2 \\
2016-06-17 & 2457557    & $-$      & $-$      & $-$               & 1.4       & 3.3    &  1.5      & \DIFdelbeginFL \DIFdelFL{0.9 }
\DIFdelendFL 0.6      & 1.3       & \DIFdelbeginFL \DIFdelFL{2.3 }
\DIFdelendFL 1.1 & 2.5 & 2.2 \\
2016-07-24 & 2457594    & 3.2      & 2.6      & 3.2               & 1.7       & 3.7    &  1.4      & \DIFdelbeginFL \DIFdelFL{0.9 }
\DIFdelendFL 0.6      & 1.4       & \DIFdelbeginFL \DIFdelFL{1.6 }
\DIFdelendFL 0.6 & 1.6 & 1.3 \\
2016-09-26 & 2457658    & 3.1      & 2.4      & 2.5               & 0.9       & 3.6    &  1.3      & 0.7      & \DIFdelbeginFL \DIFdelFL{0.7 }
\DIFdelendFL 1.3       & \DIFdelbeginFL \DIFdelFL{1.5 }
\DIFdelendFL 1.1 & 2.3 & 2.3 \\
2016-10-05 & 2457667    & $-$      & $-$      & $-$               & 0.8       & 3.3    &  1.4      & \DIFdelbeginFL \DIFdelFL{1.0 }
\DIFdelendFL 0.6      & 1.6       & \DIFdelbeginFL \DIFdelFL{1.4 }
\DIFdelendFL 1.1 & 2.5 & 2.1 \\
2016-10-13 & 2457675    & $-$      & $-$      & $-$               & 0.8       & 3.3    &  1.4      & 0.6      & \DIFdelbeginFL \DIFdelFL{0.6 }
\DIFdelendFL 1.1       & \DIFdelbeginFL \DIFdelFL{2.5 }
\DIFdelendFL 1.2 & 2.2 & 1.9 \\
2016-11-17 & 2457710    & 4.2      & 2.5      & 2.6               & 0.9       & 3.3    &  1.3      & \DIFdelbeginFL \DIFdelFL{0.7 }
\DIFdelendFL 0.5      & 1.4       & \DIFdelbeginFL \DIFdelFL{1.5 }
\DIFdelendFL 1.2 & 2.5 & 2.1 \\
2016-12-04 & 2457727    & $-$      & 3.0      & 2.3               & 0.8       & 3.2    &  1.3      & \DIFdelbeginFL \DIFdelFL{0.9 }
\DIFdelendFL 0.5      & 1.3       & \DIFdelbeginFL \DIFdelFL{1.3 }
\DIFdelendFL 0.7 & 2.2 & 1.8 \\
2016-12-08 & 2457731    & 3.6      & 2.8      & 2.3               & 0.7       & 3.0    &  1.6      & \DIFdelbeginFL \DIFdelFL{0.6 }
\DIFdelendFL 0.7      & 1.4       & \DIFdelbeginFL \DIFdelFL{1.6 }
\DIFdelendFL 0.9 & 2.4 & 2.0 \\
2017-04-28 & 2457872 & 3.1  & 1.0  & 2.2   & 1.1  & 3.3  &  1.3 & \DIFaddFL{$-$      }& \DIFaddFL{$-$       }& \DIFaddFL{$-$ }& \DIFaddFL{$-$ }& \DIFaddFL{$-$ }\\
2017-07-03 & 2457938 & 1.9  & 1.6  & 2.1     & 1.1   & 3.4  & 1.2  & 0.7  & 1.0 & 0.5 & 1.8 & 1.7 \\
2017-08-22 & 2457985 & $-$  & \DIFaddFL{$-$      }& \DIFaddFL{$-$               }& 0.8  & 3.2  &  1.3  &   $-$       & 1.2   & 0.8 & 2.3 & 2.1 \\
\DIFaddendFL 2018-08-13 & 2458344    & 3.4      & $-$      & 2.2               & 0.9       & 3.2    &  1.3      & \DIFdelbeginFL \DIFdelFL{1.1 }
\DIFdelendFL 0.4      & 1.4       & \DIFdelbeginFL \DIFdelFL{1.4 }
\DIFdelendFL 0.9 & 2.3 & 1.7 \\
2018-11-06 & 2458429    & 3.8      & 1.9      & 2.4               & 1.6       & 3.3    &  1.5      & \DIFdelbeginFL \DIFdelFL{0.8 }
\DIFdelendFL 0.6      & 1.6       & \DIFdelbeginFL \DIFdelFL{1.1 }
\DIFdelendFL 1.0 & 1.9 & 1.7 \\
2018-11-21 & 2458444    & $-$      & $-$      & $-$               & $-$       & $-$    &  $-$      & $-$      & $-$       & $-$ & $-$ & $-$ \DIFdelbeginFL 
\DIFdelFL{$-$ }
\DIFdelFL{$-$ }\DIFdelendFL \\
2018-12-12 & 2458465    & 4.4      & 2.9      & 2.1               & 0.6       & 3.7    &  1.6      & \DIFdelbeginFL \DIFdelFL{0.7 }
\DIFdelendFL 0.5      & 1.5       & 0.9 & \DIFdelbeginFL \DIFdelFL{0.9 }
\DIFdelendFL 2.0 & 2.6 \\
2018-12-25 & 2458478    & 3.9      & 3.3      & 2.3               & 0.7       & 3.2    &  1.5      & \DIFdelbeginFL \DIFdelFL{0.9 }
\DIFdelendFL 0.6      & 1.7       & \DIFdelbeginFL \DIFdelFL{$-$ }
\DIFdelendFL 0.9 & 2.3 & 1.8 \\
2019-10-28 & 2458785    & $-$      & $-$      & $-$               & 0.8       & 2.7    &  1.4      & 0.6      & 1.5   & 0.5 & 1.4  & 1.5  \\
2021-06-14 & 2459380   & 4.9  & 2.1   & 2.1    & 0.7   & 3.3 & 1.5 & $-$    & 1.1       & 0.5 & 1.6  & 1.6  \\
\hline

\end{tabular}
\end{table*}

\subsubsection{O I line}

    O I line at $\lambda$7773\AA~ is formed in the broad component regions (BCR) which corresponds to the warm
    gas in the envelope or the hot photospheres above the disc in the T Tauri stars. 
    This is because the high energy state of O I $\lambda$7773\AA~ ($\sim$9 eV) prevents its formation in the
    photospheres of cool stars \citep{1992ApJS...82..247H}. 
    Thus, O I $\lambda$7773\AA~ is an indicator of disc turbulence. 
    In the previous studies on V1647 Ori and V899 Mon, \citet[][]{2013ApJ...778..116N,2015ApJ...815....4N}  have shown
    that the W$_\lambda$ of O I $\lambda$7773\AA~ varies significantly as the sources transitioned from the
    quiescent stages to the outburst stages. Therefore, continuous monitoring of O I line provides clues about
    the surrounding environment of the central source.  
    The mean value of W$_\lambda$ of O I at $\lambda$7773\AA~ during our monitoring period is estimated as 1.4$\pm$0.2 \AA~.  
    The scatter in the W$_\lambda$ values is of the order of error in the estimation of W$_\lambda$, thereby, indicating that after the second outburst,
    the inner glowing hot part of the disc has remained stable.

\subsubsection{NIR spectral features }

{\bf CO bandheads:} The CO (2$-$0) and (3$-$1) bandhead spectral absorption features starting at 2.29 $\mu$m are one of the defining characteristics of FUors \citep{1998apsf.book.....H}.
These overtone features are generally observed in many YSOs (either in emission or in absorption) and in giants.
In the FUor/EXor family of sources, these bandhead CO features are believed to originate in the inner part of the heated disc where the temperature ranges between 2000 K $<$ T $<$ 5000
K and 
density n$_H$ is $>$ 10$^{10}$ cm$^{-3}$ \citep{1991ApJ...380..617C,2011ApJ...736...72K}. 
As previously reported by 
\citet[][]{2010ATel.2854....1L}  and \citet{2011ApJ...730...80M}, the CO bandhead features in V2493 Cyg are in absorption. During our monitoring period also, we have found that the CO bandhead 
features to be in absorption. The strength of the bandheads in absorption remained almost the same during our monitoring period. The absorption feature in CO bandheads therefore implies that the 
surface temperature of the inner part of the disc remained hotter than the \DIFdelbegin \DIFdel{central young star \citep{1991ApJ...380..617C}}\DIFdelend middle layer of the 
disc 
(c.f. Figure 7 of \citealt{1991ApJ...380..617C}). \\
{\bf Metallic lines:} The metallic line of Na $\lambda$2.208 $\mu$m is not detected during our spectral monitoring. The Na $\lambda$2.208 $\mu$m if present, follows the profile of
CO bandheads in K owing to their similar ionisation potentials \citep{2009ApJ...693.1056L}.
{\bf Paschen lines:} We have also identified several Paschen lines in our NIR spectra. All the Paschen lines are found to be in absorption. The Paschen$\beta$ line is most prominent in the Paschen family and is observable in both the TIRSPEC and TANSPEC spectra. The absorption  features present in Paschen$\beta$ is one of the defining characteristics of FUors \citep{2018ApJ...861..145C}. We could detect the Paschen (6-3) spectral feature in our TANSPEC spectra.  The Paschen lines are in absorption 
\DIFdelbegin \DIFdel{and blue-shifted, indicating of the
outflowing winds from V2493 Cyg. }
\DIFdelend therefore indicating their origin from the inner disc due to accretion/wind physics \citep{2011ApJ...730...80M}.\\ 
{\bf Brackett lines:} The hydrogen Brackett (11-4) line is detected in absorption. Previously, \citet{2010ATel.2854....1L}, have reported the observed absorption feature in the Brackett 11 line.
During our monitoring period, we have observed a weak Brackett 11 absorption feature only once with TANSPEC in the spectrum obtained on 2020 October 25. This feature is also blue-shifted, possibly indicative of similar origin like that of Paschen lines. \\
{\bf 1.083 $\mu$m He I:} During our monitoring period, we have also observed the absorption feature in He I 1.083 $\mu$m. This absorption 
feature is produced due to the resonant scattering of the 1.083 $\mu$m photons by the metastable triplet state of Helium. This metastable 
state of Helium is produced by the EUV to X-ray radiations due to accretion. The absorption feature of He I 1.083 $\mu$m is attributed to outflows generated due to accretion.

\subsection{Correlation between equivalent width  `W$_{\lambda}$' of spectral lines}

\begin{figure*}
\centering
\includegraphics[width=0.95\textwidth]{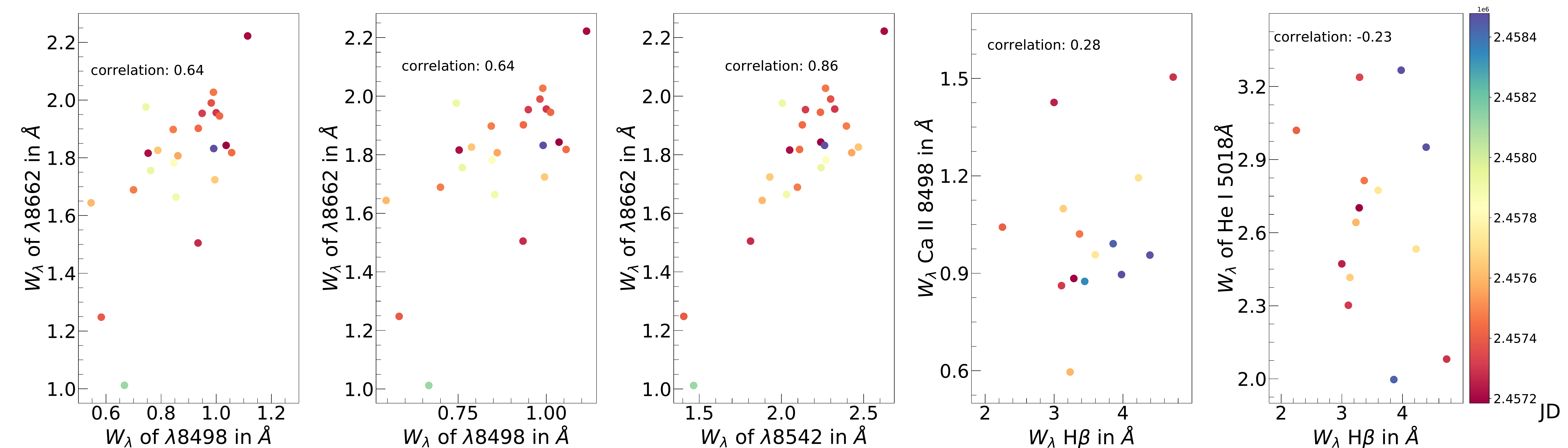}
\caption{\label{f5} Figure showing the correlation between the W$_{\lambda}$ of the different \DIFdelbeginFL \DIFdelFL{spectral lines }\DIFdelendFL spectroscopic features of V2493 Cyg. 
}
\end{figure*}

Correlation between the W$_{\lambda}$ of different spectral lines provides an important diagnostic tool to 
probe the physical association between different regions of a PMS star undergoing outburst \citep{1992ApJS...82..247H,2013ApJ...778..116N}. This in turn help us in 
constraining/refining the various physical models that are used to describe these episodically accreting
systems. Here, we have also tried to see the correlation (Pearson correlation coefficient) between the W$_{\lambda}$ of various lines and few of them are shown in Figure \ref{f5}.

The emission lines of the Ca II IRT originate from/or near the accretion funnels in a magnetospheric accretion in low mass YSOs
\citep{1998AJ....116..455M}. Therefore, these emission lines are found to be tightly correlated as seen in the case of the V1647 Ori \citep[r=0.88,][]{2013ApJ...778..116N}.
In case of V2493 Cyg, the Ca II IRT lines are in absorption.
Correlation between the absorption features of the Ca II IRT to our knowledge has not been studied before. We have calculated
the correlation between the absorption features of the Ca II IRT (between W$_{\lambda}$ of $\lambda$8498\AA~ with $\lambda$8542\AA~ or with 
$\lambda$8662\AA) and found that the coefficients are moderately correlated (r$\sim$0.6).
This moderate correlation is likely due to the additional physical process of the strong outflowing winds occurring with the magnetospheric accretion.
The correlation coefficient between the W$_{\lambda}$ of $\lambda$8542\AA~ line and $\lambda$8662\AA~ line is bit tighter with r=0.86.

We have found very weak correlations
between the $\lambda$8498\AA~ line of the Ca II IRT and H$\beta$ line (r$\sim$0.2) and H$\beta$ and He I $\lambda$5015\AA~/Fe II $\lambda$5018\AA~ lines (r=$-$0.28). This implies weak physical connection between
these absorbing regions.

The Ca II IRT lines act as tracers of the accretion rate, while the O I $\lambda$7773\AA~ is used as a proxy to monitor
the disc turbulence. As both the lines originate in the BCR regions of the T Tauri stars \citep{1992ApJS...82..247H}, we
investigated the correlation between these lines and did not find any.  This could be due to no connection between the absorbing regions. We also did not find any correlation between the Na I D lines and the Fe II $\lambda$ 6496\AA~ which implies that the regions from which they originate are independent of each other.

\subsection{Medium Resolution Echelle Spectra}

\begin{figure*}
\centering
\includegraphics[width=0.95\textwidth]{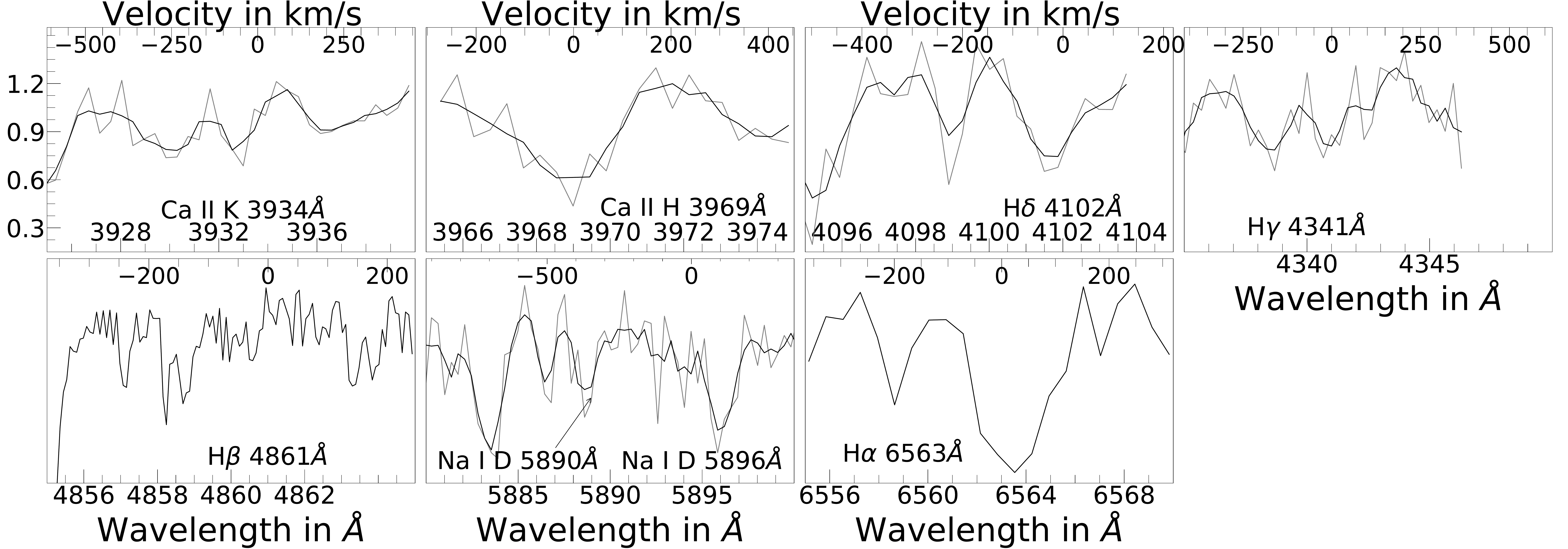}
\caption{\label{mres1} MRES spectra showing \DIFdelbeginFL \DIFdelFL{blueshifted }\DIFdelendFL \DIFaddbeginFL blue-shifted absorption lines that are tracing the
outflowing winds. The gray plot represents the original spectra and the darker plot is the 3
pixel smoothed spectra of the original spectra.}
\end{figure*}

The medium resolution echelle spectra obtained from MRES displayed blue-shifted absorption features that trace
the outflowing disc winds. We can identify the Ca II H and K lines ($\lambda$3934\AA~ and $\lambda$3969\AA~), 
\DIFdelbegin \DIFdel{H$\varepsilon$ $\lambda$3970\AA~, }\DIFdelend 
H$\delta$ $\lambda$4102\AA~, H$\gamma$ $\lambda$4341\AA~,
H$\beta$, Na I D,  and H$\alpha$ lines, etc. 
The Ca II H 
\DIFdelbegin \DIFdel{line is found to be tracing an outflowing wind velocity
of $\sim-$450 km/s, while }\DIFdelend 
\DIFaddbegin \DIFadd{and }\DIFaddend the Ca II K line is tracing a wind velocity of around $-$75 km/s. The outflowing wind velocity as traced by the hydrogen Balmer series lines, 
\DIFdelbegin \DIFdel{H$\varepsilon$, }\DIFdelend 
H$\delta$, H$\gamma$ is \DIFdelbegin \DIFdel{$\sim-$150 km/s,
}\DIFdelend 
$-$150 km/s and $-$350 km/s, respectively. The wind velocity traced by H$\beta$ is $-$250 km/s. H$\alpha$ line 
is found to be in absorption. The absorption feature in H$\alpha$ observed using MRES is consistent with the
absorption feature that we observed in our HFOSC spectra from 2019 October 28 onwards. Both the Na I D lines at 
$\lambda$5890$\AA$ and $\lambda$5896$\AA$ are blue-shifted and trace wind velocities of $\sim-155$ km/s and
$-469$ km/s respectively. Figure \ref{mres1} shows the plot of the outflowing wind features obtained from the  MRES spectrum.

\begin{figure}
\centering
\includegraphics[width=0.45\textwidth]{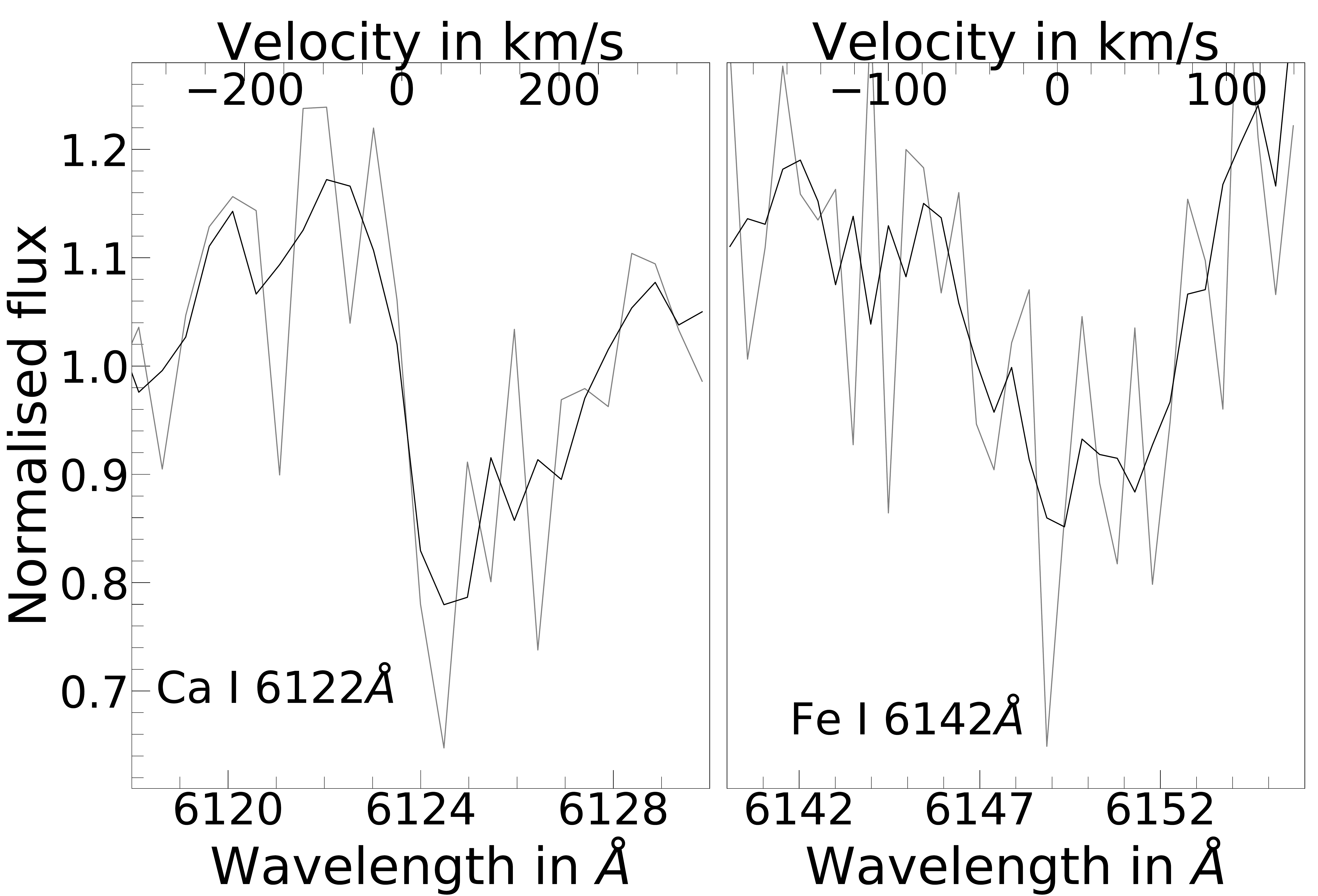}
\includegraphics[width=0.45\textwidth]{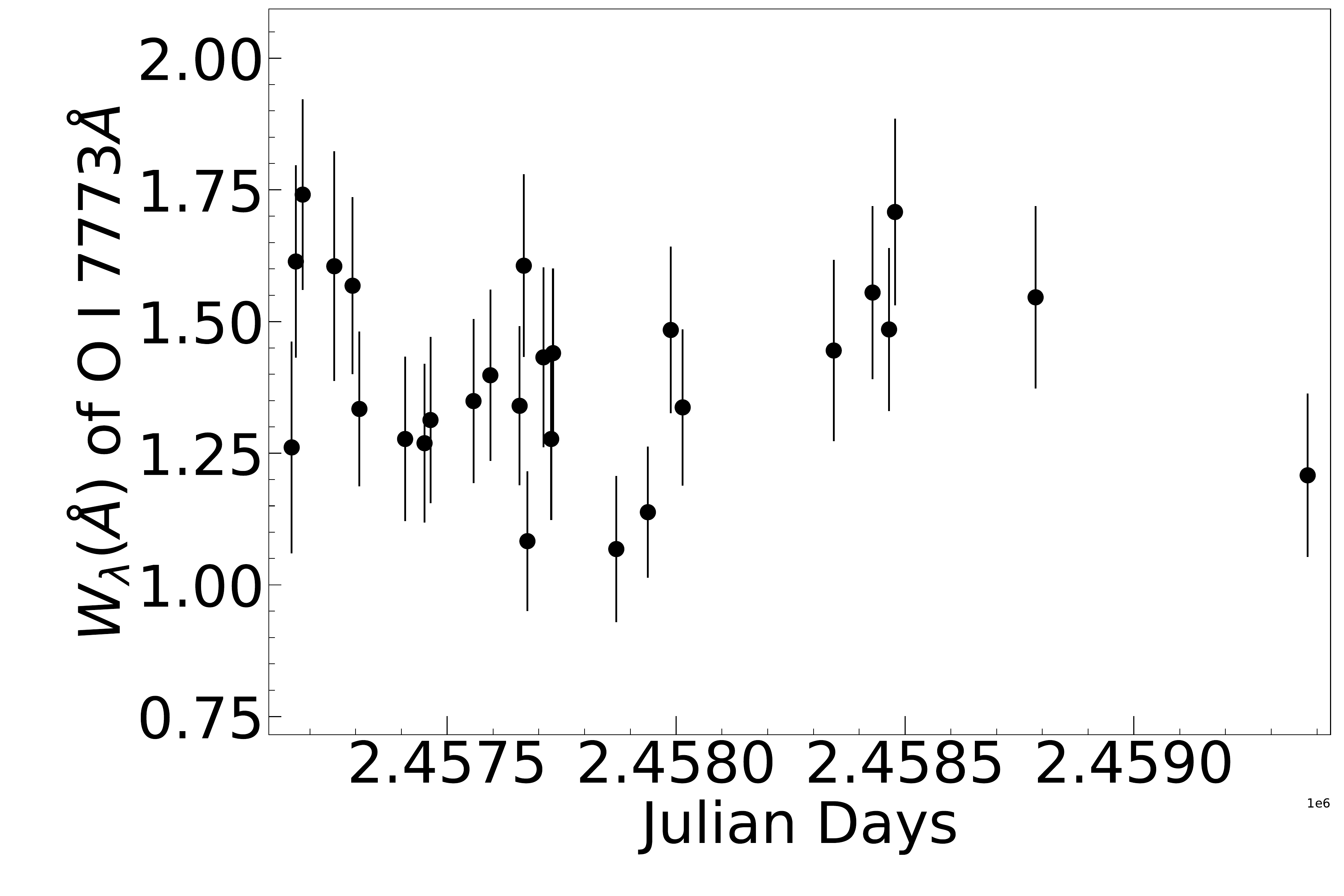}
\caption{\label{mres2} Top panel shows the MRES absorption lines of Ca I $\lambda$6122\AA~ and
Fe I $\lambda$6142\AA~ which show the evidence of disc rotation. The gray plot represents the 
original spectra and the darker plot is the 3 pixel smoothed spectra of the original spectra.
Bottom panel shows the  evolution of the W$_{\lambda}$  of O I $\lambda$7773\AA~ line which
acts as a tracer for disc turbulence.} 
\end{figure}

We also identify two neutral metallic lines Ca I $\lambda$6122\AA~ and Fe I $\lambda$6142\AA~ in our 
MRES spectrum (cf. top panel of Figure \ref{mres2}). Previously, \citet{2015ApJ...807...84L} had identified these two lines during their high
resolution spectroscopic monitoring using HET-HRS and attributed the presence of these lines as evidence
for disc rotation. They further concluded from their spectroscopic monitoring about the steady rebuilding of the inner disc post outburst. Our detection of these two neutral metallic lines also points towards similar conclusion. 
 Bottom panel of Figure \ref{mres2} shows the evolution of the O I $\lambda$7773\AA~ during our monitoring period. 
 \DIFdelbegin \DIFdel{Variation of the W$\lambda$ }\DIFdelend 
 The variation of W$_{\lambda}$ of the O I $\lambda$7773\AA~ \DIFdelbegin \DIFdel{line indicates towards the 
 ongoing disc turbulence in V2493 Cyg.  }\DIFdelend is within the error bars, which therefore possibly implies towards the 
 stabilisation of the disc post the second outburst. The W$_{\lambda}$ of O I 7773\AA~ is believed to be an indicator of disc turbulence \citep{1992ApJS...82..247H}.

\subsection{Outflow wind velocities }  \label{outflow_velocity}

Strongly blue-shifted absorption lines are one of the defining characteristics of bona-fide FUors. The wind features observed in V2493 Cyg 
varied with time during our monitoring period. We have calculated the outflow wind velocities of V2493 Cyg from several absorption  lines, i.e., 
H$\beta$, He I $\lambda$5015\AA~/Fe II $\lambda$5018\AA~, Mg I $\lambda$5167\AA~/Fe II 5169\AA~,  Ba II $\lambda$5853\AA~, Na I D,
Fe II $\lambda$6496\AA~, K I $\lambda$7699\AA~ and the Ca II IRT lines \citep{1998apsf.book.....H}, and these are
listed in the Table \ref{tab:velocity_log}.
We have found variation in the shape of the H$\beta$ line profile during our monitoring period.During our monitoring period, the H$\beta$ 
line was not detected on 2016 June 17 but it re-appeared on 2016 July 24. H$\beta$ was undetected again on 2016 October 13 before re-appearing 
again on 2016 November 17. The absorption profile of H$\beta$ almost 
disappeared in 2016 December 04 while it re-appeared prominently again on 2016 December 08. Similar behaviour in the H$\beta$ line evolution is 
again observed on 2018 December 12. The variation in the line profile of H$\beta$ coincided with the line profile variations of H$\alpha$, 
therefore likely pointing towards a common origin. The shape of
the line profile of He I $\lambda$5015\AA~/Fe II $\lambda$5018\AA~ also varied during our monitoring period. The He I $\lambda$5015\AA~/Fe II $\lambda$5018\AA~ displayed absorption profile till 2017 April 28. The absorption profile disappeared from our spectra on 2017 September 17 till
the end of our monitoring period upto 2021 June 14. We interpret this disappearance with caution. The disappearance maybe due to the spectral 
feature becoming too faint to be detectable in our spectrum.

During our monitoring period, the mean velocities of the Ca II IRT lines are $-$77$\pm$35, $-$87$\pm$35 and 
$-$88$\pm$37 km/s respectively. The mean velocities of H$\beta$, He I $\lambda$5015\AA~/Fe II $\lambda$5018\AA, Mg I $\lambda$5167\AA~/Fe II 5169\AA~, 
Ba II $\lambda$5853\AA~, Na I D, Fe II $\lambda$6496\AA~ and K I $\lambda$7699\AA~
lines come out to be $-$86$\pm$46, $-$115$\pm$51, $-$50$\pm$49, $-$62$\pm$38,
$-$67$\pm$38, $-$103$\pm$66 and $-$91$\pm$55 km/s respectively. The typical error in 
estimation of the velocities is $\sim$\DIFdelbegin \DIFdel{7 }\DIFdelend 25 km/s (c.f \citealt{2022ApJ...926...68G}), hence the large scatters in the velocities imply intrinsic variations
of the wind velocity as traced by the different lines. This observation is also supported by the H$\alpha$ line profile variations as described in Section \ref{halpha}.

\begin{table*}
\centering
\caption{The variation of the wind velocity (in km s$^{-1}$) as calculated from the various lines identified in our study during
our monitoring period. The error in estimating the velocity is $\sim$25 km s$^{-1}$. The error is estimated
by taking three independent measurements of the wavelength center and then calculating the standard
deviation of the velocity which is calculated from the wavelength shift. }
\label{tab:velocity_log}
\begin{tabular}{c@{ }c@{ }c r@{ }r@{ }r@{ }r@{ }r@{ }r@{ }r@{ }r@{ }r@{  }r@{  } r@{ }}
\hline
Date       &  Julian Date   & H$\beta$    & He I/Fe II    & Mg I/Fe II       & Ba II     & Na I D   & Fe II    &  K I     &     &     Ca II IRT             &  \\
           &                &$\lambda$4861&$\lambda$5015/$\lambda$5018&$\lambda$5167/$\lambda$5169&$\lambda$5853&$\lambda$5890/6&$\lambda$6496&$\lambda$7699& $\lambda$8498 & $\lambda$8542 & $\lambda$8662 \\
           \hline
2015-05-17 &  2457160.267523 & $-$        & $-$         & $-$           & $-$20   & $-$72   & $-$68   & $-$117  & $-$82   & $-$91    &  $-$89       \\
2015-05-26 &  2457169.346111 & $-$        & $-$         & $-$           & $-$21   & $-$66   & $-$51   & $-$46   & $-$98   & $-$111   &  $-$114      \\
2015-06-10 &  2457184.332384 & $-$66      & $-$71       & $-$29         & $-$     & $-$14   & $-$70   & $-$84   & $-$94   & $-$101   &  $-$112      \\
2015-08-18 &  2457253.342766 & $-$137     & $-$110      & $-$103        & $-$     & $-$54   & $-$65   & $-$94   & $-$97   & $-$129   &  $-$142      \\
2015-09-27 &  2457293.279664 & $-$84      & $-$40       & $-$14         & $-$82   & $-$78   & $-$12   & $-$86   & $-$61   & $-$65    &  $-$72       \\
2015-10-12 &  2457308.150961 & $-$53      & $-$79       & $-$40         & $-$     & $-$34   & $-$20   & $-$30   & $-$50   & $-$65    &  $-$77       \\
2015-11-10 &  2457337.234954 & $-$165     & $-$193      & $-$45         & $-$70   & $-$84   & $-$12   & $-$     & $-$     & $-$      &  $-$         \\
2016-01-20 &  2457408.048056 & $-$106     & $-$154      & $-$33         & $-$     & $-$75   & $-$82   & $-$99   & $-$106  & $-$107   &  $-$124      \\
2016-03-03 &  2457450.448137 & $-$        & $-$         & $-$           & $-$     & $-$13   & $-$     & $-$17   & $-$47   & $-$53    &  $-$72       \\
2016-03-15 &  2457463.448137 & $-$15      & $-$75       & $-$18         & $-$13   & $-$79   & $-$51   & $-$67   & $-$88   & $-$95    &  $-$94       \\
2016-06-17 &  2457557.416366 & $-$        & $-$         & $-$           & $-$28   & $-$78   & $-$188  & $-$199  & $-$155  & $-$169   &  $-$175      \\
2016-07-24 &  2457594.171053 & $-$106     & $-$162      & $-$16         & $-$29   & $-$86   & $-$121  & $-$91   & $-$25   & $-$      &  $-$10       \\
2016-09-26 &  2457658.136389 & $-$48      & $-$37       & $-$22         & $-$18   & $-$113  & $-$40   & $-$36   & $-$     & $-$      &  $-$        \\
2016-10-05 &  2457667.278981 & $-$        & $-$         & $-$           & $-$46   & $-$56   & $-$93   & $-$31   & $-$     & $-$      &  $-$        \\
2016-10-13 &  2457675.187049 & $-$        & $-$         & $-$           & $-$103  & $-$134  & $-$169  & $-$184  & $-$105  & $-$104   &  $-$100      \\
2016-11-17 &  2457710.142685 & $-$135     & $-$231      & $-$12         & $-$158  & $-$156  & $-$50   & $-$49   & $-$25   & $-$32    &  $-$30       \\
2016-12-04 &  2457727.101065 & $-$        & $-$161      & $-$83         & $-$57   & $-$56   & $-$48   & $-$115  & $-$84   & $-$73    &  $-$91       \\
2016-12-08 &  2457731.166157 & $-$71      & $-$105      & $-$64         & $-$96   & $-$52   & $-$85   & $-$53   & $-$58   & $-$51    &  $-$57       \\
2018-08-13 &  2458344.283194 & $-$85      & $-$148      & $-$59         & $-$130  & $-$22   & $-$130  & $-$124  & $-$78   & $-$72    &  $-$81       \\
2018-11-06 &  2458429.140617 & $-$101     & $-$140      & $-$15         & $-$61   & $-$56   & $-$32   & $-$51   & $-$11   & $-$36    &  $-$68       \\
2018-12-12 &  2458465.125059 & $-$24      & $-$68       & $-$67         & $-$     & $-$97   & $-$56   & $-$37   & $-$19   & $-$23    &  $-$17       \\
2018-12-25 &  2458478.104021 & $-$40      & $-$73       & $-$10         & $-$114  & $-$57   & $-$16   & $-$27   & $-$     & $-$      &  $-$56       \\
2019-10-28 &  2458785.225682 & $-$        & $-$         & $-$           & $-$85   & $-$31   & $-$209  & $-$172  & $-$115  & $-$111   &  $-$116      \\
2017-04-25 &  2457869.32894  & $-$68      & $-$89       & $-$62         & $-$51   & $-$67   & $-$194  & $-$160  & $-$131  & $-$134   &  $-$139      \\
2017-04-28 &  2457872.27083  & $-$90      & $-$110      & $-$44         & $-$57   & $-$27   & $-$128  & $-$     & $-$     & $-$      &  $-$         \\
2017-07-03 &  2457938.32894  & $-$        & $-$         & $-$29         & $-$31   & $-$17   & $-$126  & $-$36   & $-$     & $-$      &  $-$       \\
2017-08-22 &  2457988.32894  & $-$        & $-$         & $-$           & $-$47   & $-$174  & $-$256  & $-$175  & $-$73   & $-$81   &  $-$100      \\
2017-09-17 &  2458014.32894  & $-$198     & $-$         & $-$245        & $-$95   & $-$57   & $-$221  & $-$179  & $-$78   & $-$118  &  $-$99       \\
2021-06-14 &  2459379.53727  & $-$55      & $-$142      & $-$61         & $-$28   & $-$48   & $-$118  & $-$108  & $-$102  & $-$107  &  $-$87       \\

\hline       
\end{tabular}
\end{table*}

\section{Discussion and Conclusion} \label{pt4}

V2493 Cyg is a bonafide FUor which underwent its first outburst in the summer of 2010 and then
transitioned to a short intermediate quiescent stage before re-brightening to maximum brightness in 2012 April. 
We have carried out our near simultaneous spectro-photometric monitoring from 2013 September 27 till 2021 June 14. During our monitoring period, particularly in the time period spanning between 2015 and 2019, V2493 Cyg
dimmed by $\sim$0.6 mag in V-band from its peak V-band magnitude on 2010 August at an average rate of $\sim$27.6 $\pm$ 5.6 mmag/yr. The average value of the decay rate for V2493 Cyg is of the same order as that of FU Ori and BBW 76 at 14mmag/yr and 23 mmag/yr, respectively \citep{2014prpl.conf..387A}. The similar timescales of the decay rates therefore possibly point towards the occurrence of similar relaxing 
phenomenon in V2493 Cyg as that in FU Ori and BBW 76. 
This is typical behaviour of FU Ori type objects post outburst. As the inner disc of FU Ori type stars depletes, and it takes around $\sim$100 years \DIFdelbegin \DIFdel{to replenish again to trigger another outburst }\DIFdelend \DIFaddbegin for the inner part of the disc to be completely depleted 
\citep{2014prpl.conf..387A}.
Our simultaneous optical $V$, $R_C$ and $I_C$ band photometry revealed significant color changes in the Phase 6
compared to the previous phases with V$-$R$_C$ color and V$-$I$_C$ color becoming redder by 0.14 and 0.32 mag,
respectively. This feature is also evident in the multi-epoch SEDs where the slope at the optical regime \DIFdelbegin \DIFdel{became
}\DIFdelend has become steeper indicating reddening of the optical colors. \DIFdelbegin \DIFdel{Investigating this color change by the method of reddening invariant color analysis points towards that this reddening is not due to the change in the extinction events but due to a gradual decline of the physical process (accretion rate) that had triggered the second outburst. Combining our NIR photometric data with the previously published NIR data revealed two distinct clustering of V2493 Cyg in the NIR CMD planewith one clustering corresponding to Phase 1 and 2 and the other corresponds to Phase 6 of the LC. In the NIR CC diagram, we note the excursion of }\DIFdelend 
The physical origins of the optical color evolution of V2493 Cyg was further studied using the recently developed theoretical models based on the optical CMD parameter space by \citep{2022ApJ...936..152L}. The isomass curves divides the regions of the optical CMD plane into regions based on the nature of accretion. The evolution of optical colors during our monitoring reveals that V2493 is still accreting via the boundary layer accretion, the so called “FUor regime”. Therefore, the reddening of the optical colors is likely due to the expansion of the emitting region around V2493 Cyg. Similar conclusion has been derived by \citet{2021Symm...13.2433S} in their study. The NIR color evolution also follows the trend of the optical colors which possibly hints towards a similar conclusion as that of the optical colors.

The most interesting photometric evolution of V2493 Cyg is its MIR color evolution. While the optical colors became redder in Phase 6, we have found that the MIR colors $W1-W2$, obtained from the 
NEOWISE survey, became bluer.  
The blueing of the MIR colors as observed in the case of V2493 Cyg can be possibly attributed to the brightening of the disc due to the boundary layer accretion \citep{2022ApJ...936..152L}.

        Combining our spectroscopic data with the previously published data by
        \citet{2015ApJ...807...84L}, we see that there is a gradual transition from the P-Cygni profile to 
        absorption feature in the Ca II $\lambda$8498 \AA~ line. The remaining two lines of the Ca II IRT 
        \DIFdelbegin \DIFdel{is }\DIFdelend are also found to be in absorption during our monitoring period. Such transition to the absorption
        features points towards \DIFdelbegin \DIFdel{the dominance of the outflowing wind over the physical process that triggered
        the second outburst. Our period analysis of the W$_\lambda$ of the Ca II IRT revealed weak
        periodicities of $\sim$ 12.3 days, 9.5 days and 10.2 days respectively. Such periodicities possibly
        implies at the changing of the continuum levels and is also observed in the small scale magnitude
        variations in the photometric I$_C$ band light curve. Similar conclusions have been made in the case of
        another well studied source V1647 Ori by \citet{2013ApJ...778..116N}.
        This view is also supported by the evolution of the H$\alpha$ line.}\DIFdelend the heavy outflow regime during our monitoring period.
        \DIFaddend Combining the results of the
        evolution of H$\alpha$ line during our monitoring period with that of \citet{2011ApJ...730...80M}, we
        find that the emission feature evolved into Type II-B, Type-II R, \DIFdelbegin \DIFdel{P Cygni }\DIFdelend P-Cygni profile and absorption
        features as categorised by the H$\alpha$ atlas of \citet{1996A&AS..120..229R}. Theoretical modelling 
        of these variety of line profiles by \citet{1990ApJ...349..168H,2006MNRAS.370..580K} reveals that the observed profiles is a result 
        of the outflowing winds modulated by variable wind acceleration.  
        The temperature of the systems also varies in between 5000 K to 7500 K in the aforementioned
        different models. These variations can also be observed in the R-band LC of V2493 Cyg.
        We have also studied the evolution of the O I $\lambda$7773 
        \AA~ line and found that the scatter in its W$_{\lambda}$ is of the order of the error in estimation
        of W$_{\lambda}$. This implies at possible stabilization of the disc after second outburst. Previous
        studies on V1647 Ori and V899 Mon by \citet{2013ApJ...778..116N,2015ApJ...815....4N} and Gaia 20eae by \citet{2022ApJ...926...68G} have 
        revealed that this line acts as tracer of disc turbulence. Any sudden
        variations of the W$_{\lambda}$ of 
        O I $\lambda$7773\AA~ line point towards the possible transition between active and quiescent
        states. This view is also enhanced by the presence of the neutral metallic lines of calcium and iron
        at $\lambda$6122 and $\lambda$6142\AA~ respectively as observed in the
        MRES spectrum.  These lines have attributed as signatures of disc rotation by \citet{2015ApJ...807...84L} which they stated as evidence 
        for the steady re-building of the inner disc post the second outburst.
        The NIR spectra of V2493 Cyg consist of the blue-shifted absorption features indicative of the outflowing winds from the source. The 
        \DIFdelbegin \DIFdel{two }\DIFdelend most notable spectral features that were not detected
        in the spectrum of V2493 Cyg is the Br$\gamma$ 2.16 $\mu$m\DIFdelbegin \DIFdel{and the He I 1.083 $\mu$m}\DIFdelend . Br$\gamma$ is an important tracer of the accretion rate and is typically 
        observed in highly accreting T Tauri stars,
        though it is absent in many FUors \citep{2015ApJ...815....4N}. In the NIR spectrum, we have detected the He I 1.083 $\mu$m feature. The 
        He I 1.083 $\mu$m acts a proxy indicator for the outflowing winds generated due to the enhanced accretion rate \citep{2022ApJ...926...68G}.  Non-detection of \DIFdelbegin \DIFdel{these two lines likely indicate }\DIFdelend the Br$\gamma$ line in emission further reinforces the notion that V2493 Cyg is 
        likely accreting via boundary layer accretion which is in conformity with the evolution of V2493 Cyg \DIFdelbegin \DIFdel{away from the 
        accretion dominant regime. }\DIFdelend in the 
        optical CMD plane. The NIR spectroscopic features of CO bandheads display almost no variations during our monitoring. Therefore, it can
        be empirically stated that there has been no appreciable change in accretion rate post second outburst as the strength of the
        bandheads is directly related to the accretion rate \citep{1991ApJ...380..617C}. 

\acknowledgments

We thank the anonymous reviewer for valuable comments which greatly improved the scientific content of the paper.
We thank the staff at the 1.3m DFOT and 3.6m DOT, Devasthal (ARIES), for their co$-$operation during observations.
It is a pleasure to thank the members of 3.6m DOT team and IR astronomy group at TIFR for their support
during TANSPEC observations. TIFR$-$ARIES Near Infrared Spectrometer (TANSPEC) was built in collaboration with TIFR,
ARIES and MKIR, Hawaii for the DOT. 
We thank the staff of IAO, Hanle and CREST, Hosakote, that made these observations
possible. The facilities at IAO and CREST are operated by the Indian
Institute of Astrophysics. We also thank the staff at NARIT for the successful observations. 
We acknowledge with thanks the variable star observations from the AAVSO International Database contributed by observers worldwide and used in this research.
SS, NP, RY  acknowledge  the support of the Department of Science  and Technology,  Government of India, under project No. DST/INT/Thai/P-15/2019.
DKO acknowledges the support of the Department of Atomic Energy, Government of India, under Project Identification No. RTI 4002.
JPN and DKO acknowledge the support of the Department of Atomic Energy, Government of India, under project Identification No. RTI 4002.

%

\vspace{5mm}
\DIFdelbegin 
\DIFdelend \DIFaddbegin \facilities{2m HCT (HFOSC \& TIRSPEC), 2.4m TNO (MRES), 3.6m DOT (TANSPEC), ZTF, AAVSO, 1.3m DFOT, PTF, AAVSO}
\DIFaddend 


\software{astropy \citep{2013A&A...558A..33A},  
          IRAF \citep{1986SPIE..627..733T,1993ASPC...52..173T}, 
          DAOPHOT-II \citep{1987PASP...99..191S}
          }
\DIFdelbegin 

\DIFdelend 

\newpage
\bibliography{v2493cyg}{}
\bibliographystyle{aasjournal}

\appendix

\section{FIGURES 5 and 6, to be kept in ONLINE form.}

\setcounter{figure}{4}

\begin{figure}
\centering
\includegraphics[width=0.95\textwidth]{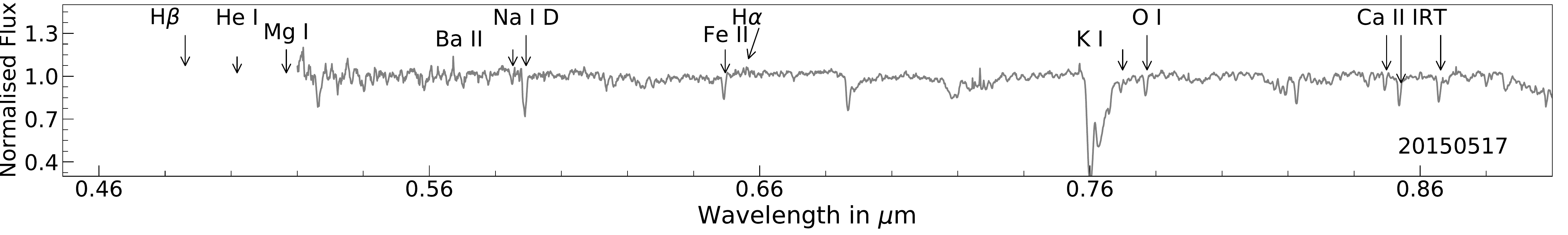}
\includegraphics[width=0.95\textwidth]{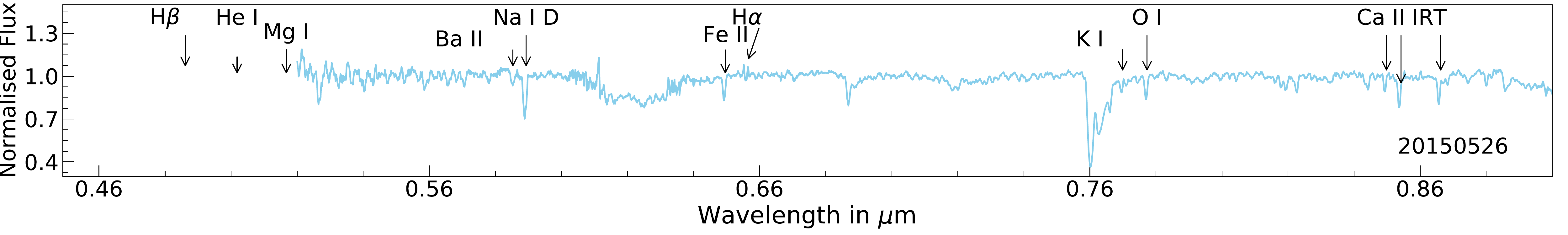}
\includegraphics[width=0.95\textwidth]{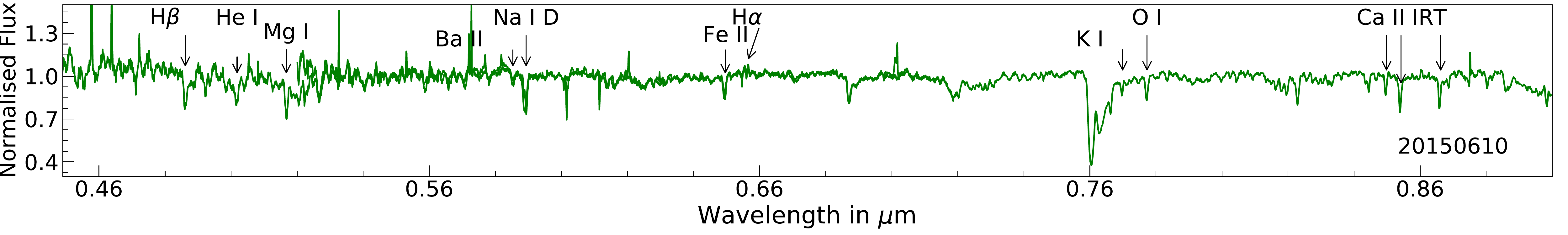}
\includegraphics[width=0.95\textwidth]{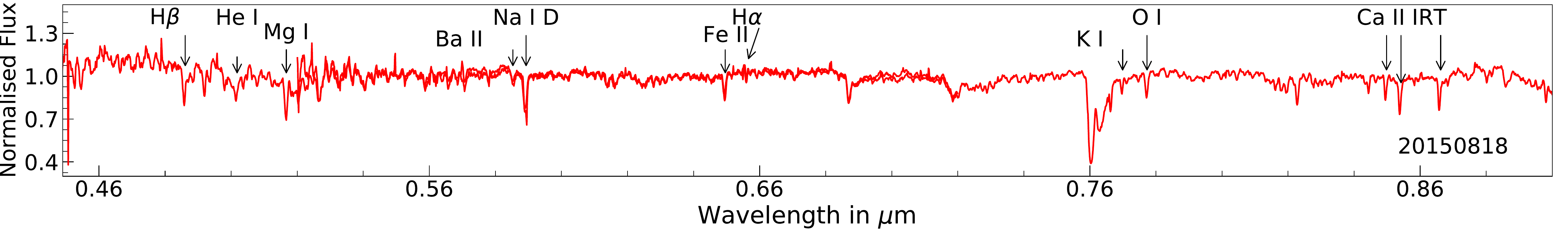}
\includegraphics[width=0.95\textwidth]{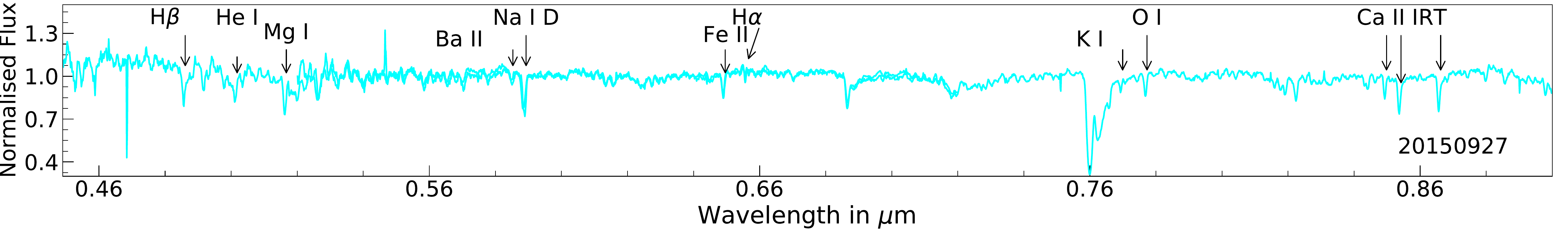}
\includegraphics[width=0.95\textwidth]{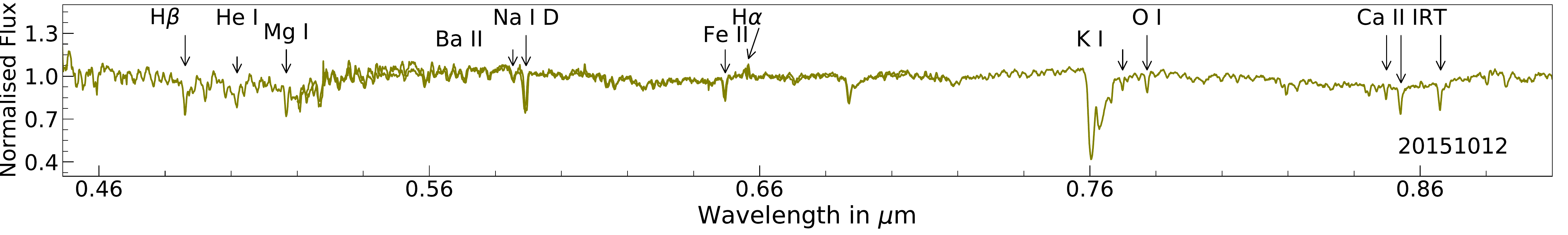}
\includegraphics[width=0.95\textwidth]{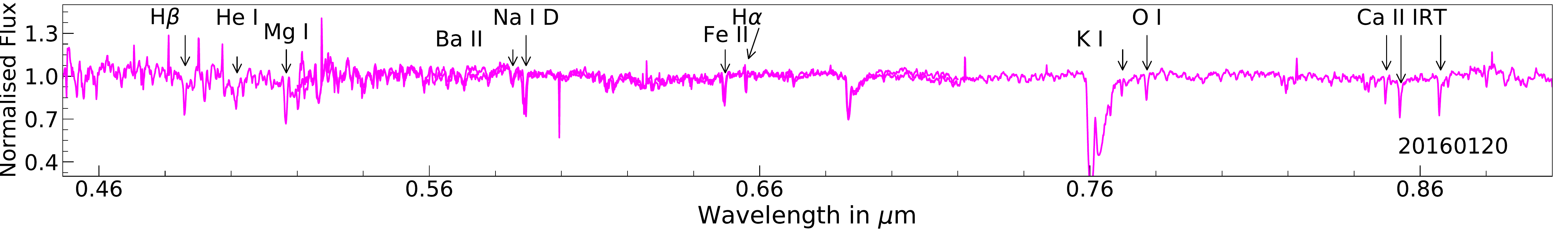}
\includegraphics[width=0.95\textwidth]{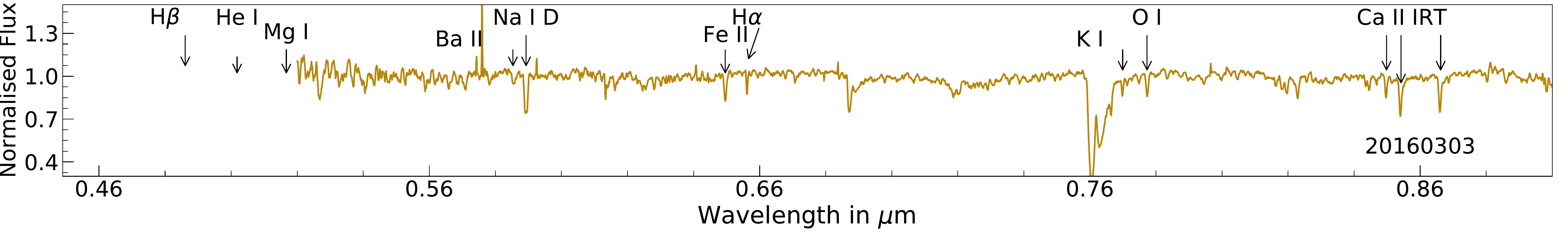}
\includegraphics[width=0.95\textwidth]{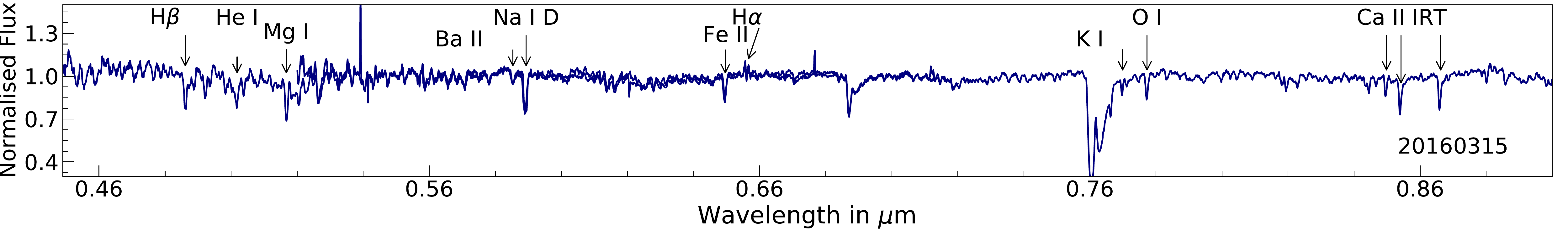}
\caption{\label{all_spectra1} Normalised spectra of V2493 Cyg of all the dates obtained during our monitoring period using HFOSC ($\sim$0.4-0.9 $\mu$m), TIRSPEC ($\sim$1.0-2.4 $\mu$m) on 2m HCT and TANSPEC ($\sim$0.65-2.4 $\mu$m) on 3.6m DOT. The lines that are used for the present study has been marked.}
\end{figure}
\setcounter{figure}{4}
\begin{figure}
\centering
\includegraphics[width=0.95\textwidth]{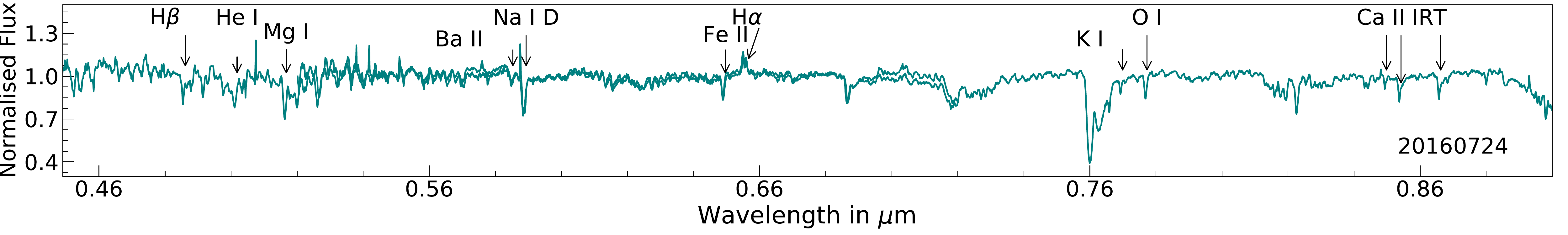}
\includegraphics[width=0.95\textwidth]{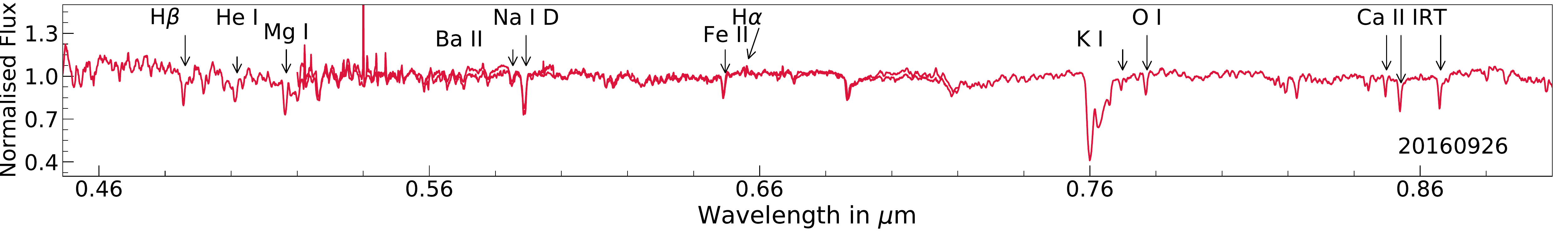}
\includegraphics[width=0.95\textwidth]{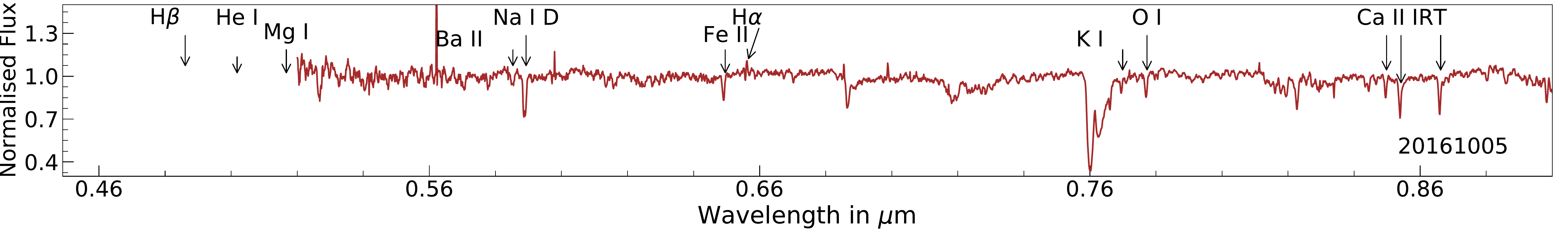}
\includegraphics[width=0.95\textwidth]{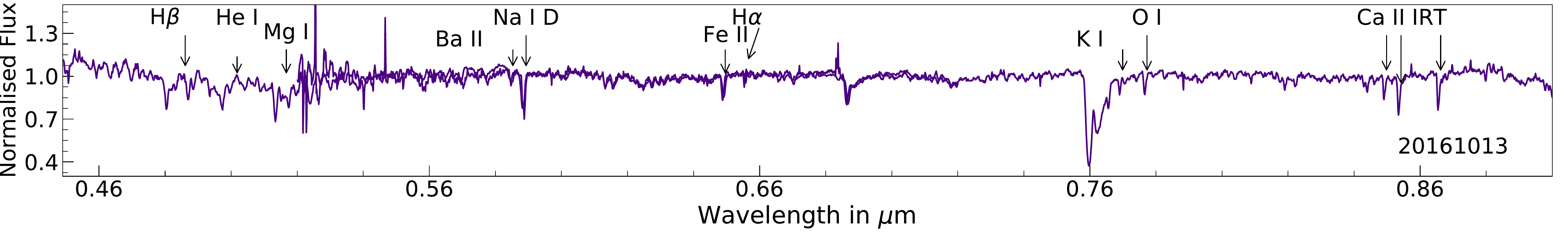}
\includegraphics[width=0.95\textwidth]{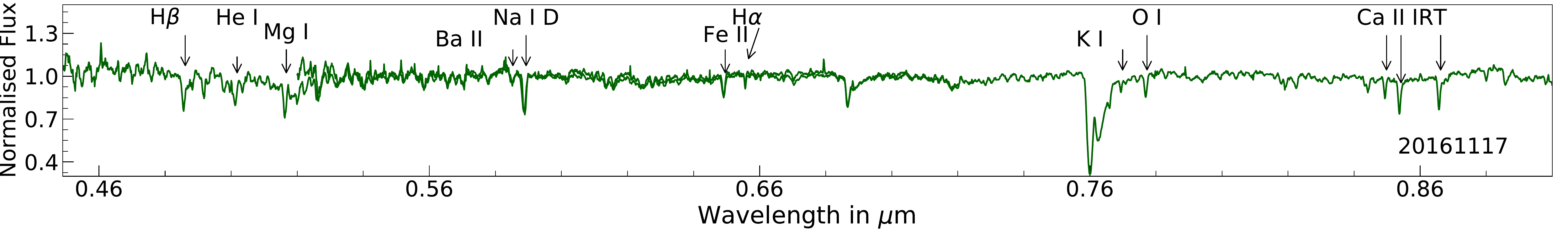}
\includegraphics[width=0.95\textwidth]{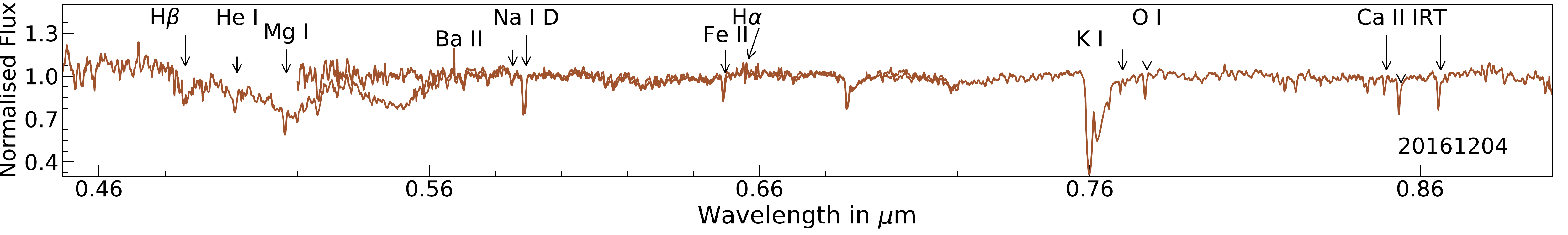}
\includegraphics[width=0.95\textwidth]{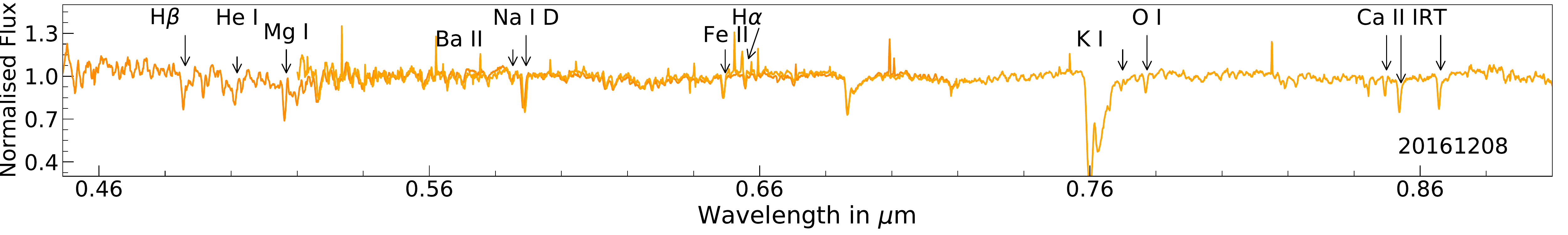}
\includegraphics[width=0.95\textwidth]{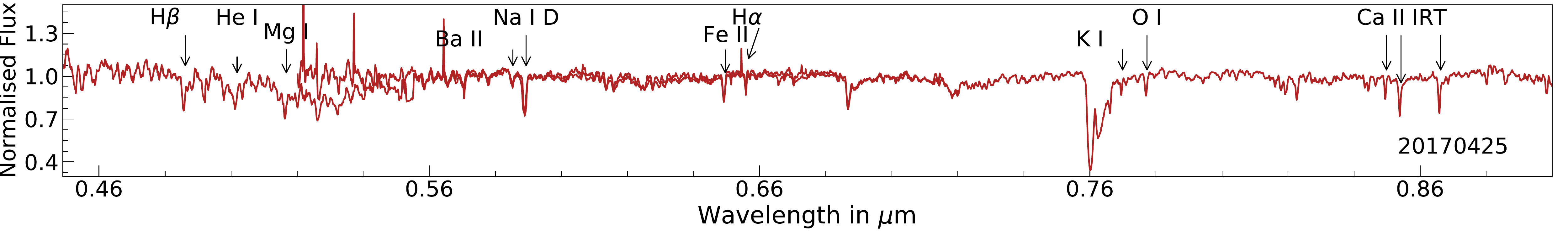}
\includegraphics[width=0.95\textwidth]{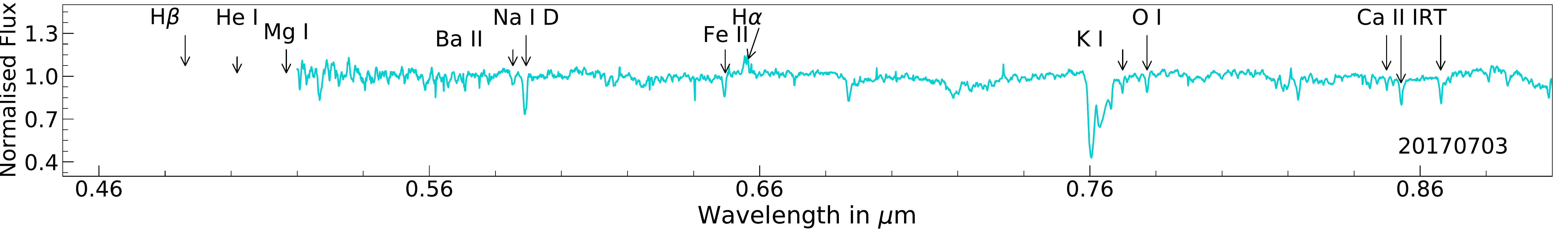}
\caption{\label{all_spectra2} Contd.}
\end{figure}

\setcounter{figure}{4}
\begin{figure}
\centering
\includegraphics[width=0.95\textwidth]{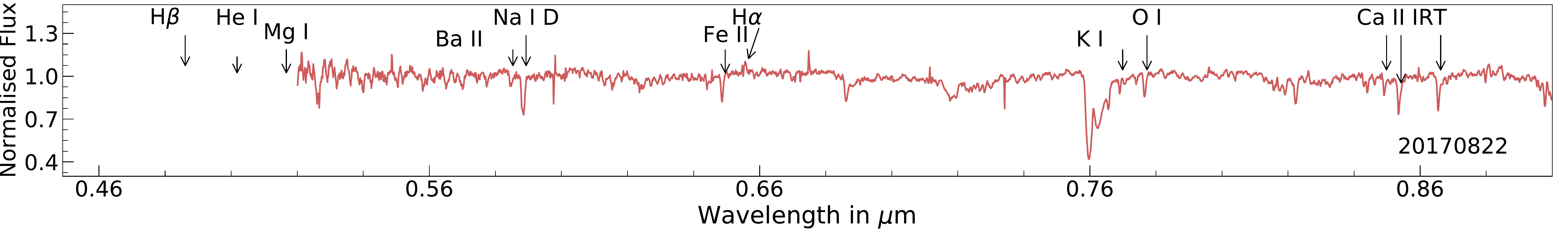}
\includegraphics[width=0.95\textwidth]{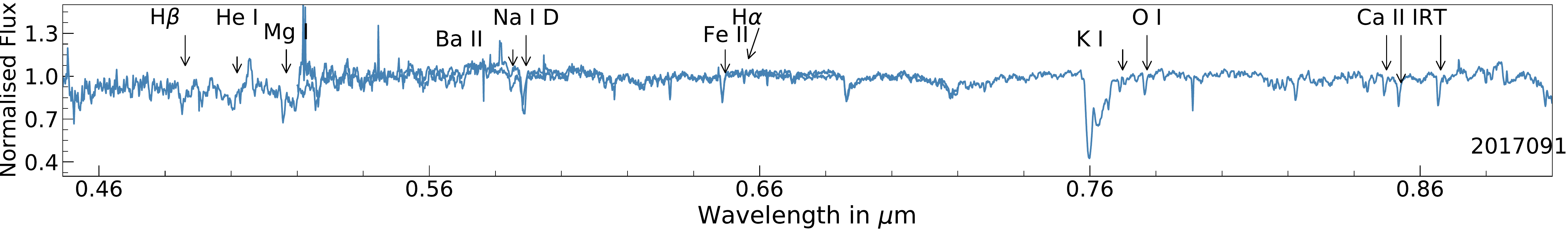}
\includegraphics[width=0.95\textwidth]{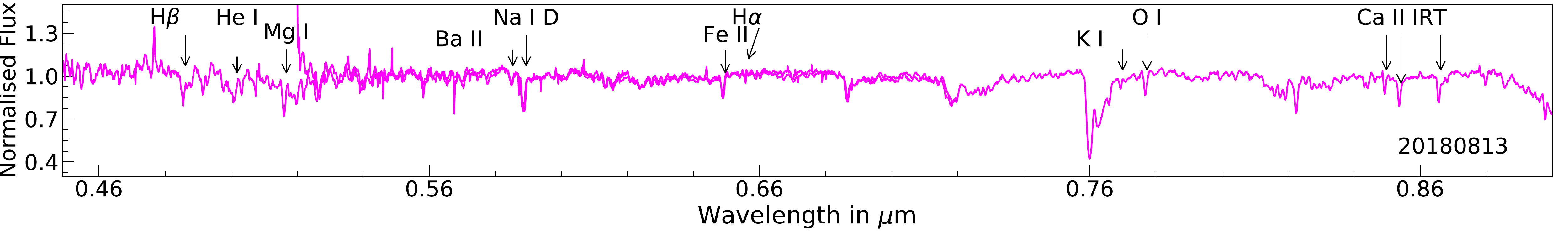}
\includegraphics[width=0.95\textwidth]{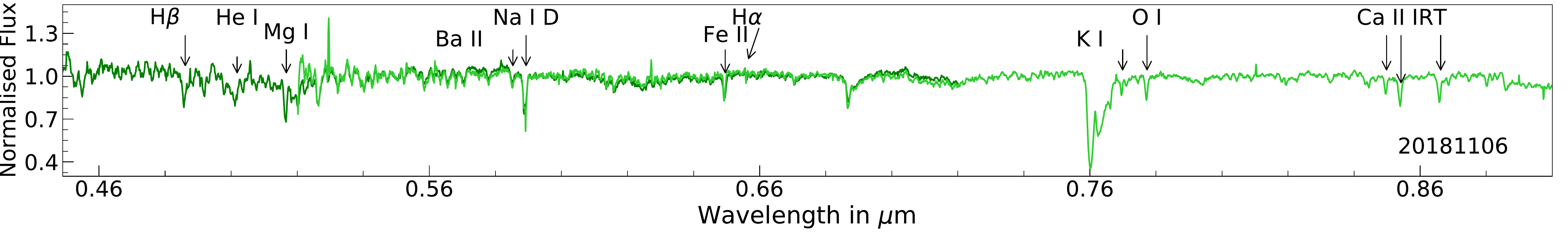}
\includegraphics[width=0.95\textwidth]{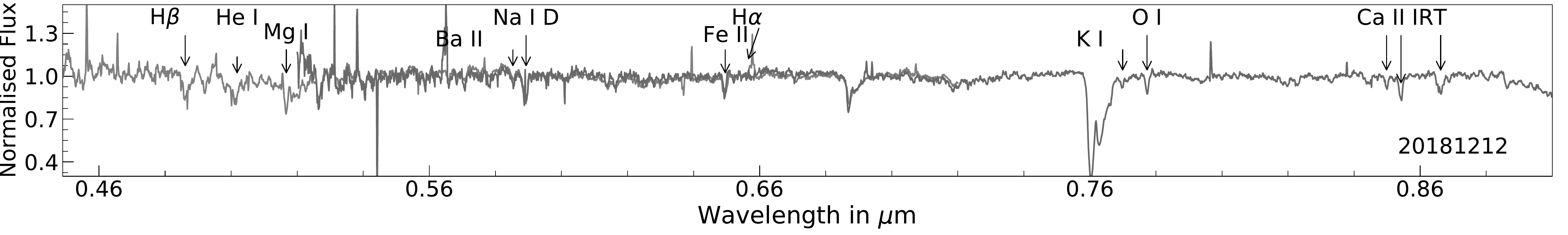}
\includegraphics[width=0.95\textwidth]{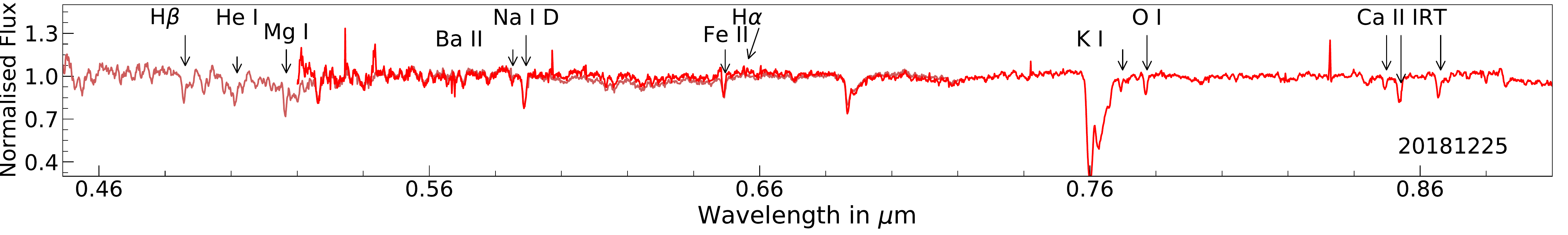}
\includegraphics[width=0.95\textwidth]{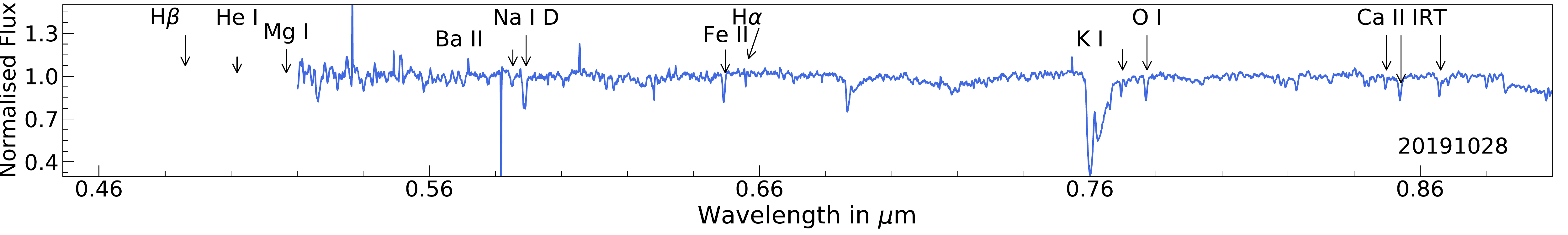}
\includegraphics[width=0.95\textwidth]{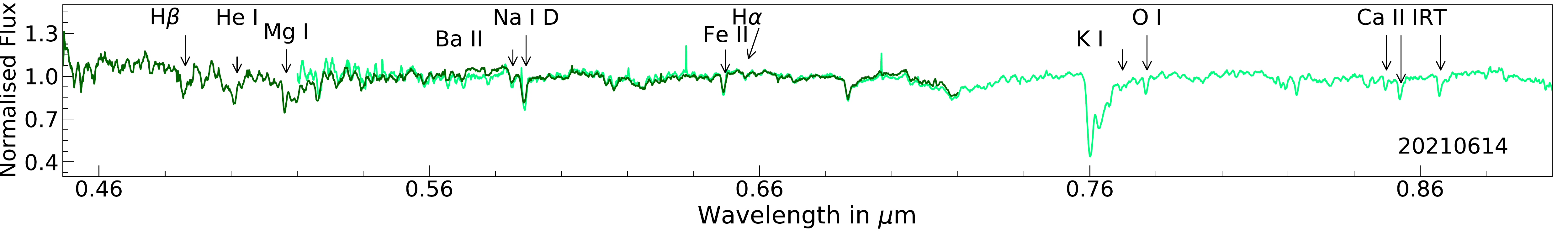}
\includegraphics[width=0.95\textwidth]{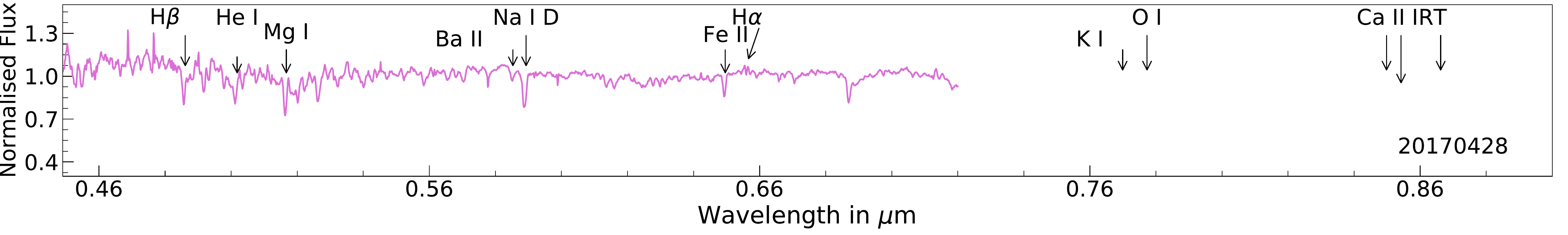}
\caption{\label{all_spectra3} Contd. }
\end{figure}

\setcounter{figure}{4}
\begin{figure*}
\centering
\includegraphics[width=0.95\textwidth]{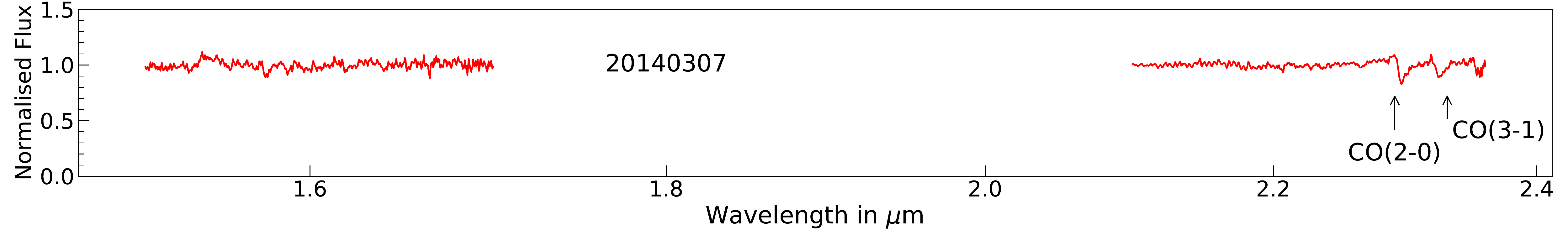}
\includegraphics[width=0.95\textwidth]{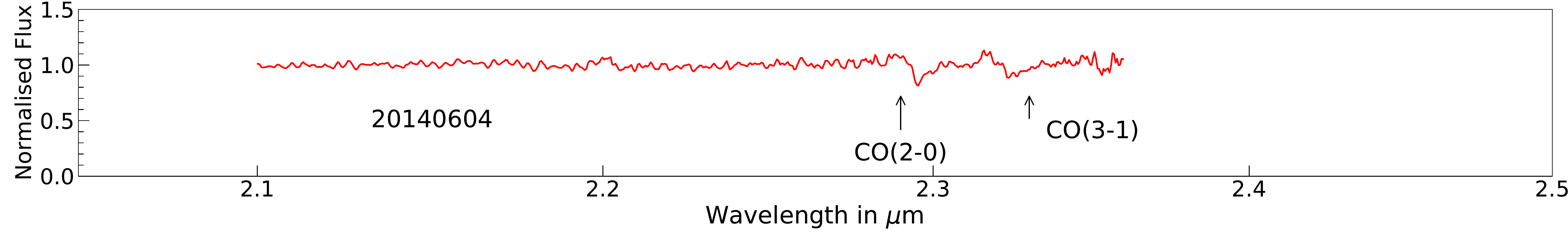}
\includegraphics[width=0.95\textwidth]{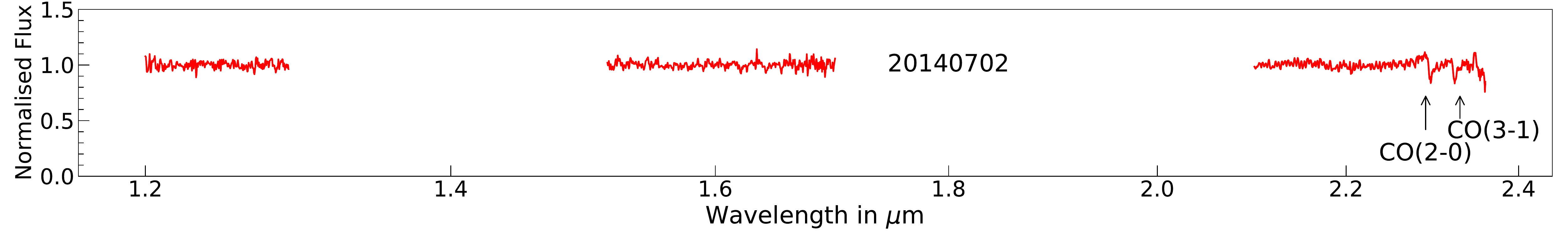}
\includegraphics[width=0.95\textwidth]{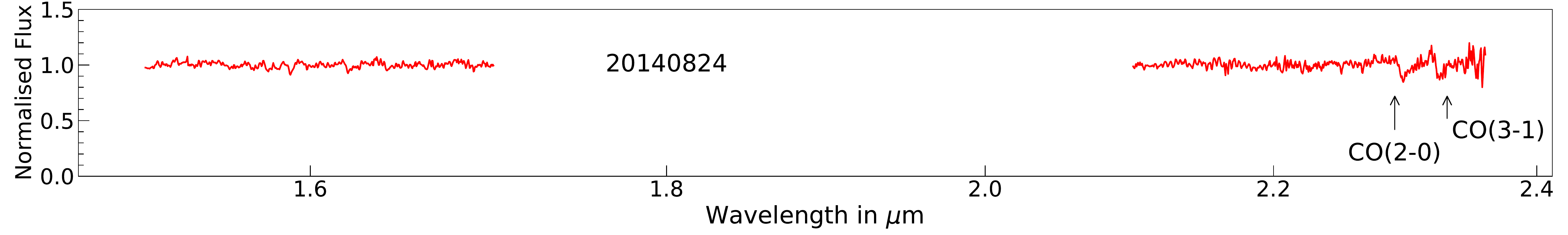}
\includegraphics[width=0.95\textwidth]{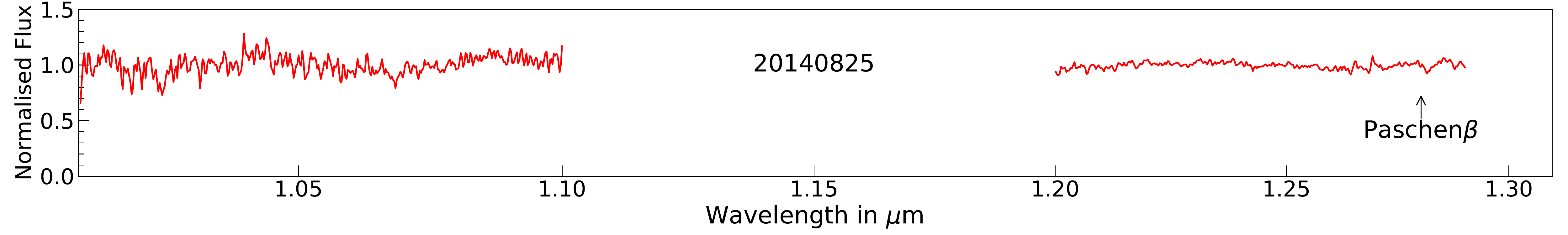}
\includegraphics[width=0.95\textwidth]{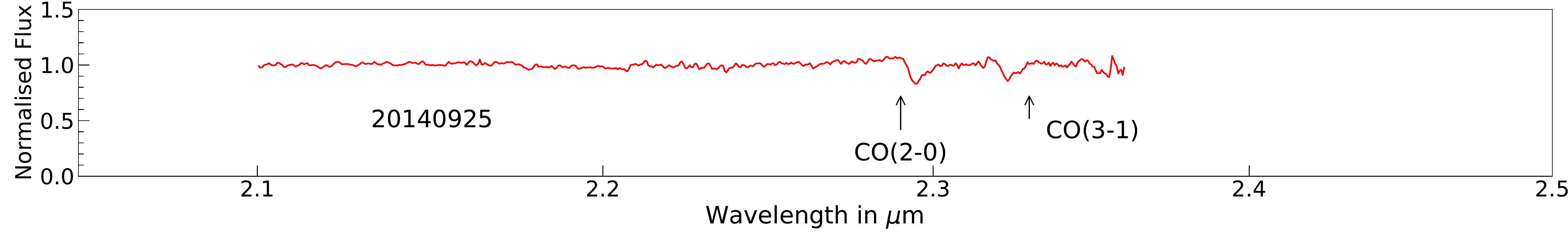}
\includegraphics[width=0.95\textwidth]{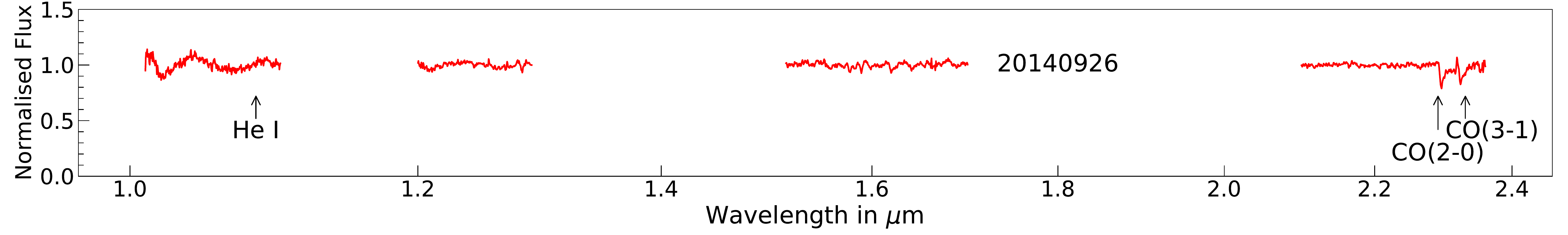}
\includegraphics[width=0.95\textwidth]{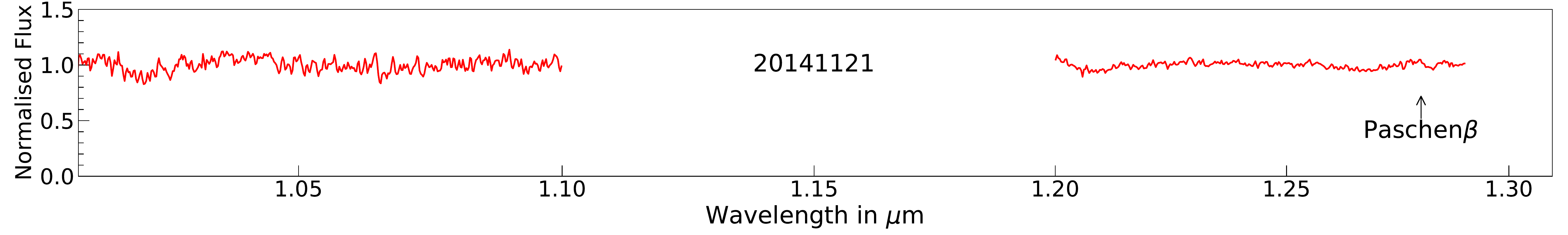}
\includegraphics[width=0.95\textwidth]{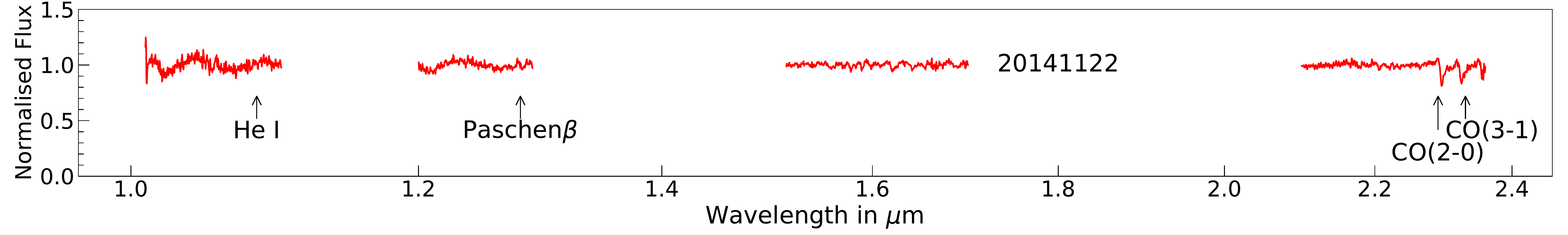}
\caption{\label{appndx3} Contd. }
\end{figure*}

\setcounter{figure}{4}
\begin{figure*}
\centering
\includegraphics[width=0.95\textwidth]{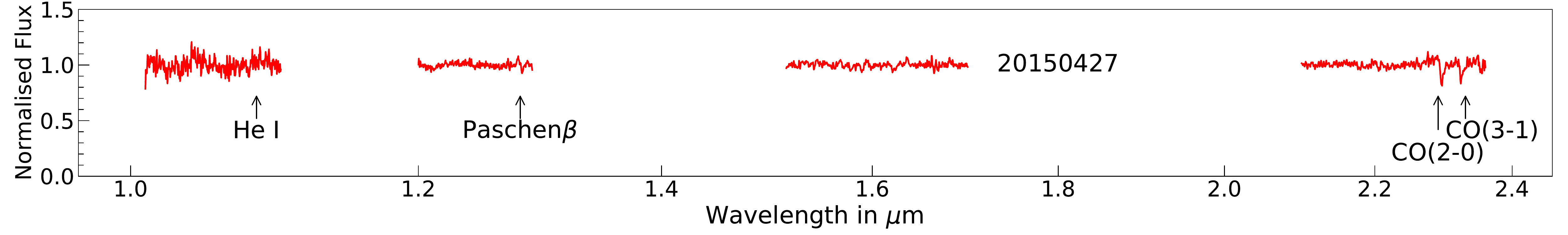}
\includegraphics[width=0.95\textwidth]{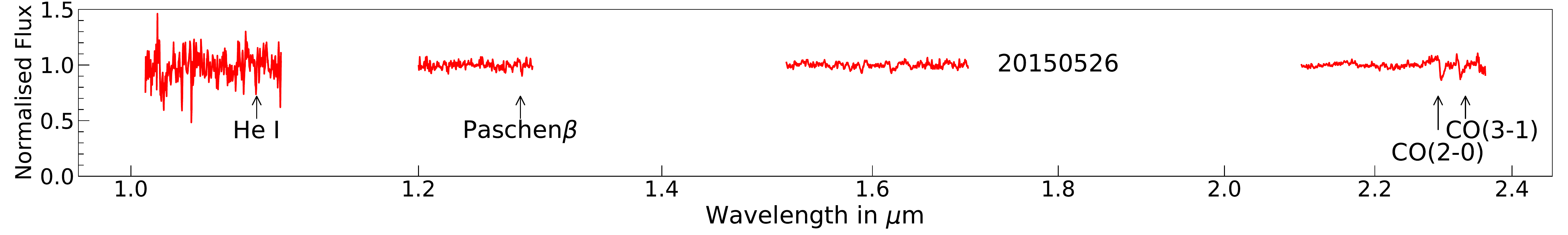}
\includegraphics[width=0.95\textwidth]{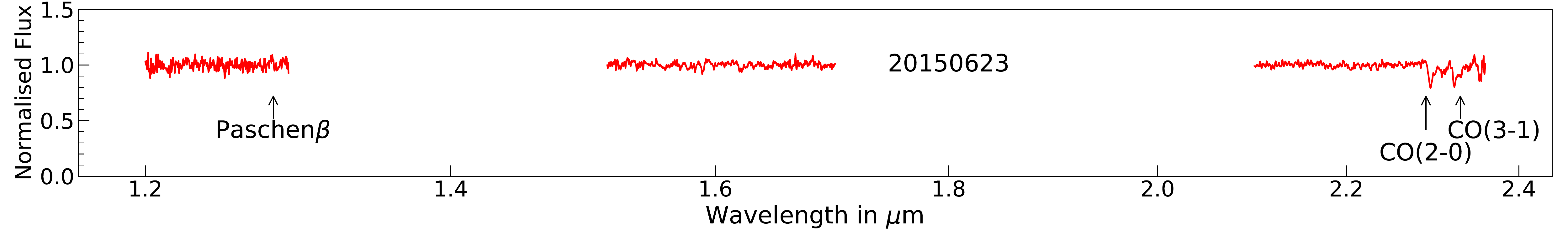}
\includegraphics[width=0.95\textwidth]{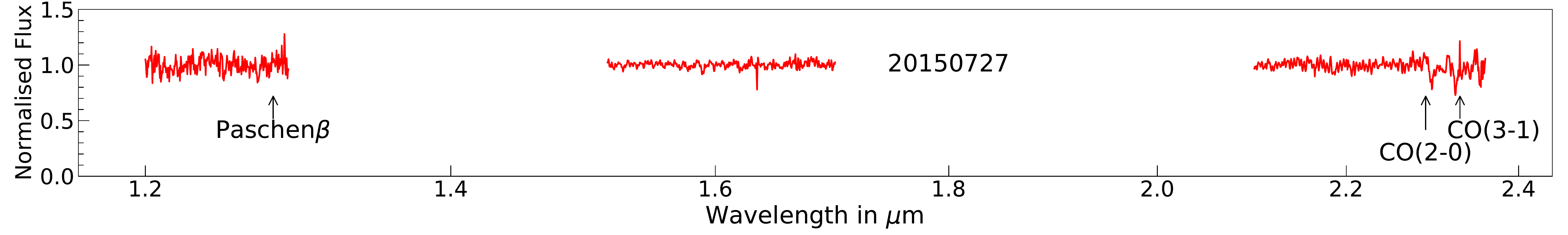}
\includegraphics[width=0.95\textwidth]{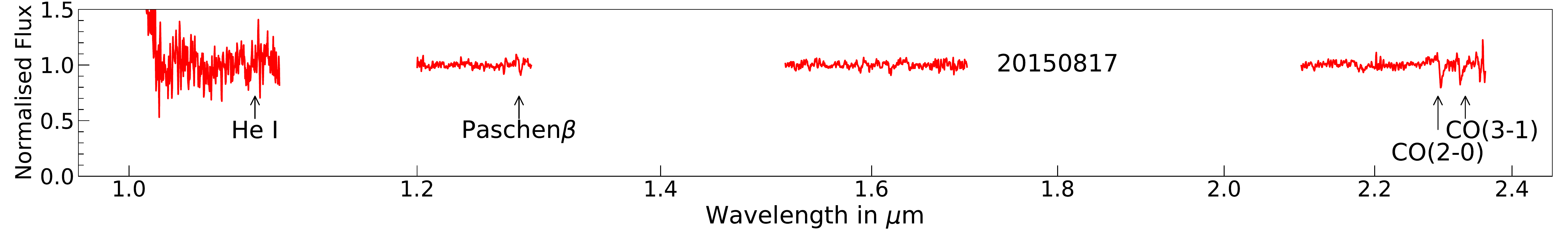}
\includegraphics[width=0.95\textwidth]{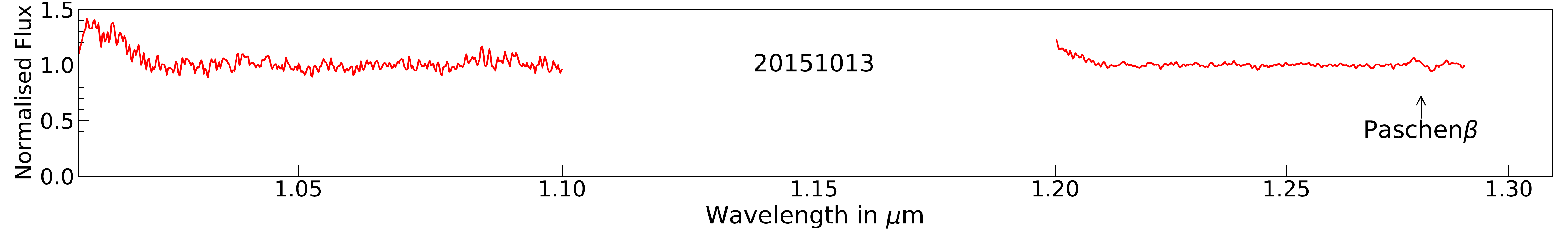}
\includegraphics[width=0.95\textwidth]{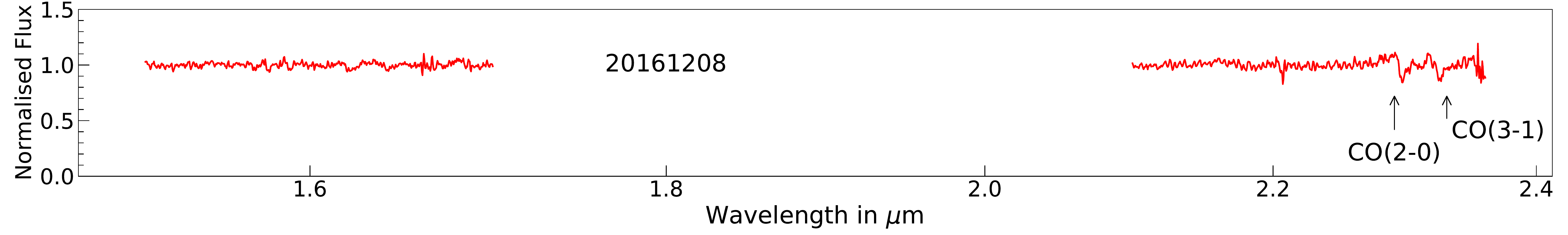}
\includegraphics[width=0.95\textwidth]{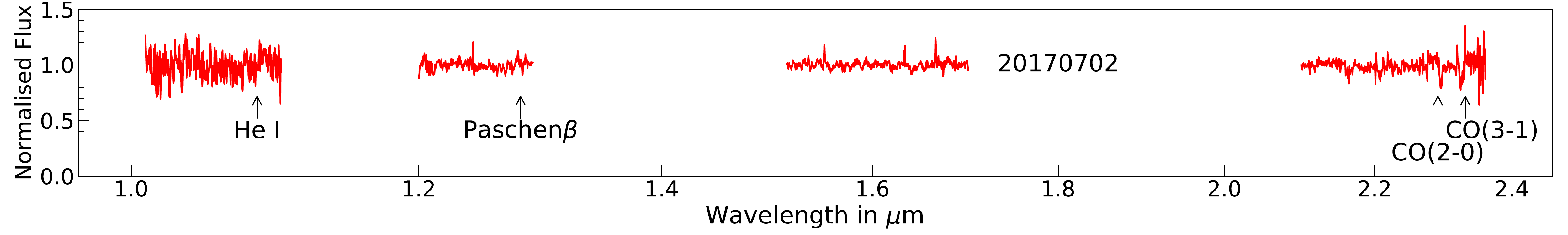}
\includegraphics[width=0.95\textwidth]{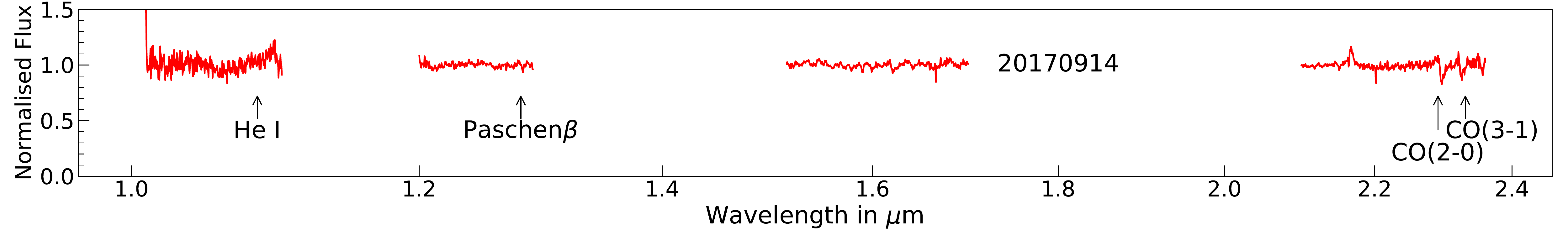}
\caption{\label{appndx4} Contd. }
\end{figure*}

\setcounter{figure}{4}
\begin{figure}
\centering
\includegraphics[width=0.95\textwidth]{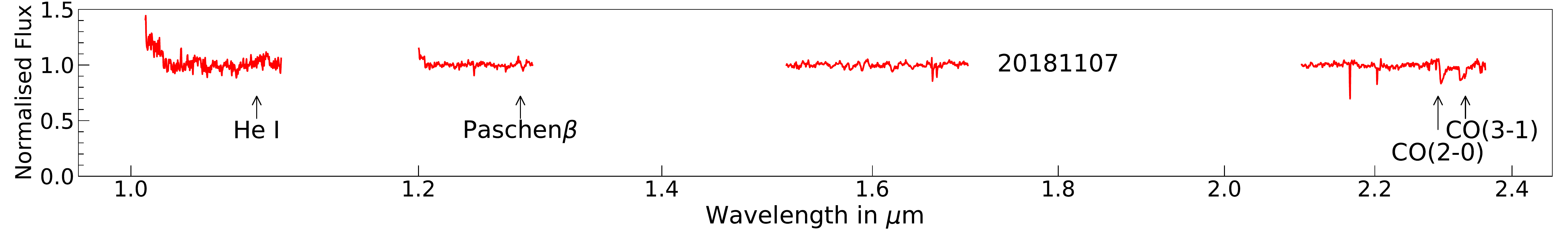}
\includegraphics[width=0.95\textwidth]{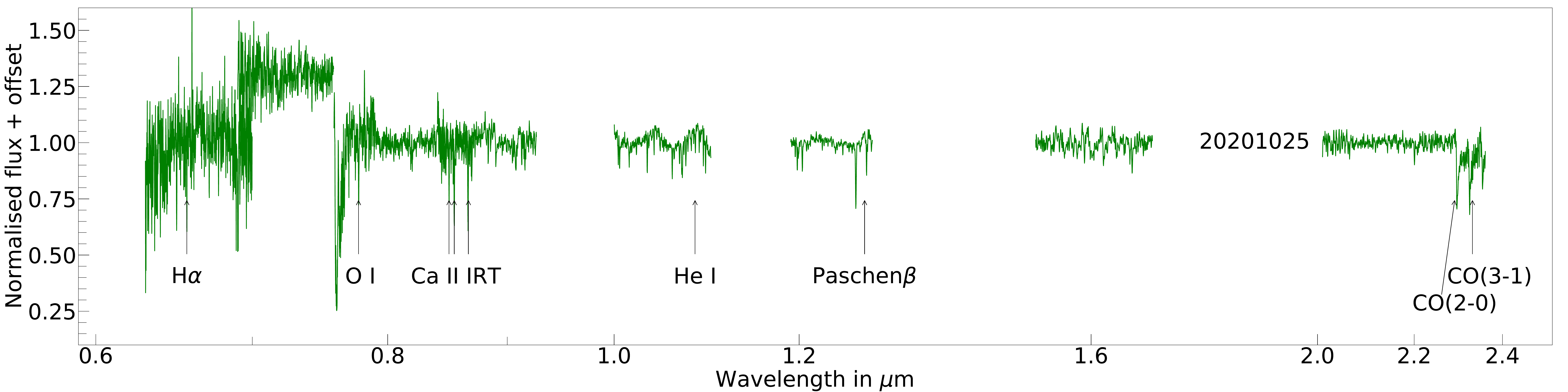}
\includegraphics[width=0.95\textwidth]{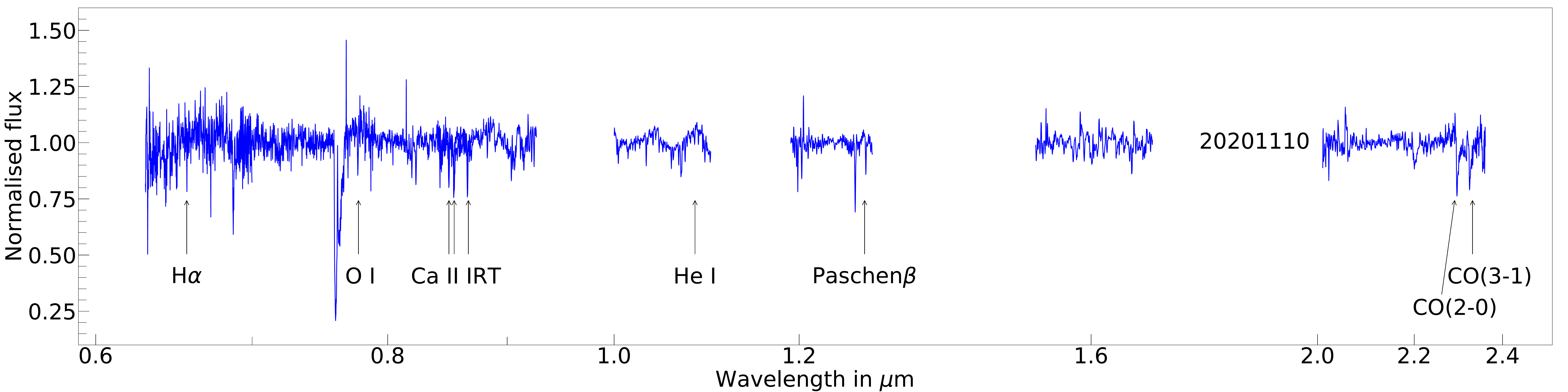}
\caption{\label{appndx5} Contd. }
\end{figure}



\setcounter{figure}{5}

\begin{figure}
\centering
\includegraphics[width=0.95\textwidth]{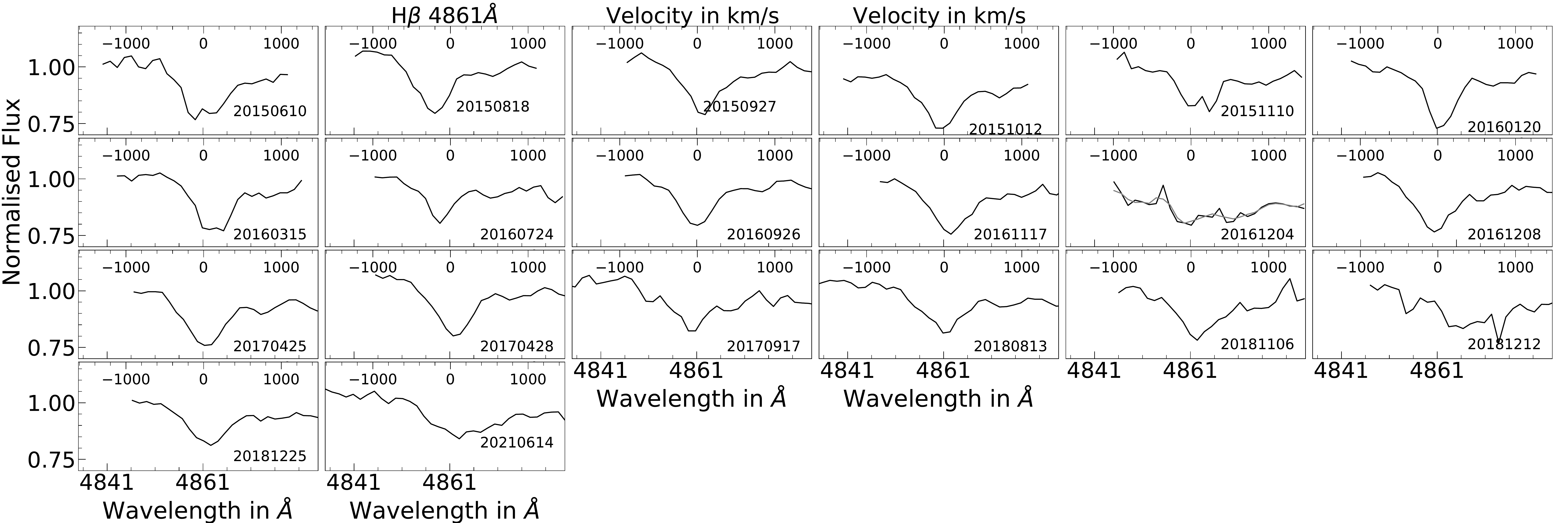}
\includegraphics[width=0.95\textwidth]{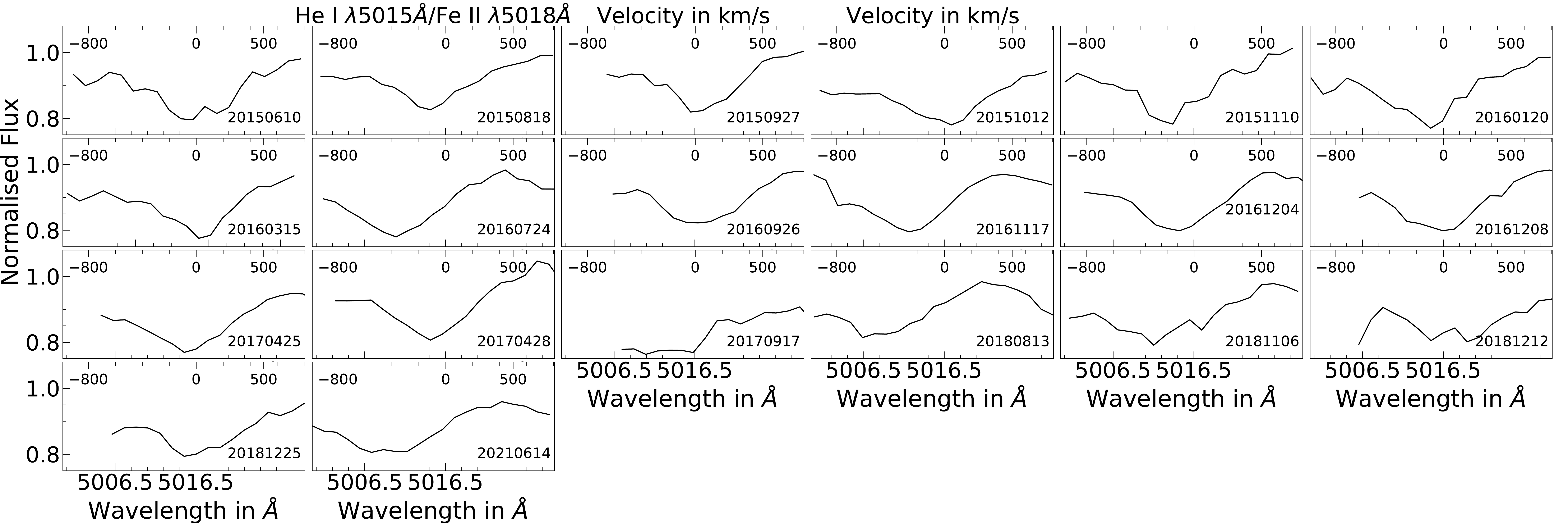}
\includegraphics[width=0.95\textwidth]{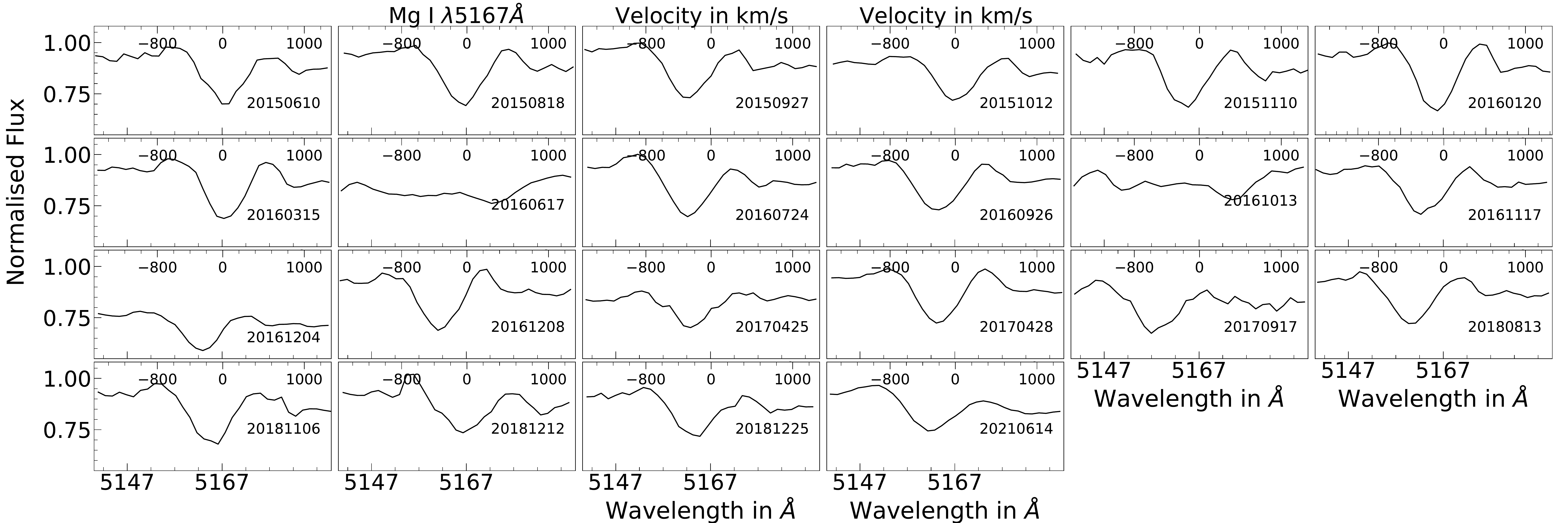}
\includegraphics[width=0.95\textwidth]{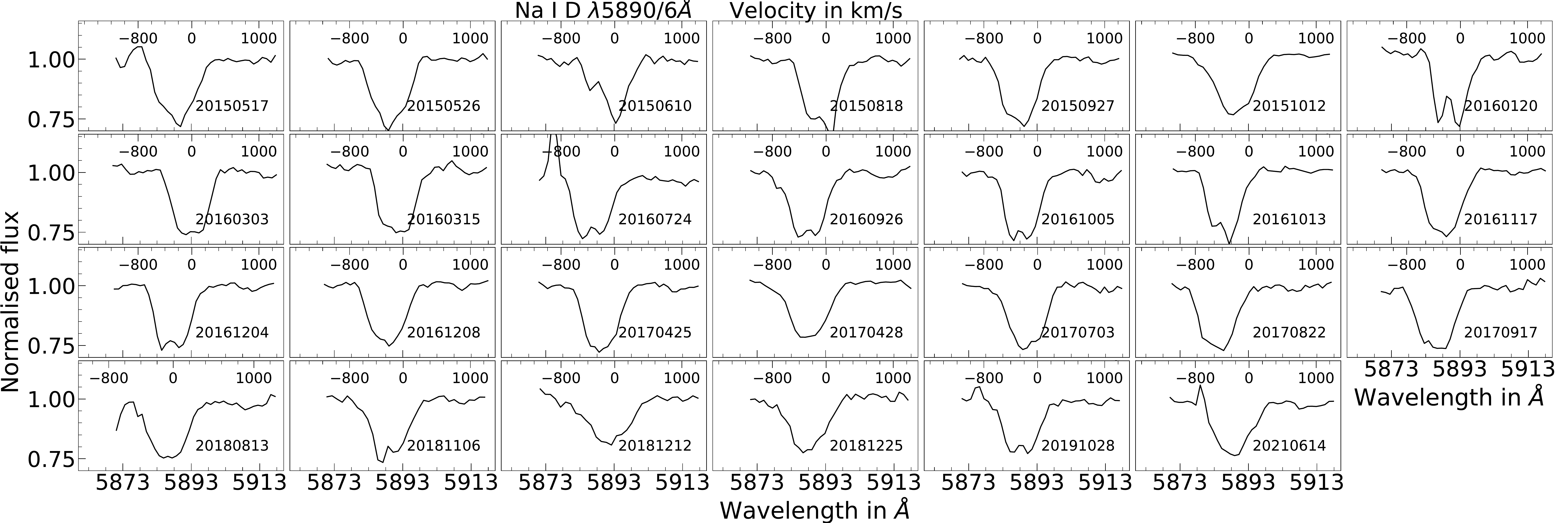}
\caption{\label{appndx1} The time evolution of the different line profiles in the spectra of V2493 Cyg during our monitoring period.
}
\end{figure}

\setcounter{figure}{5}
\begin{figure*}
\centering
\includegraphics[width=0.95\textwidth]{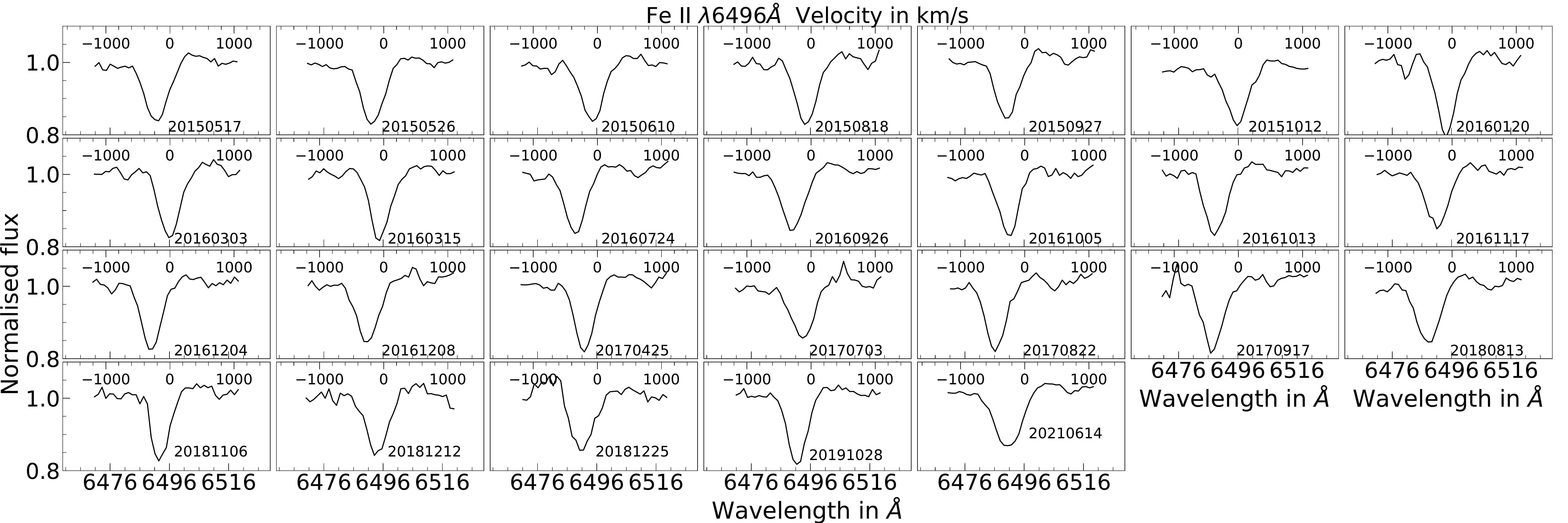}
\includegraphics[width=0.95\textwidth]{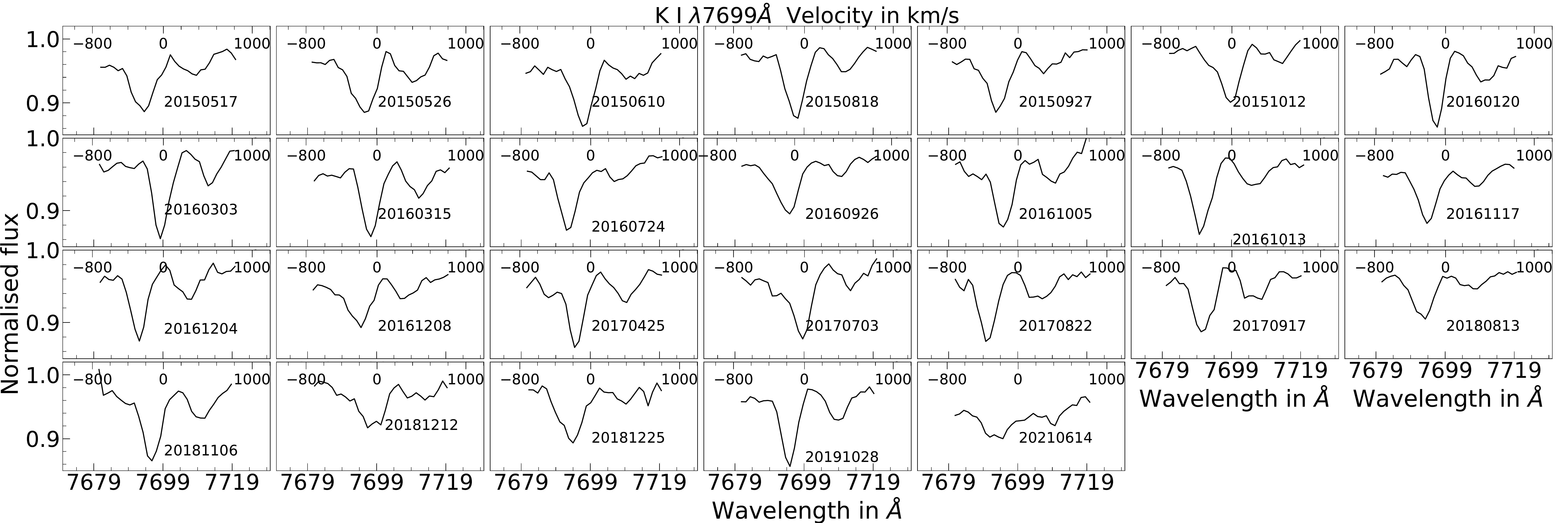}
\includegraphics[width=0.95\textwidth]{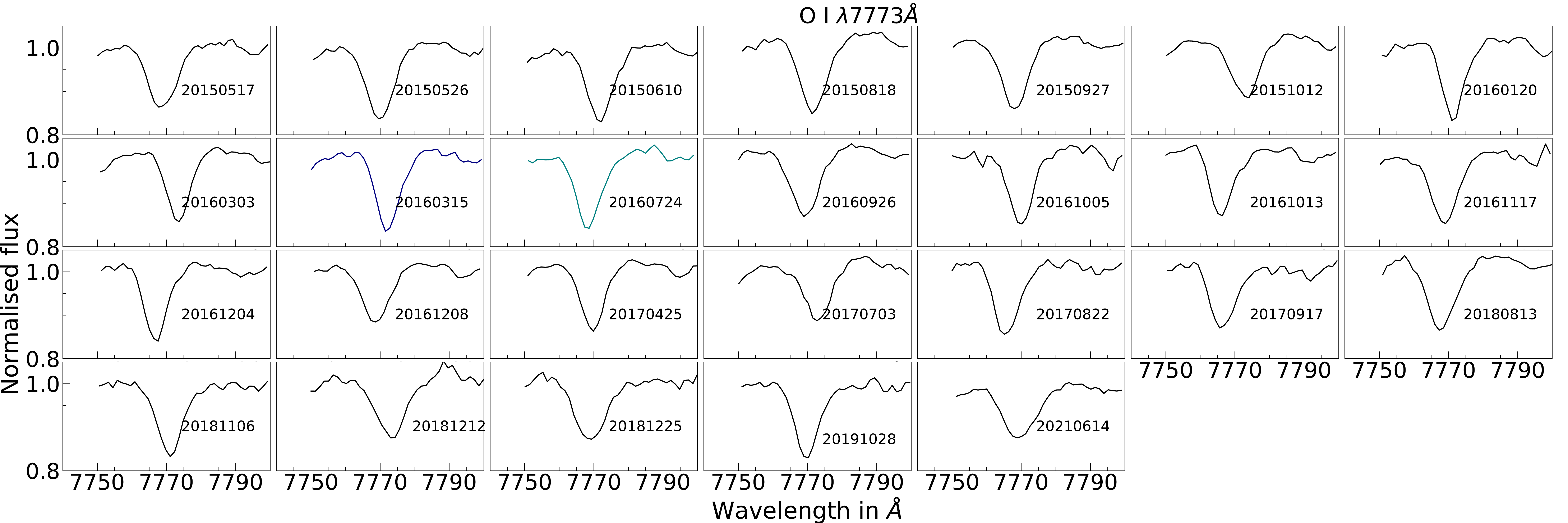}
\includegraphics[width=0.95\textwidth]{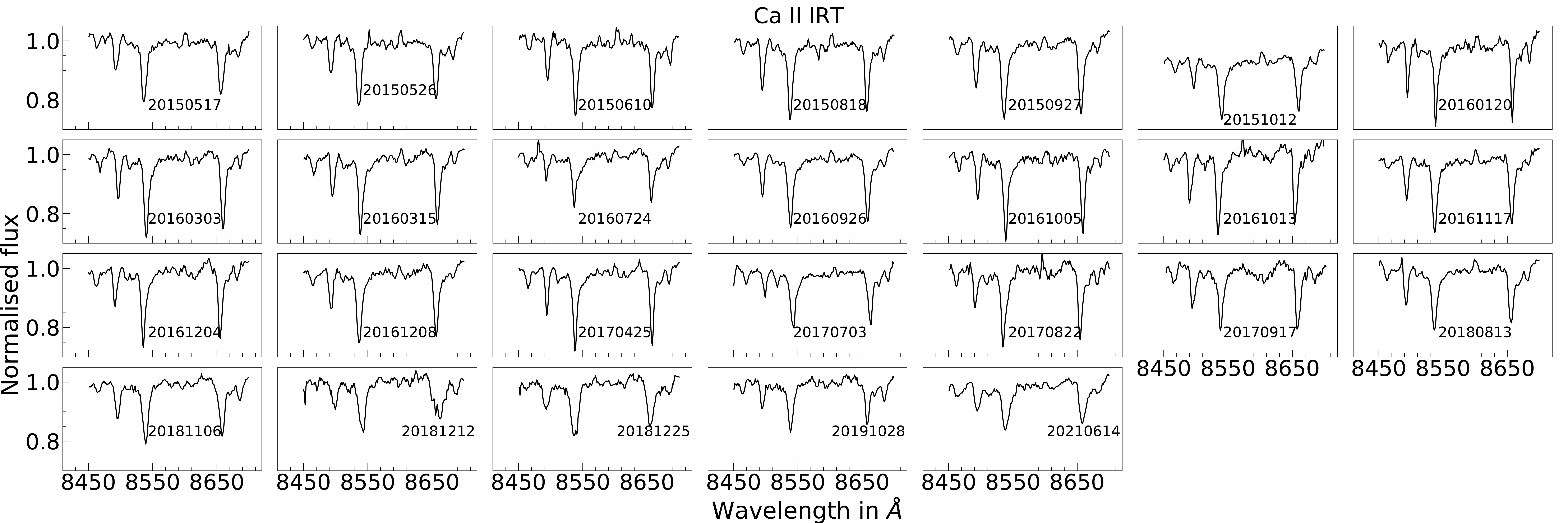}
\caption{\label{appndx2} Contd.
}
\end{figure*}

\setcounter{figure}{5}

\begin{figure}
\centering
\includegraphics[width=0.95\textwidth]{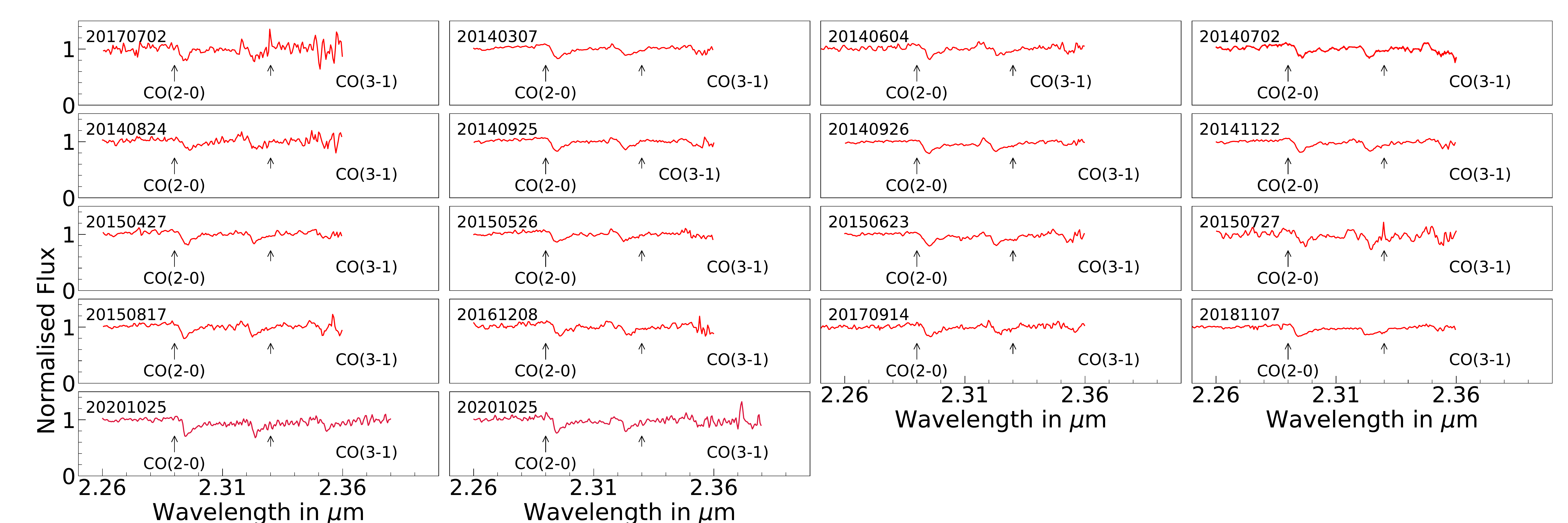}
\includegraphics[width=0.95\textwidth]{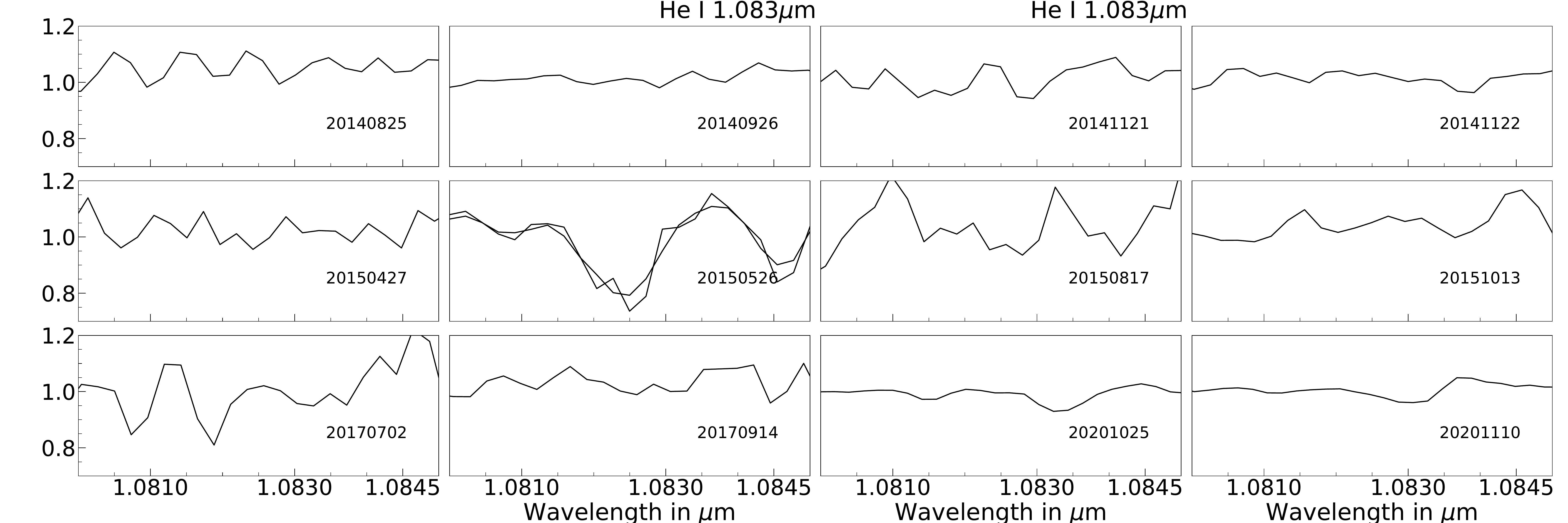}
\includegraphics[width=0.95\textwidth]{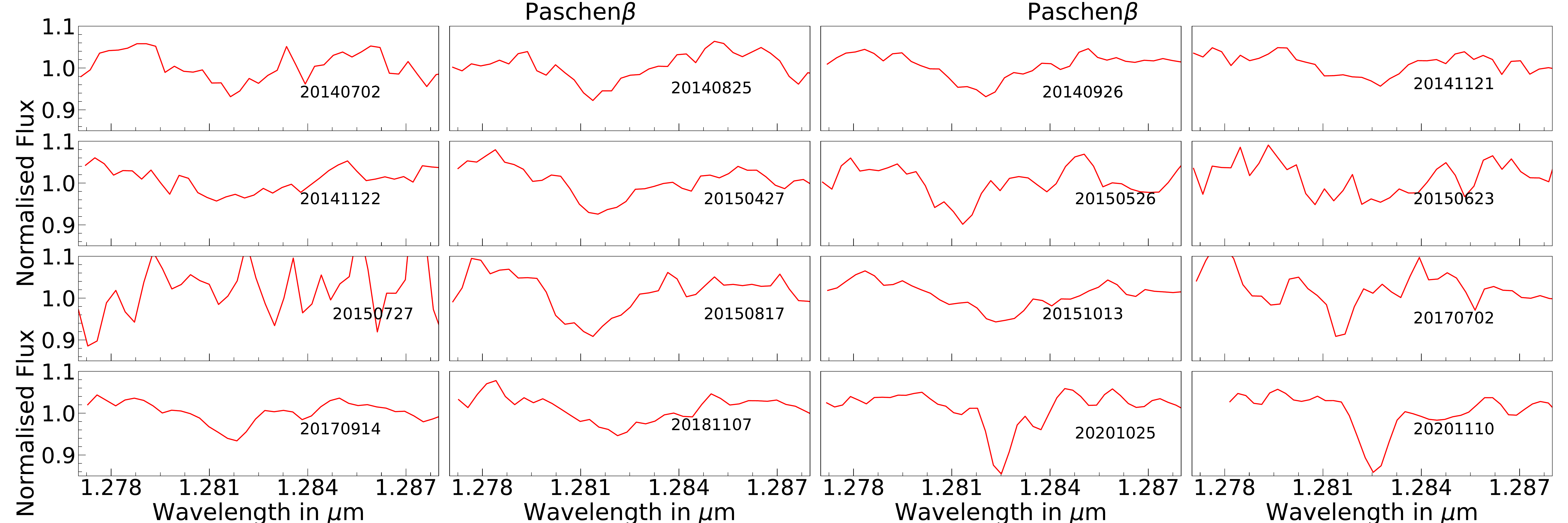}
\caption{\label{appndx5} Contd.
}
\end{figure}





\end{document}